\documentclass[preprint,11pt,superscriptaddress]{revtex4-2}
\setcitestyle{numbers,super,sort&compress}
\usepackage[main=english,american,british,UKenglish]{babel}
\usepackage{hyperref}
\usepackage{graphicx}
\usepackage{subcaption}
\usepackage{threeparttable}
\usepackage{booktabs}
\usepackage{amsmath}
\allowdisplaybreaks
\usepackage[margin=1in]{geometry}
\usepackage{array}
\usepackage{makecell}
\usepackage{caption}
\AtBeginDocument{\renewcommand{\selectlanguage}[1]{}}

\begin{document}
\title{Controlling the phase behaviour of ultraconfined water via bilayer graphene stacking}
\author{Yixuan Pu}
\affiliation{
    Department of Physics and Astronomy, University College London, 7-19 Gordon St, London WC1H 0AH, UK
}
 \affiliation{
    Thomas Young Centre and London Centre for Nanotechnology, 9 Gordon St, London WC1H 0AH, UK
}
\author{Benjamin X. Shi}
\affiliation{Initiative for Computational Catalysis, Flatiron Institute, 160 5th Avenue, New York, NY 10010, USA}
\author{Pavan Ravindra}
\affiliation{Department of Chemistry, Columbia University, 3000 Broadway, New York, NY 10027, USA.}
\author{Chris Pickard}
\affiliation{
Department of Materials Science \& Metallurgy, University of Cambridge, 27 Charles Babbage Road, Cambridge, UK
}
\affiliation{
Advanced Institute for Materials Research, Tohoku University 2-1-1 Katahira, Aoba, Sendai, Japan
}
\author{Angelos Michaelides}
\email{am452@cam.ac.uk}
\affiliation{Yusuf Hamied Department of Chemistry, University of Cambridge, Lensfield Road, Cambridge, CB2 1EW, UK}
\affiliation{
Lennard-Jones Centre, University of Cambridge, Trinity Ln, Cambridge, CB2 1TN, UK
}
\author{Venkat Kapil}
\email{v.kapil@ucl.ac.uk}
\affiliation{
    Department of Physics and Astronomy, University College London, 7-19 Gordon St, London WC1H 0AH, UK
}
\affiliation{
    Thomas Young Centre and London Centre for Nanotechnology, 9 Gordon St, London WC1H 0AH, UK
}
\begin{abstract}
\noindent Water confined within nanoscale capillaries exhibits phase behaviour and transport properties that differ substantially from bulk, and these effects are commonly interpreted as consequences of geometric confinement and reduced dimensionality.
Here we show that confinement topology alone is insufficient to predict the behaviour of nanoconfined water.
Using machine learning interatomic potentials with first-principles accuracy, we compute the density–temperature phase diagram of water confined within bilayer graphene nanocapillaries and compare AA and AB stacking arrangements, which differ only by a lateral shift of~1.4~\AA{}.
Despite this minimal structural change, AA stacking can stabilise different ice polymorphs, can increase the melting temperature by more than 100 K, can enhance proton transfer, and alters the onset of superionic behaviour relative to AB stacking.
We trace these effects to stacking-induced changes in the hydrogen-bond network associated with modifications to the lateral free energy landscape and neighbouring O--O separations.
Our results demonstrate that even subtle atomistic variations in the confining walls can qualitatively reshape the physical and chemical behaviour of nanoconfined water, with implications for the interpretation and control of fluids under angstrom-scale confinement.
\end{abstract}
\maketitle
\newpage
\section{Introduction}
\noindent A growing body of evidence demonstrates that the properties of water when confined in nanometre-sized capillaries deviate markedly from those of bulk water~\cite{faucher_critical_2019, aluru_fluids_2023, trushin_structure_2025,bocquet_nanofluidics_2020}.
Experiments report anomalies including enhanced and selective transport~\cite{majumder_enhanced_2005, whitby_fluid_2007, holt_fast_2006, peng_publisher_2018}, low interfacial friction~\cite{falk_molecular_2010, secchi_massive_2016}, high osmotic pressure gradients~\cite{siria_giant_2013, siria2017new}, and altered dielectric properties~\cite{fumagalli_anomalously_2018, wang_-plane_2025}.
These properties often depend on the size and chemical nature of the confining material and as such provide opportunities to \textit{engineer} properties of aqueous systems, placing nanoconfined water at the forefront of technologies such as water desalination and purification~\cite{surwade_water_2015, radha_molecular_2016, esfandiar_size_2017, abraham_tunable_2017}, drug delivery~\cite{chakravarty_delivery_2010, geng_stochastic_2014}, and clean energy~\cite{siria2017new, dong_situ_2018, zuo_near-frictionless_2023}.
A central quantity underlying this range of properties is the phase behaviour of nanoconfined water, which has attracted widespread interest from experiments and atomistic simulations. \\
\noindent To date, investigations of the phase behaviour of nanoconfined water have largely treated nanoconfinement as a \textit{topological confinement} effect, i.e., as arising from restricting the motion of water to lower dimensions.
Accordingly, experiments and simulations have focused on identifying how confinement width and dimensionality alter the stable phases of water.
On the experimental side, transmission electron microscopy of water encapsulated between graphene sheets suggested a dense “square phase” of ice that can form mono-, bi-, and tri-layer structures~\cite{algara-siller_square_2015, zhou_observation_2015,algara-siller_algara-siller_2015, wang_wang_2015}.
More recently, scanning probe and nanoscale nuclear magnetic resonance techniques have identified a solid--liquid transition near ambient conditions for water confined by hexagonal boron nitride and a hydrophilic diamond surface, indicating that there exists a critical confinement width of around 2 nm that can trigger changes in the phase behaviour of water~\cite{zheng_experimental_2026}.
On the simulation side, molecular dynamics (MD) driven by empirical force fields predicted a variety of novel monolayer phases and anomalies in phase transitions~\cite{koga_freezing_1997, han_phase_2010, zhao_highly_2014, zhao_ferroelectric_2014, sobrino_fernandez_mario_aa-stacked_2015, zhu_compression_2015, zhu_two-dimensional_2016, corsetti_enhanced_2016, zubeltzu_continuous_2016, raju_phase_2018}, though the results depended strongly on the force field~\cite{li_replica_2019}.
Subsequently, first-principles studies based on density functional theory (DFT) identified unconventional metastable phases comprised of pentagonal, square, and rhombic networks~\cite{chen_two_2016, corsetti_structural_2016}, with the thermodynamically stable phases sensitive to the DFT functional used~\cite{chen_evidence_2016}.
More recently, employing quantum Monte Carlo to identify an appropriate DFT functional and machine-learning potentials trained to DFT~\cite{kapil_first-principles_2022}, the temperature--pressure phase diagram was mapped at first-principles level accuracy.
These simulations uncovered rich behaviours, including a non-monotonic pressure-dependent melting temperature and hexatic and superionic phases at high pressure and temperature~\cite{kapil_first-principles_2022, lin_temperature-pressure_2023, jiang_rich_2024,kapil_author_2025}, and further provided insight into the dielectric response and autoionisation of nanoconfined water~\cite{ravindra_quasi-one-dimensional_2024, dufils_origin_2024, advincula_how_2025, ravindra_nuclear_2026}. \\
\noindent This body of work establishes that the phase behaviour of water can be controlled by restricting the motion of water to lower dimensions.
However, an alternative hypothesis is that nanoconfinement is not fully determined by geometry alone, and that the chemical nature and atomistic morphology of the confining material can themselves influence the phase diagram.
This hypothesis is well motivated by studies of interfacial water, where the nature of the surface is known to alter the structure and hydrogen-bonding of water~\cite{bjorneholm_water_2016}.
More recently, spectroscopy on aqueous electrolytes confined between graphene and CaF\textsubscript{2} sheets showed that interfacial effects dominate at nanometre separations, while geometric confinement alters the hydrogen-bond network only below $\sim$8~\AA~\cite{wang_interfaces_2025}.
This suggests that beyond \textit{topological confinement}, interfacial effects can also influence confined water.
With this in mind, recently, experimental techniques have advanced to enable exquisite control over the stacking of two-dimensional (2D) heterostructures, whose electronic properties are well known to depend sensitively on the relative arrangement of their constituent monolayers~\cite{frisenda_recent_2018,zhang_twist_2021,fox_stacking_2024}.
This provides a timely opportunity to investigate whether surface morphology can similarly be used to control the phase behaviour of nanoconfined water, for instance by modifying the stacking arrangement of the graphene nanocapillary walls.\\
\noindent In this work, we aim to understand how the morphology of the confining material influences the phase diagram of nanoconfined water using atomistic simulations with first principles accuracy driven by machine learning interatomic potentials (MLIPs).
We focus on the simplest nanocapillary, \textit{viz.} water confined by bilayer graphene.
This is an instructive system because graphene is conventionally regarded as atomically smooth and thus its morphology is not expected to strongly perturb the phase behaviours of nanoconfined water.
We investigate water of different densities in
nanocapillaries formed by AA- and AB-stacked graphene.
We find that while AB stacking yields a phase diagram broadly consistent with that predicted for water under an atomistically smooth confining potential \cite{kapil_first-principles_2022}, AA stacking alters the stable phases and raises the melting temperature by up to $\sim$100~K.
Stacking also affects the dynamics of water and its autoionisation in an interesting way: compared to AB stacking, AA stacking slows water self-diffusion but accelerates proton mobility.
We trace these differences to stacking-induced changes in hydrogen bonding, which provides a unifying microscopic explanation that is also observable in the phonon density of states (DOS) of the confined ice phases.
Our results demonstrate that subtle atomistic variations in the confining walls can qualitatively alter the behaviour of nanoconfined water, with implications for the interpretation and control of fluids under angstrom-scale confinement.
\begin{figure*}[htbp]
\includegraphics[width=\textwidth]{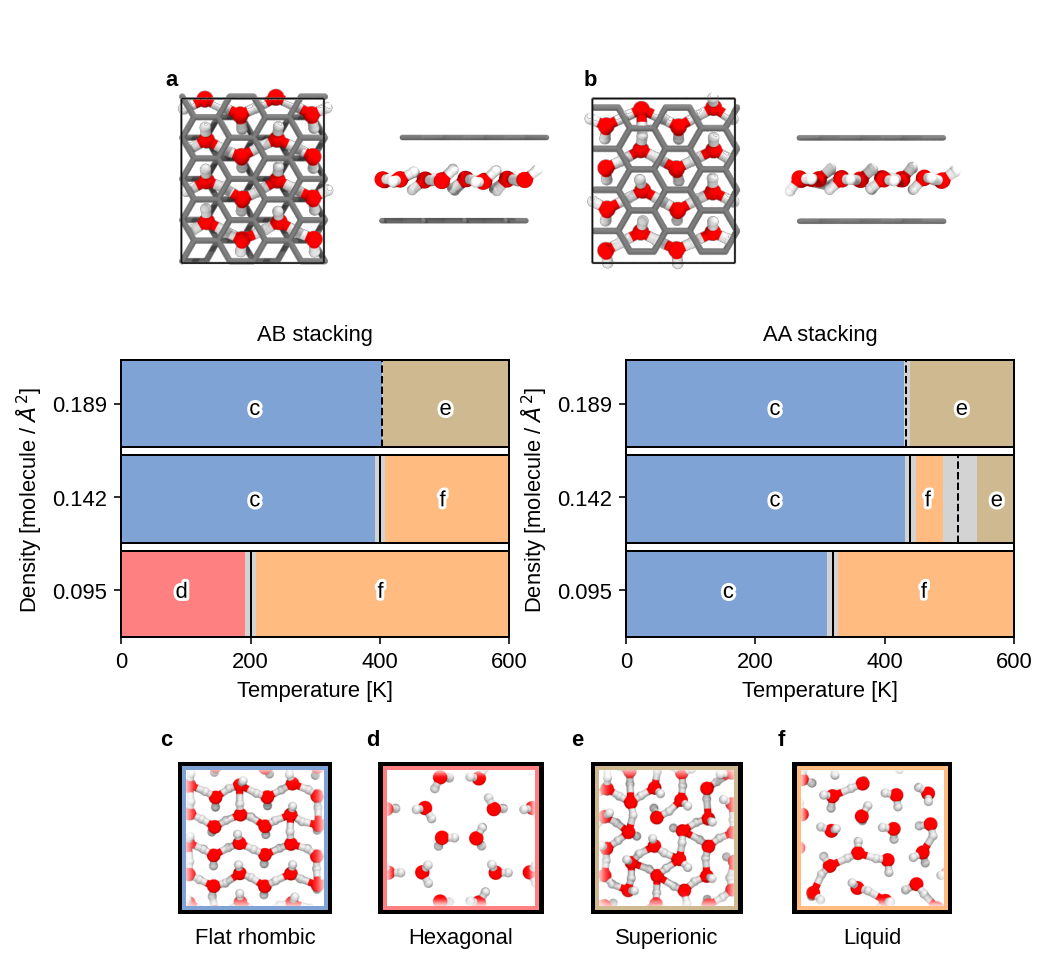}
\caption{
\textbf{Strong stacking dependence on the phase diagram of nanoconfined water.} The temperature-density phase diagram of water confined in AB-stacked (left) and  AA-stacked (right) nanocapillaries.
Transition temperatures are shown as vertical black lines.
Solid lines indicate a first-order phase transition while dashed lines indicate a continuous phase transition.
The grey regions indicate  uncertainties associated with phase transition temperatures.
\textbf{a,b}, Schematic illustration of top and side view of water molecules confined in AA- (\textbf{a}) and AB-stacked (\textbf{b}) graphene nanocapillaries.
\textbf{c-f}, The snapshots show stable structures in different parts of the phase diagrams. Diagram of the flat rhombic (\textbf{c}), hexagonal (\textbf{d}), superionic (\textbf{e}) and liquid phases (\textbf{f}) are shown with oxygen atoms in red, carbon atoms in dark grey, and hydrogen atoms in light grey. The graphene layers are omitted for clarity to better visualise the arrangement of the water molecules. }
\label{fig:phase_diagram}
\end{figure*}
\section{Results and discussion}
\noindent We explore the phase behaviour of water confined in 5~\AA{}-wide bilayer graphene nanocapillaries using a rigorous simulation framework that combines hybrid-DFT-trained MLIPs, large-scale global structure searches~\citep{pickard_ab_2011} with DFT validation, MD~\citep{rahman_correlations_1964}, and path-integral MD~\citep{parrinello_study_1984}.
\noindent \subsection{Graphene stacking alters the stability of nanoconfined ice phases}
\noindent The resulting temperature-density phase diagrams for AB- and AA-stacked nanocapillaries are shown in Fig.~\ref{fig:phase_diagram}. To begin, we analyse the impact of stacking on the thermodynamically stable ice phases observed in the low-temperature region of the phase diagrams.
We first consider the AB-stacked nanocapillary, corresponding to the natural stacking arrangement of bilayer graphene, before discussing the AA-stacked nanocapillary.
In each case, we examine three density regimes chosen to facilitate comparison with the low-, intermediate-, and high-pressure regions explored previously for water under uniform confinement at the same 5~\AA{} confinement width in Ref.~\citenum{kapil_first-principles_2022}. \\
\noindent The lowest density considered here corresponds to the simulated densities associated with near-ambient lateral pressures in Ref.~\citenum{kapil_first-principles_2022}.
In this regime, AB stacking stabilises a hexagonal phase.
The intermediate density is representative of the densities encountered in the few-GPa regime of Ref.~\citenum{kapil_first-principles_2022}, which was used there to model the lateral pressures expected for water encapsulated between graphene sheets in experiments~\cite{algara-siller_square_2015}.
Here, AB stacking stabilises a flat-rhombic phase.
The highest density considered here corresponds to densities associated with lateral pressures higher than those explored by~\citet{kapil_first-principles_2022} and potentially overlapping with those considered by~\citet{jiang_rich_2024}.
At this density, the flat-rhombic phase remains stable.
Interestingly, these results are consistent with the phases observed for water under uniform confinement, i.e., when water--graphene interactions are represented by a laterally uniform external potential.
The hexagonal phase in the low-density regime and the flat-rhombic phase at intermediate density are equivalent to those observed at low and intermediate pressures in Ref.~\citenum{kapil_first-principles_2022}, while the persistence of the flat-rhombic phase at high density is consistent with the high-pressure phase observed in Ref.~\citenum{jiang_rich_2024}. \\
\noindent We next consider the phase behaviours of water in an AA-stacked bilayer graphene nanocapillary.
While this capillary differs from the canonical AB-stacked bilayer graphene only by a global lateral translation of one graphene layer by $\sim$1.42~\AA{}, it is sufficient to alter the stable phases.
Interestingly, the flat-rhombic phase is identified as the stable phase in all three density regimes.
Thus
we observe a qualitative difference between AA and AB stacking in the low-density regime: AB favours the hexagonal phases whereas AA favours the flat-rhombic phase.
In contrast, at higher densities, stacking does not qualitatively change the nature of the stable phases, aside from small structural differences.
As shown in Table~\ref{tab:OO_distance}, in the intermediate-density regime, the O--O distances associated with the flat-rhombic phase are smaller by around 0.02~\AA{}, whereas this difference becomes negligible in the high-density regime.
\subsection{AA stacking increases the melting temperature relative to AB stacking}
\noindent We next investigate how the phase behaviour of nanoconfined water at finite temperature is influenced by the stacking of bilayer graphene, starting with solid--liquid phase transitions.
Solid--liquid transition temperatures are identified by analysing the temperature dependence of the ensemble-averaged potential energy, the self-diffusion coefficient of oxygen atoms, and the 2D spatial distribution functions of oxygen atoms from MD simulations for discontinuities or abrupt changes.
The temperature dependence of the ensemble-averaged potential energy and oxygen self-diffusion coefficient is shown in Fig.~\ref{fig:U_D_plot}, while the 2D spatial distribution functions of oxygen atoms are shown in Figs.~\ref{fig:2DAA08}--\ref{fig:2DAB16}.
The inferred transition temperatures are indicated by solid lines in the two phase diagrams in Fig.~\ref{fig:phase_diagram}. \\
\noindent In the low-density regime, the most stable phase within AB-stacked bilayer graphene is the hexagonal phase.
As can be seen in Fig.~\ref{fig:phase_diagram} (left panel), this phase melts at 200~K which is in good agreement with the melting temperature of monolayer water under a uniform confining potential~\cite{kapil_first-principles_2022}.
Here, the transition is not accompanied by a clear discontinuity in the potential energy, but by an abrupt change in the spatial distribution function of oxygen atoms (Fig.~\ref{fig:2DAB08} of the supporting information (SI)).
In the intermediate-density regime, the most stable phase within AB-stacked bilayer graphene is the flat-rhombic phase.
The melting temperature is 400\,K and is accompanied by a sudden change in the ensemble-averaged potential energy.
This temperature is nearly 60\,K higher than that of monolayer water under a uniform confining potential.
At high density, the most stable phase within AB-stacked bilayer graphene is the flat-rhombic phase.
We do not observe a discernible change in the structure of the oxygen atoms with increasing temperature, and thus no evidence of melting up to 600\,K. \\
\noindent We now consider, how these phase transitions are affected by modifying the stacking of the nanocapillary from AB to AA (Fig. 1, right panel).
In the low-density regime, the most stable phase within AA-stacked bilayer graphene is the flat-rhombic phase.
We note a large stacking-induced change in the melting temperature.
The melting temperature of this phase is estimated to be 320~K.
This temperature is $>$ 100~K higher than in AB-stacked bilayer graphene and likewise $>$ 100~K higher than that of water under uniform confinement.
The phase transition is characterised by a sharp change in the potential energy accompanied by a jump in the diffusion coefficient (Fig.~\ref{fig:U_D_plot}).
In the intermediate-density regime, the most stable phase within AA-stacked bilayer graphene is the flat-rhombic phase.
The melting temperature of this phase is 440\,K, which is higher than the 400\,K observed for AB stacking.
This suggests a reduced but diminished impact of AA stacking increasing the melting temperature.
In the high-density regime, the most stable phase within AA-stacked bilayer graphene is the flat-rhombic phase.
Consistent with the case of AB stacking, we do not observe a discernible change in the structure of the oxygen atoms with increasing temperature, and thus melting, up to 600\,K for AA stacking.
\subsection{Graphene stacking controls ionic conductivity}
\begin{figure*}[htbp]
\centering
\includegraphics[width=\textwidth]{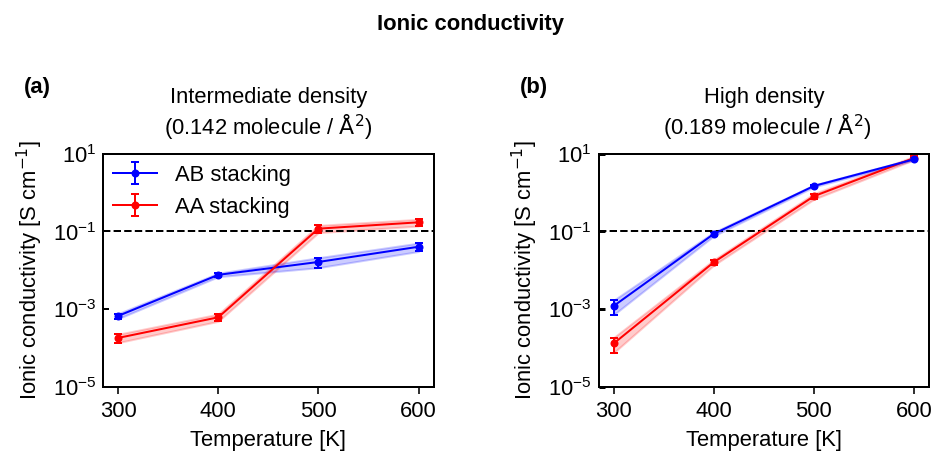}
\caption{\textbf{Stacking dependence on ionic conductivity of nanoconfined water.} \textbf{a,b}, Ionic conductivity as a function of temperature for intermediate-density phase (\textbf{a}) and high-density phase (\textbf{b}) water confined in AA- and AB-stacked graphene, indicating a superionic phase transition. The red lines represent AA stacking, the blue lines represent AB stacking. The horizontal dashed lines indicate the superionic threshold (0.1 S cm$^{-1}$).}
\label{fig:ionic_conductivity}
\end{figure*}
\noindent We next study the impact of stacking on phase transitions triggered by autoionisation in water -- specifically the transition to a superionic phase.
Incorporating quantum nuclear effects is important for characterising this phase and the associated transition.~\cite{ravindra_nuclear_2026}.
To estimate the onset temperature for superionic behaviour, we apply three criteria of increasing strictness.
First, using path-integral molecular dynamics, we compute the quantum free energy profile along a proton-transfer coordinate and identify temperatures at which the associated barrier becomes comparable to thermal fluctuations.
Second, we analyse the corresponding trajectories to verify the occurrence of  proton-transfer events.
Third, we estimate the quantum ionic conductivity using temperature-elevation path-integral coarse-graining simulations (Te PIGS)~\cite{musil_quantum_2022}, as shown in Fig.~\ref{fig:ionic_conductivity}, and identify the onset of superionic behaviour when the conductivity exceeds 0.1\,S/cm.
The inferred onset temperatures are indicated by dashed lines in the two phase diagrams in Fig.~\ref{fig:phase_diagram}. \\
\noindent As shown in Fig.~\ref{fig:PTCAA08} and ~\ref{fig:PTCAB08}, for both stackings considered, the free energy barrier along the proton-transfer coordinate at the lowest density is around six times thermal energy.
In this regime, we do not observe proton-transfer events on the ns timescale of our simulations, and therefore conclude that superionic behaviour is not present over the temperature range considered here.
At intermediate density, for both stackings, the free energy barrier along the proton-transfer coordinate decreases to around three times thermal energy (Fig.~\ref{fig:PTCAA12} and ~\ref{fig:PTCAB12}) and proton-transfer events are observed.
At high density, for both stackings, we observe pronounced proton delocalisation, with an approximately flat free energy profile along the proton-transfer coordinate (Fig.~\ref{fig:PTCAA16} and ~\ref{fig:PTCAB16}); correspondingly, we observe rapid proton-transfer events.
We thus analyse the quantum ionic conductivity in the intermediate- and high-density regimes. \\
\noindent We first consider the AB-stacked bilayer graphene nanocapillary.
As shown in Fig.~\ref{fig:ionic_conductivity}, in the intermediate density regime, the ionic conductivity increases markedly over the 300\,K--600\,K temperature range, but does not reach the superionic threshold.
This observation is consistent with the ionic conductivity of monolayer water under uniform confinement from constant lateral pressure simulations at around 3~GPa, whose equilibrium density is close to the intermediate density considered here~\cite{ravindra_nuclear_2026, coles_nanoconfined_2026}.
This system exhibits partial dissociation but does not reach the threshold for superionic behaviour.
In the high-density regime, the ionic conductivity (Fig.~\ref{fig:ionic_conductivity})  exceeds the superionic threshold above $\sim$300~K, indicating a transition to a superionic phase.
This phenomenon is also consistent with previous results under uniform confinement~\cite{kapil_first-principles_2022}, although the observed ionic conductivity is nearly an order of magnitude higher. \\
\noindent We now turn to the autoionisation-induced phase transitions in the AA-stacked bilayer graphene nanocapillary.
At intermediate density the ionic conductivity (Fig.~\ref{fig:ionic_conductivity}) increases strongly across the 300--600\,K temperature range and approaches the superionic threshold between 500\,K and 600\,K.
While we analyse the underlying origin in more detail below, these results suggest that AA stacking enhances water autoionisation relative to AB stacking at intermediate density at high temperatures.
In the high-density regime, the ionic conductivities (Fig.~\ref{fig:ionic_conductivity}) for AA and AB stacking are similar, consistent with our earlier observations of diminished sensitivity to stacking under high pressures.
\begin{figure*}[!t]
\centering
\includegraphics[width=\textwidth]{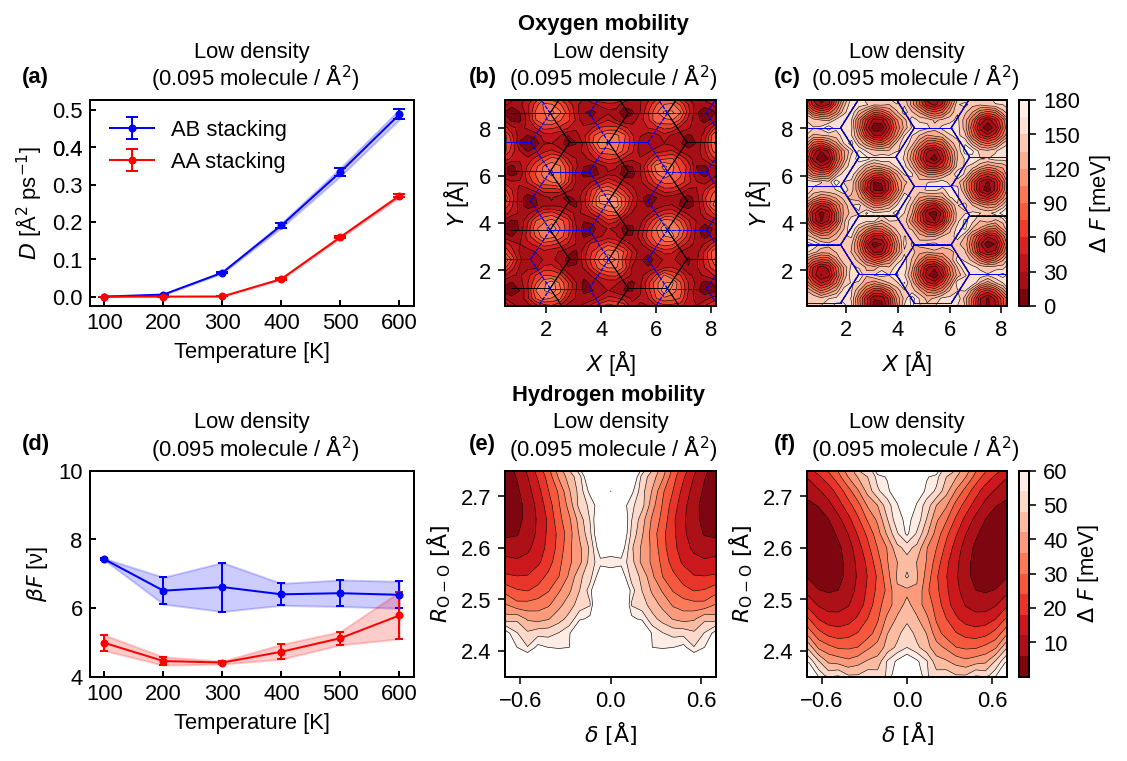}
\caption{\textbf{Graphene stacking influences both oxygen dynamics and proton transfer of nanoconfined water.} \textbf{a}, Comparison of diffusion coefficients of water confined in AB and AA-stacked nanocapillaries in the low density phase. The red line represents AA stacking, the blue line represents AB stacking. \textbf{b,c}, Free energy profiles of nanoconfined water along $x$ an $y$  calculated from the probability distribution of the position of water molecules in the low density phase at 600~K for water confined by AB stacking graphene (\textbf{b}) and AA stacking graphene(\textbf{c}). The $x$ and $y$ axes are defined in the plane of the graphene layer. The $x$ axis is chosen parallel to one set of horizontal C--C bonds, and the $y$ axis is perpendicular to $x$. Dark red indicates low energy regions, white indicates high energy regions. Larger corrugation is present on AA stacking graphene. \textbf{d}, Barrier along the proton transfer coordinate of water confined in AB and AA-stacked nanocapillaries in the low density phase at different temperatures. \textbf{e,f}, The free energy profile of proton transfer coordinate associated with O-O distance in the low density phase at 100~K for water confined by AB-stacked graphene (\textbf{e}) and AA-stacked graphene (\textbf{f}).}
\label{fig:4}
\end{figure*}
\subsection{AA stacking reduces oxygen diffusion but enhances proton diffusion}
\noindent Our investigation indicates that the stacking of bilayer graphene impacts the phase behaviour of water in interesting ways.
On the one hand, AA stacking is associated with a restrictive effect, increasing the melting temperature; on the other hand, it is associated with enhanced ionic conductivity.
To understand these trends in more depth, we investigate how the structural and dynamical properties of the oxygen and hydrogen atoms are influenced by the atomistic morphology of AA- and AB-stacked bilayer graphene. \\
\noindent We first consider oxygen diffusion.
The strongest stacking dependence is observed in the low-density regime.
As shown in Fig.~\ref{fig:4}(a), the diffusion coefficient of oxygen atoms is systematically higher -- by a factor of two to three --  in AB stacking than in AA stacking.
The same trend is observed for the intermediate-density liquid, although the difference is smaller (see Fig.~\ref{fig:Dmedium} of the SI); this analysis does not apply at high density because the oxygen atoms show negligible diffusion across the temperature range considered.
To rationalise this trend, we analyse the 2D free energy landscape associated with the lateral oxygen coordinates, folding the distribution into a reduced unit cell.
Figures~\ref{fig:4}(b) and (c) show this landscape for water in AB- and AA-stacked nanocapillaries at 600~K.
While water is mobile in both cases, in AA stacking the low-energy regions are separated by barriers of $\sim$180~meV, compared to $\sim$100~meV in AB stacking.
The higher barriers in AA stacking are associated with transient localisation and suppressed diffusion, whereas the smoother landscape in AB stacking is associated with more facile lateral motion.
This difference follows from the atomistic morphology of the nanocapillaries.
In AA stacking, carbon atoms lie directly above one another along the normal to the confinement plane, creating narrow interstitial regions that strongly repel water.
In AB stacking, the two graphene layers are laterally offset which reduces the number of strongly repulsive interstitial regions and thereby lowering lateral barriers. \\
\begin{figure*}[!t]
\centering
\includegraphics[width=\textwidth]{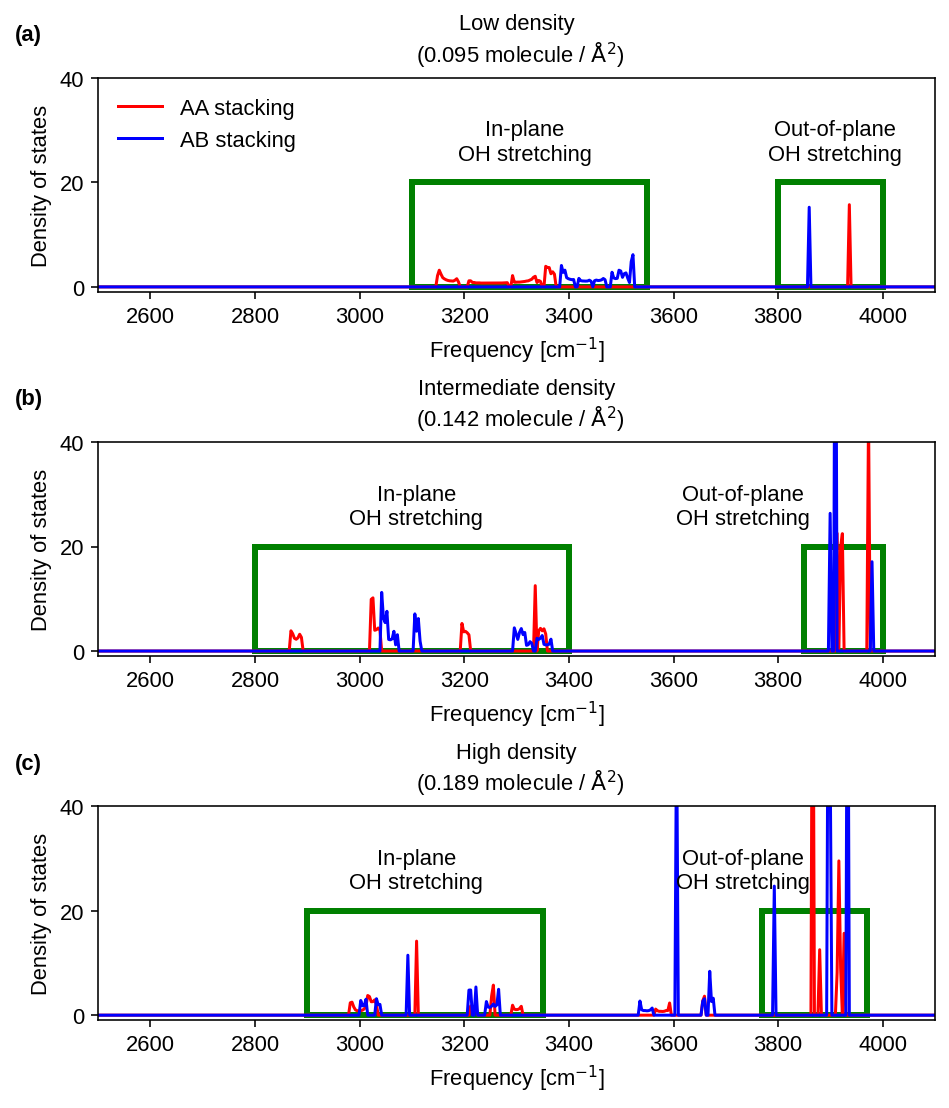}
\caption{\textbf{Phonon DOS in the O--H stretching region.}
\textbf{a--c}, Phonon DOS of water confined by rigid AA- and AB-stacked graphene sheets in the 2500--4000~cm$^{-1}$ frequency range for low-density (\textbf{a}), intermediate-density (\textbf{b}) and high-density phases (\textbf{c}).
Red and blue lines correspond to AA and AB stacking, respectively.
Green rectangles indicate the in-plane and out-of-plane O--H stretching regions.}
\label{fig:phonon}
\end{figure*}
\noindent We now turn to proton delocalisation.
Here, the opposite trend is observed, albeit originating from the same atomistic morphology.
At low and intermediate densities, AA stacking systematically reduces the barrier along the proton-transfer coordinate relative to AB stacking, whereas at high density the barrier becomes negligible for both stackings.
This is illustrated for the low-density regime in Fig.~\ref{fig:4}(d), which reports the temperature-scaled barrier along the proton-transfer coordinate for AA and AB stacking.
To understand this effect, we evaluate the joint probability distribution of the proton-transfer coordinate and the O--O distance.
We observe that AA stacking is associated with shorter average O--O distances in Fig.~\ref{fig:4}(e,f), which are known to correlate strongly with enhanced proton transfer.
As shown in Fig.~\ref{fig:LDFlow}--~\ref{fig:LDFhigh}, the shortening of the O--O distance is most pronounced at low density, where the difference in the proton-transfer barrier is largest, and nearly vanishes at high density, where the barriers for AA and AB stacking become similar (Fig.~\ref{fig:PTCbarrier}).
This discrepancy in O--O distance is more clearly illustrated by snapshots of the liquid phase confined between AA- and AB-stacked graphene where the positions of the O atoms generally correspond to the minima in the oxygen free energy landscape. As shown in Fig.~\ref{fig:OOdistance_liquid}, the two neighbouring O atoms in AA stacking are positioned near the centres of adjacent graphene hexagons.
In AB stacking, two types of neighbouring O--O pairs are observed. One type has a geometry similar to that in AA stacking, whereas the other consists of one O atom near a hexagon centre and another displaced towards the edge shared by two neighbouring hexagons. The latter geometry gives a longer O--O distance.
This anisotropy increases the average O--O distance in AB stacking, and thereby, the proton-transfer barrier.
These results highlight how subtle structural differences in the confining material can have pronounced and non-trivial consequences for the phase behaviour and dynamics of nanoconfined water.
\subsection{Stacking-induced hydrogen bonding provides a unifying microscopic explanation}
\noindent AA stacking increases the melting temperature, proton transfer, and ionic conductivity, while reducing the oxygen diffusion coefficient relative to AB stacking.
Taken together, these trends suggest that stacking affects nanoconfined water in a non-obvious way.
We therefore seek a microscopic explanation that can account for all of these observations. \\
\noindent To this end, we analyse the phonon DOS of the confined ice phases.
We focus on the high-frequency O--H stretching modes associated with in-plane hydrogen bonding and compare water confined in AA- and AB-stacked graphene.
Stacking also affects the low-frequency part of the phonon DOS. However, assigning these features to in-plane and out-of-plane oxygen fluctuations is more challenging (see Fig.~\ref{fig:phonon_SI}), so we restrict our analysis to the O--H stretching modes.
As shown in Fig.~\ref{fig:phonon}, the frequencies of the in-plane O--H stretching modes are on average lower for water confined in AA-stacked graphene than in AB-stacked graphene at all densities.
As shown in Table~\ref{tab:phonon}, the frequency difference for confined water is most pronounced at low density, reaching about 150~cm$^{-1}$, and decreases at higher density.
These results provide strong and experimentally verifiable evidence that AA stacking strengthens hydrogen bonding relative to AB stacking.
This strengthening is associated with the shorter neighbouring O--O distances in AA stacking, as illustrated by snapshots of the AA-stacked flat-rhombic and AB-stacked hexagonal phases in Fig.~\ref{fig:OOdistance}.
The geometric origin of this difference was discussed in the previous section using liquid-phase snapshots.
Stronger hydrogen bonding in AA stacking increases the cohesive energy (Table~\ref{tab:cohesive_energy}) of the ice phases, stabilising the flat-rhombic phase and raising the melting temperature.
The resulting more strongly connected hydrogen-bond network also restricts the lateral motion of water molecules, reducing oxygen diffusion, while enhanced proton sharing lowers the barrier to proton transfer and promotes proton delocalisation.
This provides a unifying microscopic explanation for the stacking dependence behaviour found in this work.
\section{Conclusions}
\noindent We investigate the density--temperature phase diagram of water confined by explicit bilayer graphene nanocapillaries.
By contrasting the phase behaviour under AB and AA stacking to those obtained under a laterally uniform confining potential~\cite{kapil_first-principles_2022}, we provide a detailed picture of how the atomistic morphology of the nanocapillary, at fixed slit width, can modulate the properties of nanoconfined water.
A seemingly minor lateral displacement of $\sim$1.42~\AA{} between the graphene layers leads to pronounced changes in both physical and chemical behaviour, including $\sim$100 K differences in melting temperature and a lowering of the onset temperature for superionic behaviour in the intermediate density regime.
In AA stacking, carbon atoms lie directly above one another along the normal to the lateral plane, creating repulsive regions and a more corrugated lateral free energy landscape.
Relative to AB stacking, this corrugation is associated with shorter O--O distances and stronger hydrogen bonding, thereby increasing the cohesive energy of the confined ice phases, altering their relative stability, and raising the melting temperatures.
At the same time, it lowers the barrier along the proton-transfer coordinate, thereby enhancing proton delocalisation and altering the ionic conductivity. \\
\noindent Broadly, our work shows that nanoscale confinement is more than a purely topological effect.
While restricting the motion of water to lower dimensions can dramatically change its properties, modifying the atomistic details of the nanocapillary can also alter water's hydrogen bonding which in turn impacts its physical and chemical properties.
This highlights that experimental outcomes in nanoconfinement should be interpreted with caution, particularly when comparing different confining materials.
Observed differences in experiments may be associated not only with the confinement geometry but also with surface chemistry or modulation of hydrogen bonding induced by the atomistic morphology of the surface.
At the same time, our results suggest that bilayer-graphene stacking, in addition to the size and shape of nanocapillaries, provides an additional lever to engineer the physical and chemical properties of nanoconfined water and, more broadly, confined fluids.
Just as experimental control of bilayer-graphene twist angles has enabled fine tuning of the electronic properties of 2D materials~\cite{cao_unconventional_2018,kim_tunable_2017}, and analogous control over the relative arrangement of confining 2D layers may enable systematic tuning of the behaviour of confined fluids.
While this work considers two limiting stacking arrangements, a broad range of behaviours relevant to nanofluidics may emerge in the intermediate regime.
\newpage
\section{Methods}
\subsection{\textbf{Workflow}}
\noindent The workflow to investigate phase diagram and the behaviour of water confined in bilayer graphene can be summarised in the following four steps.
\begin{enumerate}
    \item We employ an MLIP to represent the potential energy surface of the full system.
    \item We perform random structure searches with DFT validation to find the thermodynamically stable ice phases for all densities in both stacking arrangements at 0\,K (neglecting quantum nuclear effects).
    \item We perform classical MD simulations at various temperatures to investigate the melting temperature.
    \item We perform PIMD simulations to investigate structural properties and Te PIGS simulations~\cite{musil_quantum_2022} to investigate dynamical properties at various temperatures for all densities and both stackings. These simulations are used to understand proton transfer, transport, and the onset of the superionic phase transition.
    \item To understand how the observed behaviours are linked to hydrogen bonding, we compute phonon densities of states in the confined ice phases.
\end{enumerate}
\noindent Additional computational details specific to the systems associated with the three different density regimes are provided in tables ~\ref{tab:info_low_density} - ~\ref{tab:info_high_density}  of the SI.
\subsubsection{Overview of MLIP}
\noindent We employ a previously developed revPBE0-D3 MLIP~\cite{zhang_comment_1998, adamo_toward_1999, grimme_consistent_2010} from Ref.~\citenum{advincula_protons_2025} (reported in its respective SI).
The revPBE0-D3 functional has been previously shown to yield sub-kcal/mol accuracy for water binding to carbon nanostructures~\cite{brandenburg_interaction_2019}, provides a good description of structural and dynamical properties of bulk water under ambient conditions including the radial distribution function,  diffusion coefficient and infrared (IR) and Raman spectra~\cite{marsalek_quantum_2017}, temperature of maximum density of liquid water and melting temperature of hexagonal ice at ambient pressure~\cite{cheng_ab_2019}, and the sum frequency generation spectrum of the water-air interface~\cite{kapil_first-principles_2024} at ambient conditions.
It also yields a chemically accurate description of the lattice energies of low and high density monolayer ice phases~\cite{kapil_first-principles_2022} when benchmarked against fixed-node diffusion Monte Carlo~\cite{chen_evidence_2016} and reproduces the proton-transfer potential energy profile~\cite{ravindra_nuclear_2026} when benchmarked against mechanically-embedded local coupled-cluster theory~\cite{shi_general_2022} up to chemical accuracy.
In this study we perform an additional benchmark of this functional in Table~\ref{tab:rpagwse} of the SI.
We show that this functional reproduces differences in water–graphene and water–water interactions in nanoconfined water between the AA and AB stacking configurations close to 1 kcal/mol of random phase approximation plus GW single excitations (RPA+GWSE)\cite{klimes_lattice_2016} energies that we have generated. \\
\noindent As reported in the SI of Ref.~\citenum{advincula_protons_2025}, their revPBE0-D3 MLIP which we use in this work is trained to configurations generated systematically over five generations, including different stacking configurations, water densities beyond the regimes considered in this work, protonic defect pair configurations, slit widths, temperatures, and configurations from both classical and quantum nuclear simulations.
As a result, it provides comprehensive coverage of the tested thermodynamic conditions in this study.
The dataset was used to train a two-layer MACE model~\cite{batatia_mace_2022}, with a $6\,$\AA{} cutoff for each layer and 128 equivariant messages.
Across the entire training, the MLIP achieved an energy root mean square error (RMSE) of 1.5 meV/atom and a force RMSE of 32.6 meV/\AA{}.
\subsubsection{Random structure search with DFT validation}
\noindent We apply the random structure search method pioneered by~\citet{Pickard_2011} together with DFT validation to obtain the most thermodynamically stable phases at 0\,K (neglecting quantum nuclear effects) for each density and stacking.
For each system, we generate 1000 random ``sensible" structures with water molecules between the graphene sheets 5~\AA{} apart, which satisfy C–O $>$ 2.5~\AA{}, C–H $>$ 2.08~\AA{}, O–O $>$ 2.42~\AA{}, O–H $>$ 1.40~\AA{}, and H–H $>$ 1.75~\AA{}.
When performing structure generation, the lattice parameter along the z axis was fixed, while the in-plane lattice parameters a and b and the angle $\gamma$ were allowed to vary, subject to a constant in-plane area constraint.
The graphene sheets were treated as a set of rigid atoms in a rectangular 2D lattice with side lengths of 8.56~\AA{} and 9.88~\AA{}.
The low, intermediate, and high-density phases defined in the main text, are based on the number density of water molecules.
Since the in-plane area remains fixed during structure generation, the density is uniquely specified by the number of inserted water molecules: 8, 12, and 16 for the low, intermediate, and high-density phase, respectively.
The generated structures are then optimised using the MLIP with Atomic Simulation Environment (ASE)\cite{hjorth_larsen_atomic_2017} and a Broyden–Fletcher–Goldfarb–Shanno (BFGS) quasi-Newton optimiser.
During optimisation, the cell and the graphene layer are held fixed, only the water molecules are relaxed.
The optimisation was considered converged when the maximum force component was below  $1.0 \times 10^{-3}\ \mathrm{eV}\,\mathrm{\AA}^{-1}$.
The optimised structures were ranked by their energies to identify the most stable structures for all the densities and stackings. \\
\noindent In addition, we used DFT calculations to validate the structures obtained from the random structure searches.
For the low-density regime, the searches yielded two competing types of ordered phases: hexagonal and flat-rhombic structures.
For AA stacking, only flat rhombic structures are found, but they appear in two distinct variants. We therefore selected the two lowest-energy representative of each type identified by the MLIP-based search and recomputed their energies at the revPBE0-D3 level. The lowest-energy structure identified by the MLIP-based search was also the most stable structure at the DFT level.
For AB stacking, although the MLIP search found the hexagonal and flat-rhombic structures to be nearly degenerate. We recomputed the energies of the lowest-energy flat-rhombic structure and the hexagonal structure from MLIP search at the revPBE0-D3 level. The DFT calculations clearly stabilised the hexagonal phase, which was lower in energy than the flat-rhombic phase by approximately 0.07~meV/atom.
For the intermediate- and high-density regimes, the random structure searches yielded only flat-rhombic ordered structures at low energy.
For each stacking and density, we selected the lowest-energy structure and recomputed its energy at the rev-PBE0-D3 level.
DFT calculations on these structrues confirmed the qualitative conclusion of the MLIP-based random structure searches, yielding the same flat-rhombic stable phase in both regimes.
\subsubsection{Molecular dynamics}
\noindent We perform molecular dynamics to determine the melting temperature of the ice phases obtained from random structure search.
MD simulations are performed using the i-PI code~\cite{KAPIL2019214} and the MLIP via the ASE~\cite{hjorth_larsen_atomic_2017} socket client.
All MD simulations were carried out in a common supercell of $34.22 \times 39.52 \times 30$~\AA{}.
This supercell corresponds to a 16$\times$ increase in the in-plane area relative to the reference cell and was adopted to attenuate finite-size effects.
The MD simulations are run as a function of temperature in steps of 20\,K from 100\,K to 600\,K. Temperature dependent thermodynamic properties are computed based on MD trajectories. \\
\noindent The diffusion coefficient of O atoms was evaluated using the three-dimensional velocity autocorrelation function: \[D=\frac{1}{3N}\sum_{i=1}^{N}\int_{0}^{\infty}\left\langle \mathbf{v}_i(0)\cdot \mathbf{v}_i(t)\right\rangle\,dt\] where $N$ is number of particles and $\mathbf{v}_i$ is the velocity of the $i-$th particle. Since diffusion is confined to the plane parallel to the graphene sheets, the out-of-plane contribution to the current is negligible. Although we report the diffusion coefficient in the conventional three dimensional form, we rescale the Green--Kubo relation by a factor of 1.5 to emphasise the dominant in-plane components.
This auto correlation function is estimated with the implementation in STable AutoCorrelation Integral Estimator (STACIE)~\cite{toraman_stable_2025}, which processes time correlated data automatically free from tunable hyperparameters, giving results of high reliability.
The O--O lateral distribution function (LDF) is defined as: \[
g_{\parallel}(r)=\frac{A}{N\,2\pi r\,\Delta r}\,\langle n(r, r+\Delta r)\rangle
\] where $A$ is the in-plane area, $N$ is the number of oxygen atoms, $r$ is the in-plane distance and $n(r, r+\Delta r)$ is the number of distances in the range $[r, r+\Delta r)$. This property is obtained by computing the normalised histogram of the in-plane distances between the atoms.
Finally, the 2D spatial distribution of oxygen and hydrogen atoms represents the atomic density distribution in the $xy$ plane.
For each frame of the trajectory, all O–-H vectors defined from O to H and O–-O vectors are calculated.
The $x$ and $y$ components of these vectors are then used to construct a 2D histogram, with both axes ranging from -3 to 3~\AA{}.  \\
\noindent The simulations are run in canonical (NVT) ensemble in two stages.
In the first stage, 120\,ps long simulations are performed at each temperature for each density and stacking using a timestep of 0.5\,fs.
The configurations are saved every 200 time steps.
Graphene layers are fixed in the simulations.
To get a more precise estimate of the melting temperature, longer MD simulations are performed for a subset of temperatures.
We first examined each 120\,ps trajectory and identified the lowest temperature at which the structure becomes disordered.
We then performed a 500\,ps MD simulation at a temperature 20\,K below this preliminary melting temperature continued from the 120\,ps simulations.
If the structure became disordered during the 500\,ps simulation, we concluded that it melts at that temperature or below.
We then repeated 500\,ps simulations at successively lower temperatures, in 20\,K intervals, until the structure remained ordered throughout the full 500\,ps simulation.
By examining the configurations from 120\,ps simulations in first stage and 500\,ps simulations in second stage, the melting temperature was determined from a total sampling time of 620\,ps.
To verify whether this approach can reliably predict melting temperatures, we apply it to the phase diagram of nanoconfined water in a uniform confining potential and compare the resulting melting temperatures with reference values estimated from direct coexistence simulations.
As can be seen from Fig.~\ref{fig:phase_diagram_compare}, direct MD is able to yield melting temperatures within a few tens of K of the reference at low and medium densities.
At high densities  this error increases to 80\,K.
\subsubsection{Path integral simulations}
\noindent We perform path integral simulations to study proton transfer, the superionic phase transition and transport properties.
To this extent, we perform PIMD simulations for structural properties and Te PIGS for dynamical properties in the $NVT$ ensemble.
We use the i-PI code together with an ASE socket client to run these simulations.
 The PIMD simulations are performed from 100~K to 600~K in the NVT ensemble with an interval of 100~K with 32 replicas, a 0.25\,fs timestep, and a path-integral Langevin equation-local (PILE-L) thermostat with a thermostat time constant $\tau$ = 100.
 The positions of replicas and the centroid are saved every 200 time steps.
Based on the PIMD trajectory, we calculate the probability distribution associated with the proton transfer coordinate for all the systems.
The proton transfer coordinate $\nu$ is defined as the difference of the distances of a proton with two nearest oxygen atoms $O_1$ and $O_2$: $\nu = dO_1H - dO_2H$.
Here the assignment of $O_1$ and $O_2$ is arbitrary, thus the free energy profile is symmetric about $\nu = 0$. \\
\noindent To reduce the high computational cost of calculating diffusion coefficients and ionic conductivities, we use the Te PIGS approach, which performs classical MD on the centroid free energy surface estimated at a high temperature.
The procedure starts with training a Te PIGS model.
We perform 20\,ps PIMD using 8 replicas, a timestep of 0.5\,fs, and a PILE-L thermostat with a time constant of 100\,fs for all densities and stackings at 600\,K to generate the training set.
The physical and centroid forces, saved every 20 time steps, are fed into the training.
A small MACE model (hidden irreducible representations of $64 \times 0e$ and a cutoff radius of 3.0~\AA{}), is used to learn the difference between the centroid and the physical forces.
The Te PIGS simulations are performed in the NVT ensemble at temperatures from 300 to 600\,K, in intervals of 100\,K.
The simulation lengths are 500\,ps, 2\,ns, and 1\,ns for the low-, medium-, and high-density regimes, respectively.
All simulations use a timestep of 0.5\,fs and a weak Langevin thermostat with a time constant of 1000\,fs.
The positions are sampled every 100 time steps.
The quantum ionic conductivity is computed using the Green-Kubo formula implemented in the STACIE~\cite{toraman2025stable} package:
\begin{equation}
\sigma = \frac{1}{3 V k_{\mathrm{B}} T}
\int_{0}^{\infty}
\langle \mathbf{J}(t) \cdot \mathbf{J}(0) \rangle \, dt
\end{equation}
where $\mathbf{J}(t) = \sum_{i=1}^{N} q_i \mathbf{v}_i(t)$ with $q$ being fixed oxidation numbers of +1 for H atoms and -2 for O atoms, $V$ is the volume between graphene layers, ${k_B}$ is the Boltzmann constant, and $T$ is the temperature.
Because ion transport is restricted to the plane of confinement, the conductivity obtained from the standard three-dimensional expression was multiplied by 1.5.
To find the superionic phase transition temperature, we fit an exponential function with ionic conductivity from 300\,K to 600\,K for all intermediate and high density phases, as the temperature dependence of ionic conductivity follows an Arrhenius type relation.
The temperature where the function reaches 0.1 $S/\text{cm}$  (superionic threshold) is regarded as superionic transition temperature.
Quantum diffusion coefficients are also estimated with the STACIE~\cite{toraman2025stable} code as described for the case of molecular dynamics.
\subsubsection{Phonon DOS}
\noindent We estimate the phonon DOS using the MLIP for nanoconfined ice phases to investigate how phonon modes that participate in hydrogen bonding are affected by graphene stacking.
We calculate phonon total DOS using the finite displacement method with the implementation in phonopy~\cite{togo_implementation_2023} for the full water carbon systems.
We use a displacement distance of 0.01~\AA{} and $20 \times 20 \times 2$ mesh point.
\newpage
\begin{acknowledgments}
\noindent We thank Xavier Rosas Advincula and Christoph Schran for providing insightful comments.
VK acknowledges support from UCL’s startup funds.
The Flatiron Institute is a division of the Simons Foundation.
We are grateful for computational support from the Swiss National Supercomputing Centre (CSCS) under Project s1288 on Alps and UCL Myriad High Performance Computing Facility (Myriad@UCL).
This work used computing equipment funded by the Research Capital Investment Fund (RCIF) provided by UK Research and Innovation (UKRI), and partially funded by the UCL Cosmoparticle Initiative.
A.M. acknowledges support from the European Union under the ``n-AQUA'' European Research Council project (Grant No. 101071937).
\end{acknowledgments}
\newpage

\clearpage
\section{Supporting  Information}
\setcounter{figure}{0}
\renewcommand{\thefigure}{S\arabic{figure}}
\renewcommand{\thetable}{S\arabic{table}}
\begin{figure}[h]
\centering
\includegraphics[width=1\textwidth]{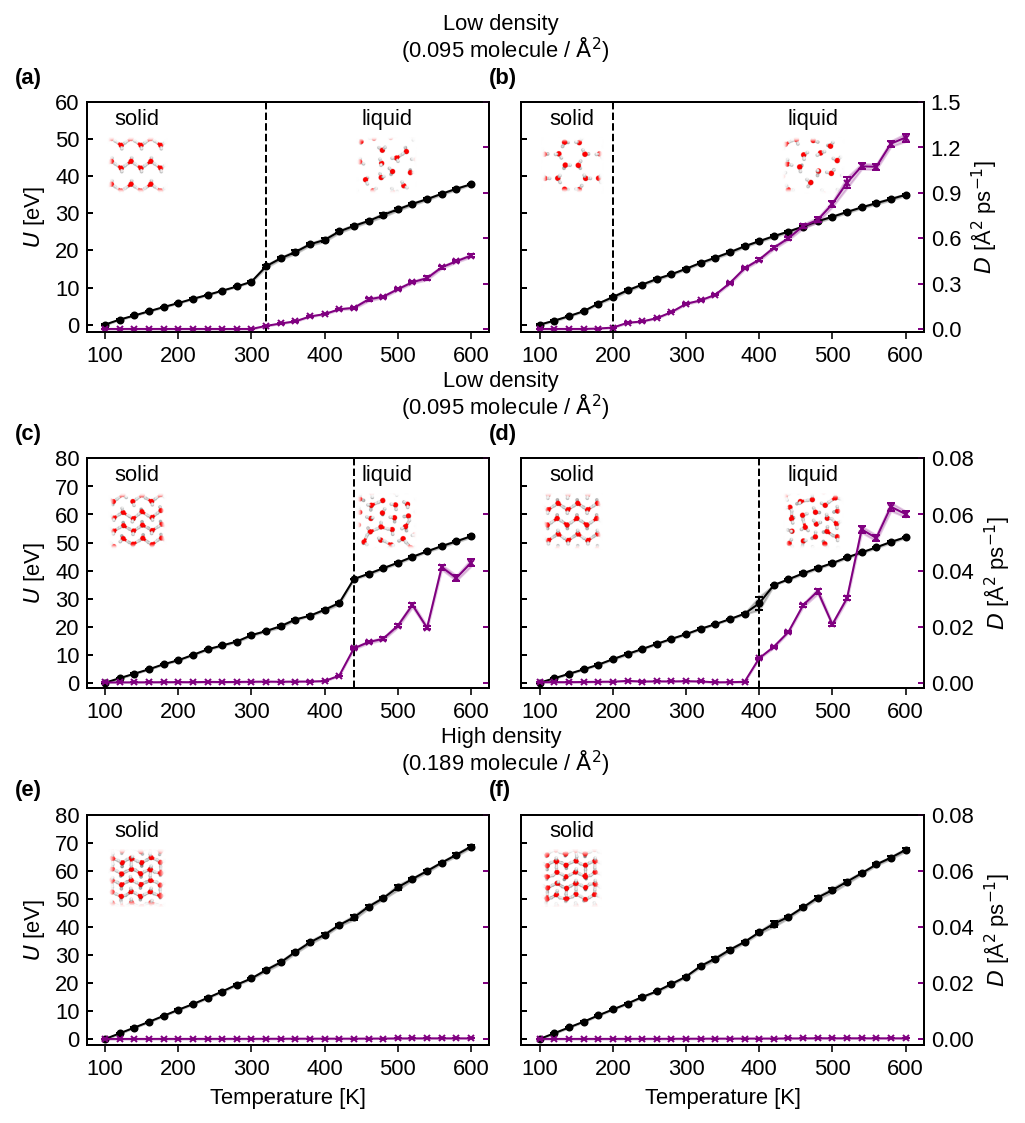}
\caption{\textbf{Temperature dependence of potential energy and oxygen diffusion. a–f, } The potential energy $U$ (black) and diffusion coefficient $D$ (violet) are plotted as a function of temperature for water confined in AA- and AB-stacked graphene. \textbf{a,b}, Low-density phase in AA stacking (\textbf{a}) and AB stacking (\textbf{b}). \textbf{c,d}, Intermediate-density phase in AA stacking (\textbf{c}) and AB stacking (\textbf{d}). \textbf{e,f}, High-density phase in AA stacking (\textbf{e}) and AB stacking (\textbf{f}). Vertical dashed lines indicate the estimated melting temperatures.}
\label{fig:U_D_plot}
\end{figure}
\clearpage
\begin{figure}[h]
\centering
\includegraphics[width=0.9\textwidth]{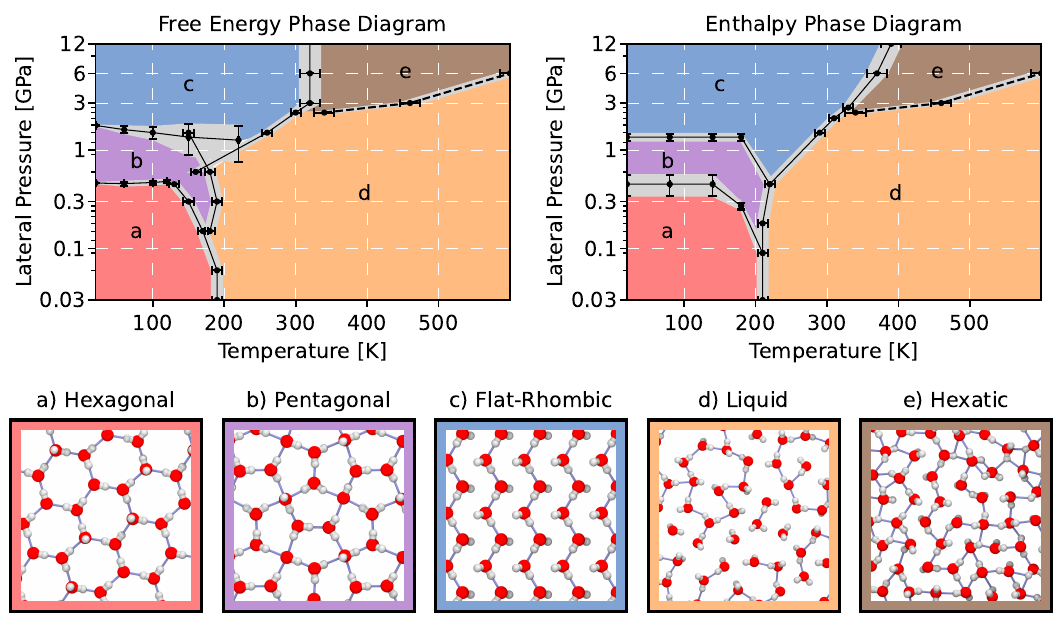}
\caption{\textbf{Benchmarking of melting temperatures estimated from MD-derived enthalpies against coexistence simulations.} The pressure–temperature phase diagram of monolayer water under uniform confining potential from Ref.~\citenum{kapil_first-principles_2022} for a 5\AA{} confinement width estimated with two different methods. Left: The order-order phase boundaries are determined using thermodynamic integration, and the order-disorder phase boundaries were determined from coexistence simulations. Right: The order-order phase boundaries are determined using the numerical enthalpy ($H = U + pV$) computed directly from the MD simulations to compare the relative stabilities of different ordered phases, and the order-disorder phase boundaries were determined from the metastability range of a solid phase. The superionic phase is omitted.  \textbf{a--e}, The structures correspond to the hexagonal, pentagonal, flat-rhombic, liquid, and hexatic phases, respectively.}
\label{fig:phase_diagram_compare}
\end{figure}
\clearpage
\begin{figure}[h]
\centering
\includegraphics[width=\textwidth]{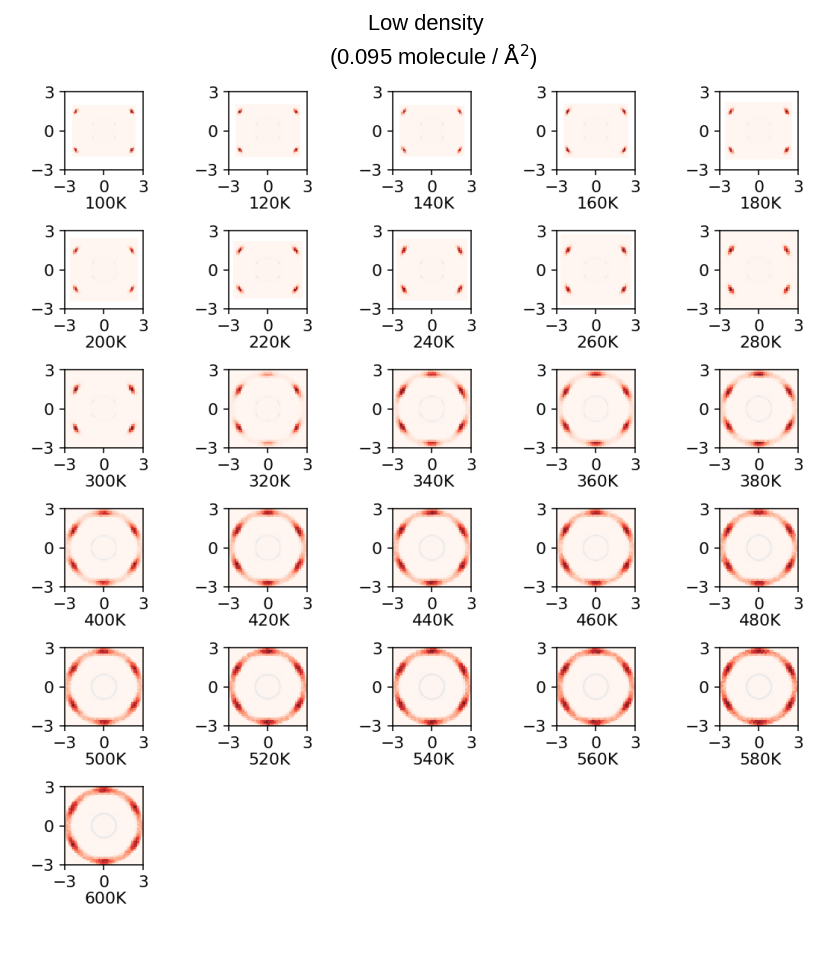}
\caption{\textbf{Temperature evolution of the 2D oxygen and hydrogen spatial distributions in the low-density AA-stacked system.} The 2D spatial distributions of oxygen and hydrogen atoms were calculated from MD trajectories at temperatures from 100 to 600~K.
Red and blue regions indicate the oxygen and hydrogen distributions, respectively.}
\label{fig:2DAA08}
\end{figure}
\begin{figure}[h]
\centering
\includegraphics[width=\textwidth]{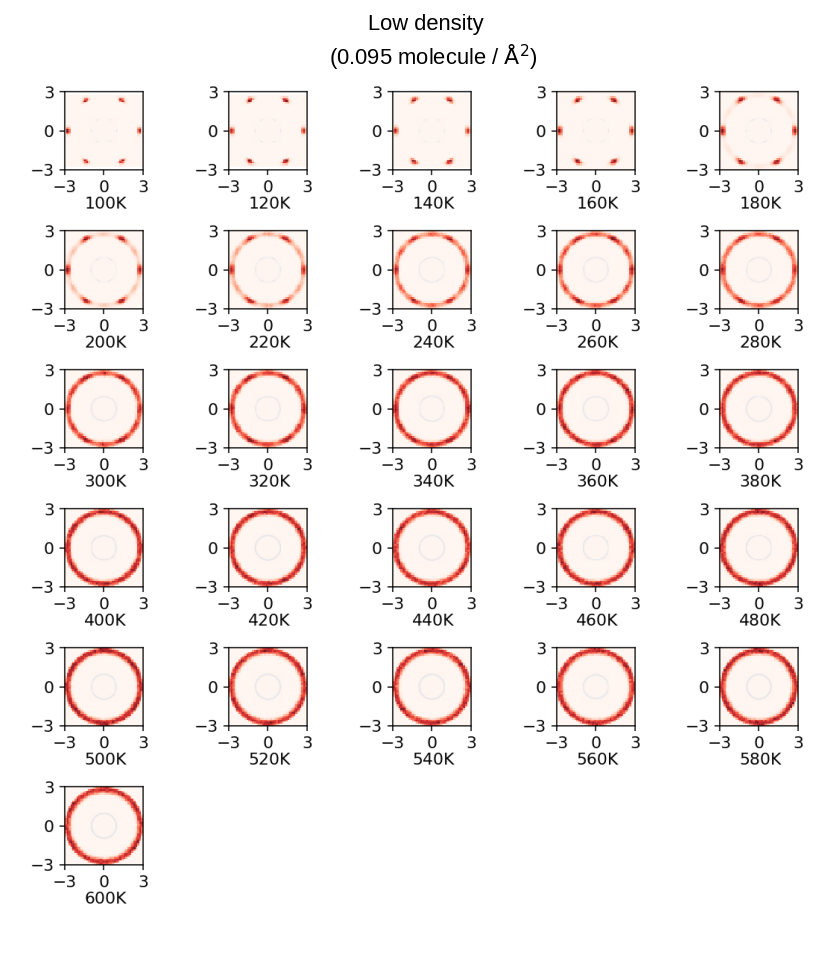}
\caption{\textbf{Temperature evolution of the 2D oxygen and hydrogen spatial distributions in the low-density AB-stacked system.} The 2D spatial distributions of oxygen and hydrogen atoms were calculated from MD trajectories at temperatures from 100 to 600~K.
Red and blue regions indicate the oxygen and hydrogen distributions, respectively.}
\label{fig:2DAB08}
\end{figure}
\clearpage
\begin{figure}[h]
\centering
\includegraphics[width=\textwidth]{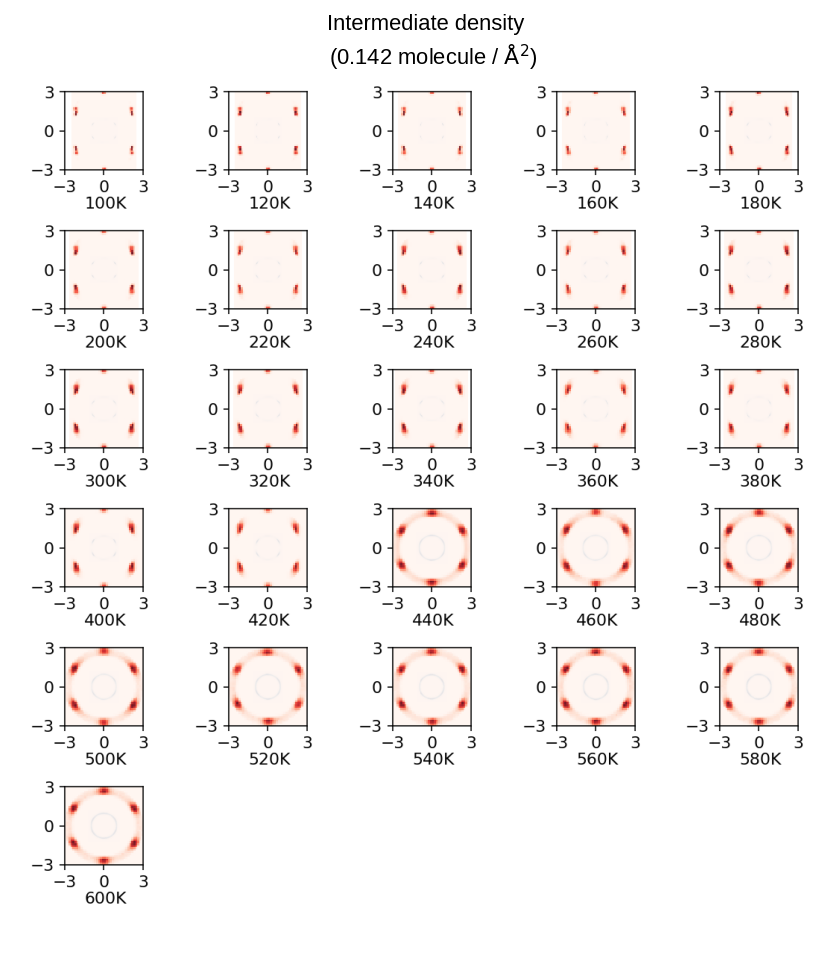}
\caption{\textbf{Temperature evolution of the 2D oxygen and hydrogen spatial distributions in the intermediate-density AA-stacked system.} The 2D spatial distributions of oxygen and hydrogen atoms were calculated from MD trajectories at temperatures from 100 to 600~K.
Red and blue regions indicate the oxygen and hydrogen distributions, respectively.}
\label{fig:2DAA12}
\end{figure}
\begin{figure}[h]
\centering
\includegraphics[width=\textwidth]{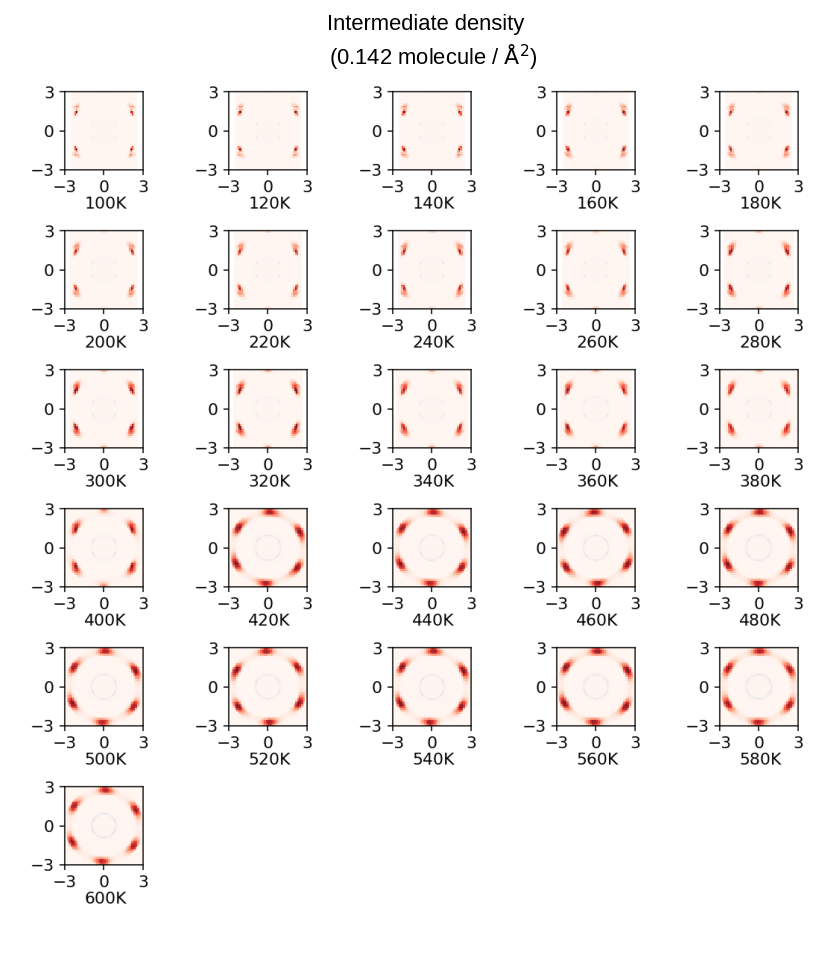}
\caption{\textbf{Temperature evolution of the 2D oxygen and hydrogen spatial distributions in the intermediate-density AB-stacked system.} The 2D spatial distributions of oxygen and hydrogen atoms were calculated from MD trajectories at temperatures from 100 to 600~K.
Red and blue regions indicate the oxygen and hydrogen distributions, respectively.}
\label{fig:2DAB12}
\end{figure}
\clearpage
\begin{figure}[h]
\centering
\includegraphics[width=\textwidth]{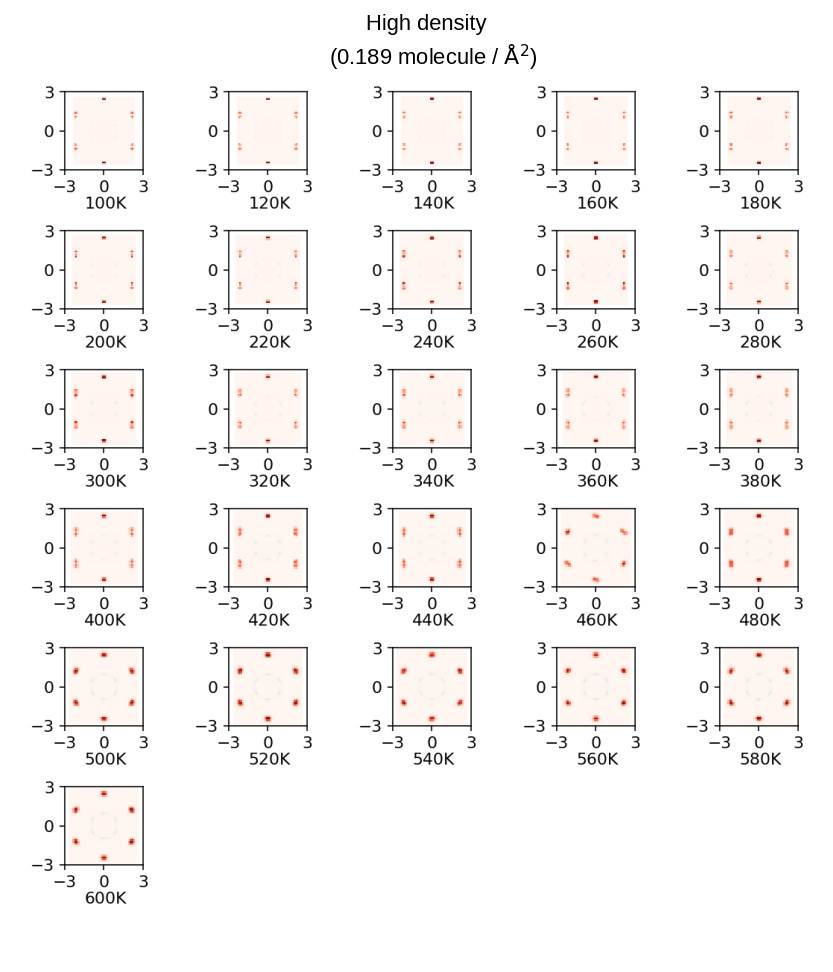}
\caption{\textbf{Temperature evolution of the 2D oxygen and hydrogen spatial distributions in the high-density AA-stacked system.} The 2D spatial distributions of oxygen and hydrogen atoms were calculated from MD trajectories at temperatures from 100 to 600~K.
Red and blue regions indicate the oxygen and hydrogen distributions, respectively.}
\label{fig:2DAA16}
\end{figure}
\begin{figure}[h]
\centering
\includegraphics[width=\textwidth]{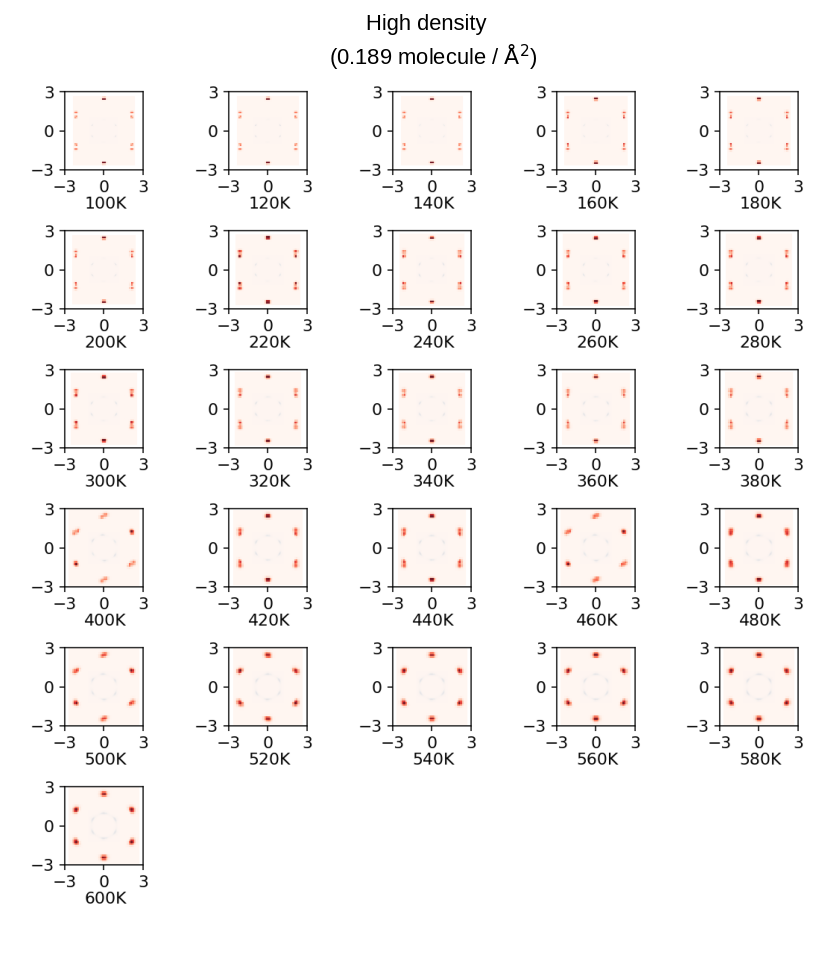}
\caption{\textbf{Temperature evolution of the 2D oxygen and hydrogen spatial distributions in the high-density AB-stacked system.} The 2D spatial distributions of oxygen and hydrogen atoms were calculated from MD trajectories at temperatures from 100 to 600~K.
Red and blue regions indicate the oxygen and hydrogen distributions, respectively.}
\label{fig:2DAB16}
\end{figure}
\clearpage
\begin{figure}[h]
\centering
\includegraphics[width=\textwidth]{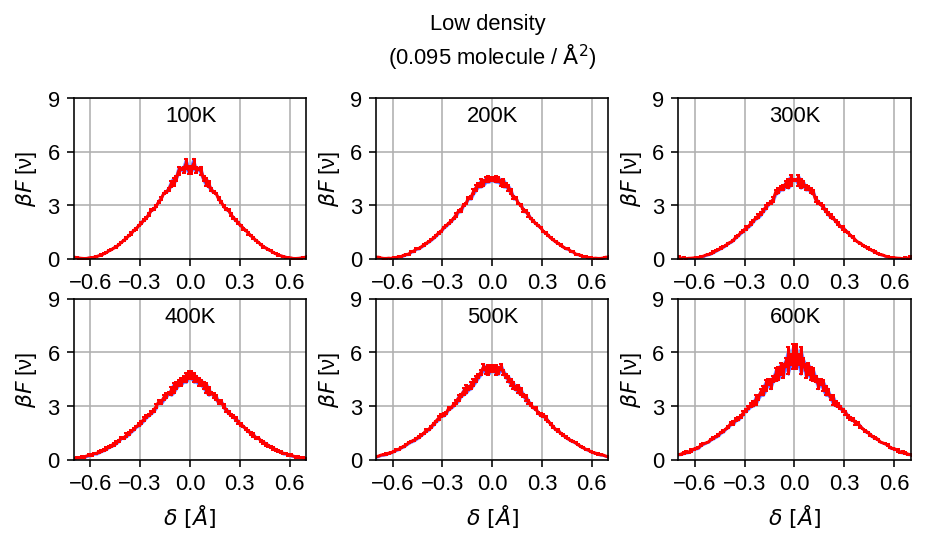}
\caption{\textbf{Temperature dependence of the proton-transfer free energy profile for low-density water confined in AA-stacked graphene.} free energy profiles, $\beta F$, along the proton-transfer coordinate $\delta$ were calculated from PIMD simulations at temperatures from 100 to 600~K.}
\label{fig:PTCAA08}
\end{figure}
\begin{figure}[h]
\centering
\includegraphics[width=\textwidth]{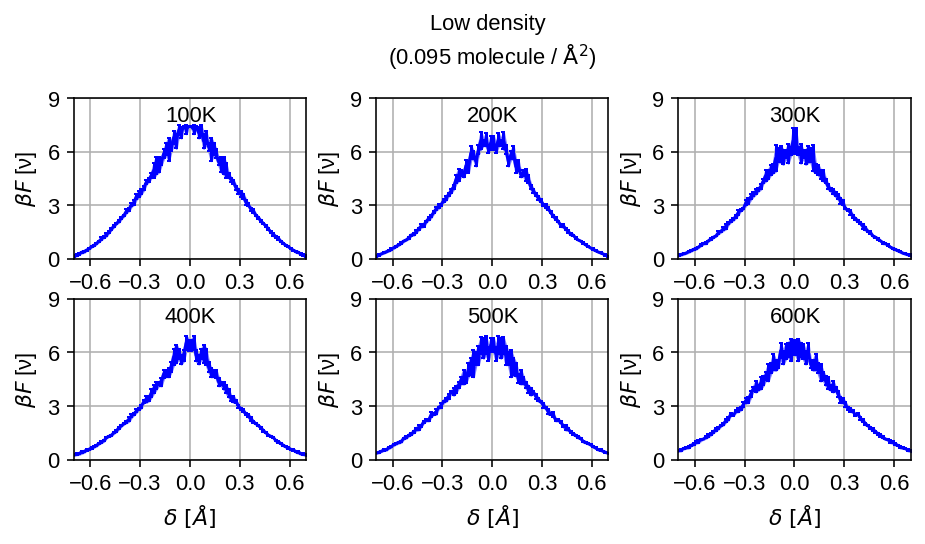}
\caption{\textbf{Temperature dependence of the proton-transfer free energy profile for low-density water confined in AB-stacked graphene.} free energy profiles, $\beta F$, along the proton-transfer coordinate $\delta$ were calculated from PIMD simulations at temperatures from 100 to 600~K.}
\label{fig:PTCAB08}
\end{figure}
\clearpage
\begin{figure}[h]
\centering
\includegraphics[width=\textwidth]{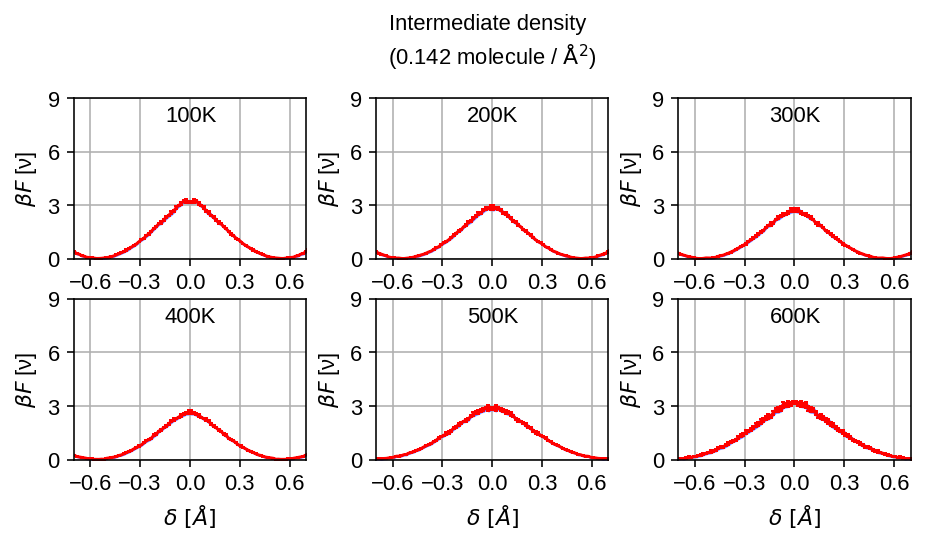}
\caption{\textbf{Temperature dependence of the proton-transfer free energy profile for intermediate-density water confined in AA-stacked graphene.} free energy profiles, $\beta F$, along the proton-transfer coordinate $\delta$ were calculated from PIMD simulations at temperatures from 100 to 600~K.}
\label{fig:PTCAA12}
\end{figure}
\begin{figure}[h]
\centering
\includegraphics[width=\textwidth]{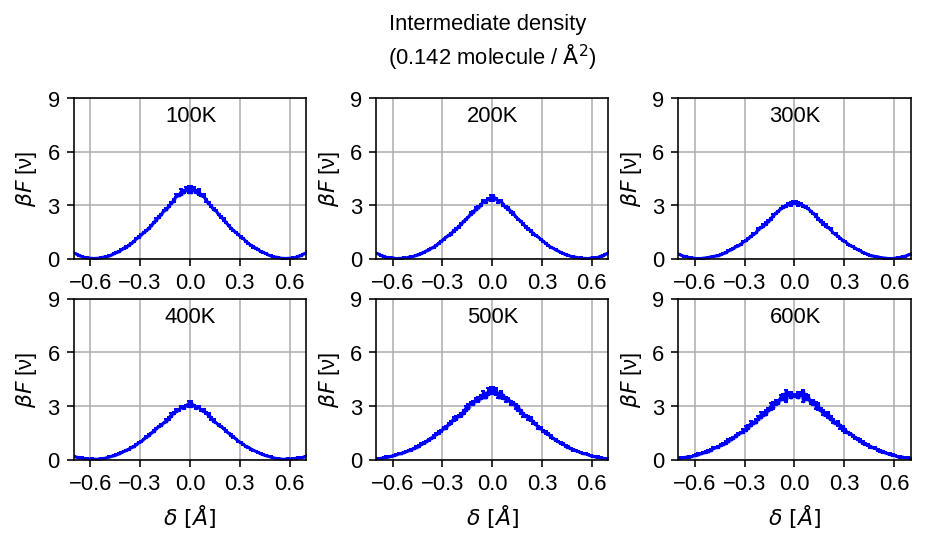}
\caption{\textbf{Temperature dependence of the proton-transfer free energy profile for intermediate-density water confined in AB-stacked graphene.} free energy profiles, $\beta F$, along the proton-transfer coordinate $\delta$ were calculated from PIMD simulations at temperatures from 100 to 600~K.}
\label{fig:PTCAB12}
\end{figure}
\clearpage
\begin{figure}[h]
\centering
\includegraphics[width=\textwidth]{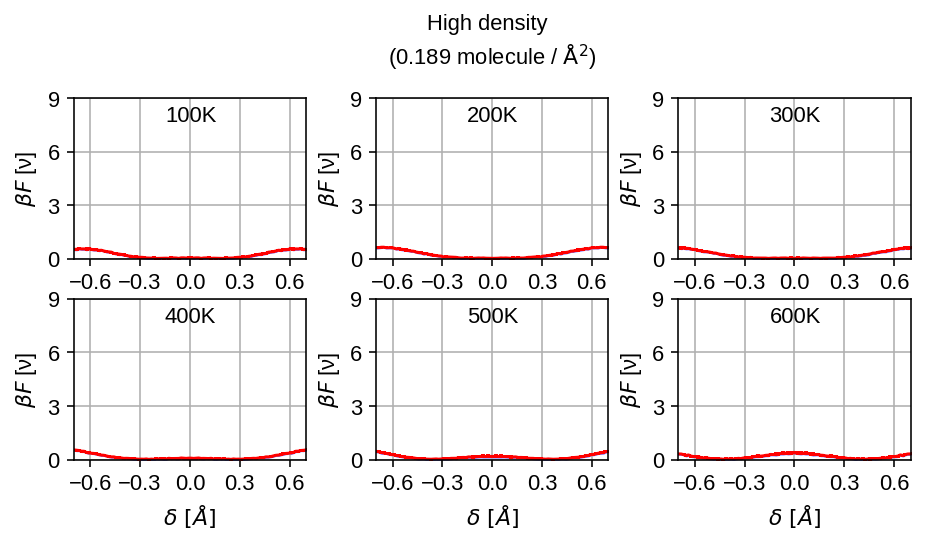}
\caption{\textbf{Temperature dependence of the proton-transfer free energy profile for high-density water confined in AA-stacked graphene.} free energy profiles, $\beta F$, along the proton-transfer coordinate $\delta$ were calculated from PIMD simulations at temperatures from 100 to 600~K.}
\label{fig:PTCAA16}
\end{figure}
\begin{figure}[h]
\centering
\includegraphics[width=\textwidth]{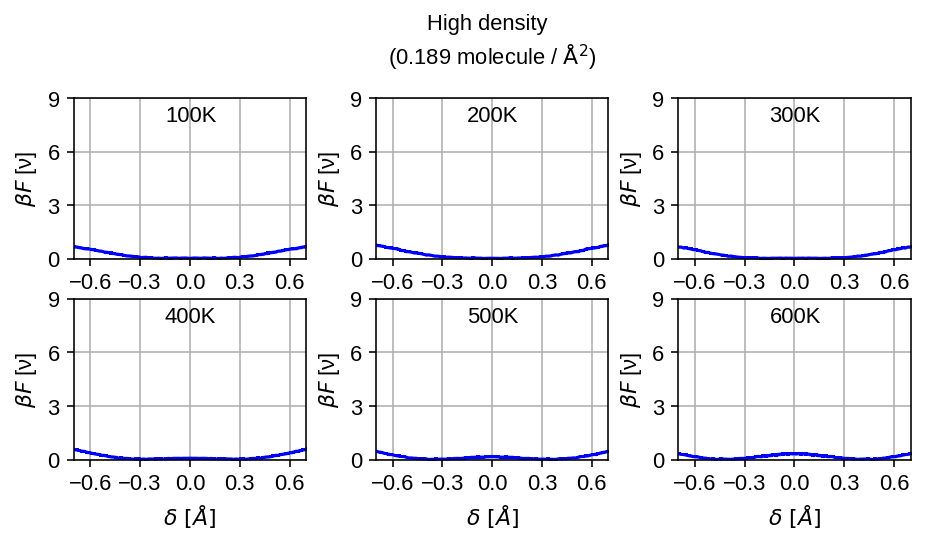}
\caption{\textbf{Temperature dependence of the proton-transfer free energy profile for high-density water confined in AB-stacked graphene.} free energy profiles, $\beta F$, along the proton-transfer coordinate $\delta$ were calculated from PIMD simulations at temperatures from 100 to 600~K.}
\label{fig:PTCAB16}
\end{figure}
\clearpage
\begin{figure}[h]
\centering
\includegraphics[width=0.5\textwidth]{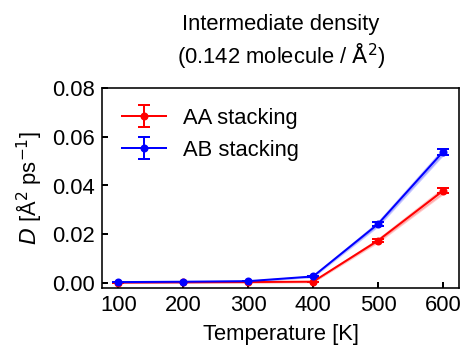}
\caption{\textbf{Temperature dependence of oxygen diffusion coefficient in the intermediate-density phase.}
Oxygen diffusion coefficients for water confined in AA- and AB-stacked graphene nanocapillaries were calculated from PIMD trajectories.
Red and blue lines represent AA and AB stacking, respectively.}
\label{fig:Dmedium}
\end{figure}
\clearpage
\begin{figure}[h]
\centering
\includegraphics[width=\textwidth]{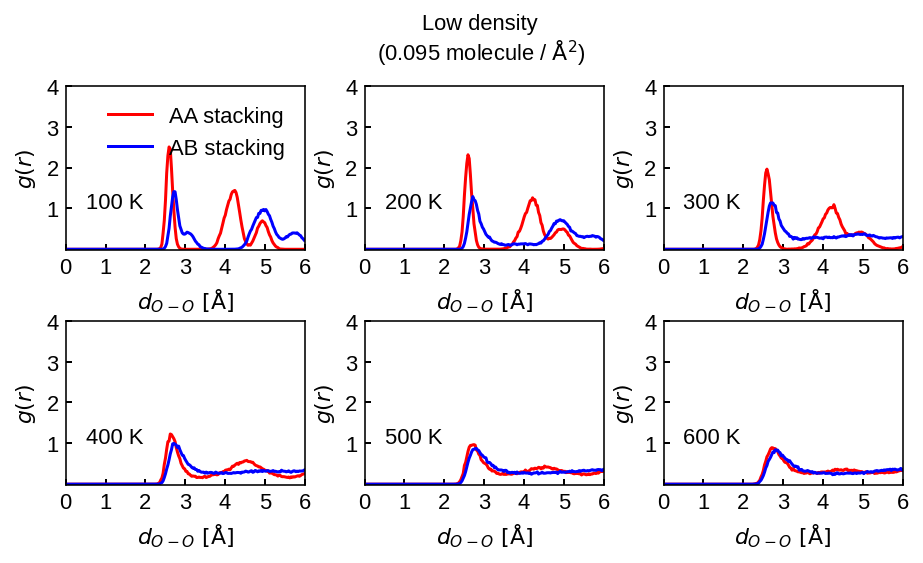}
\caption{\textbf{Temperature dependence of the O--O lateral distribution function in the low-density phase.}
O--O lateral distribution functions, $g(r)$, for water confined in AA- and AB-stacked graphene were computed from PIMD trajectories at temperatures from 100 to 600~K.
Red and blue lines represent AA and AB stacking, respectively.}
\label{fig:LDFlow}
\end{figure}
\begin{figure}[h]
\centering
\includegraphics[width=\textwidth]{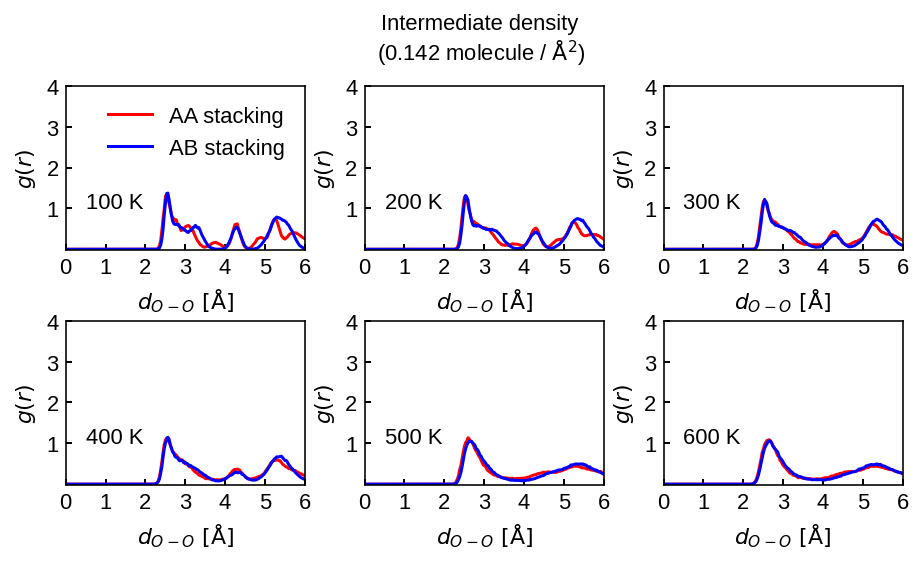}
\caption{\textbf{Temperature dependence of the O--O lateral distribution function in the intermediate-density phase.}
O--O lateral distribution functions, $g(r)$, for water confined in AA- and AB-stacked graphene were computed from PIMD trajectories at temperatures from 100 to 600~K.
Red and blue lines represent AA and AB stacking, respectively.}
\label{fig:LDFmedium}
\end{figure}
\begin{figure}[h]
\centering
\includegraphics[width=\textwidth]{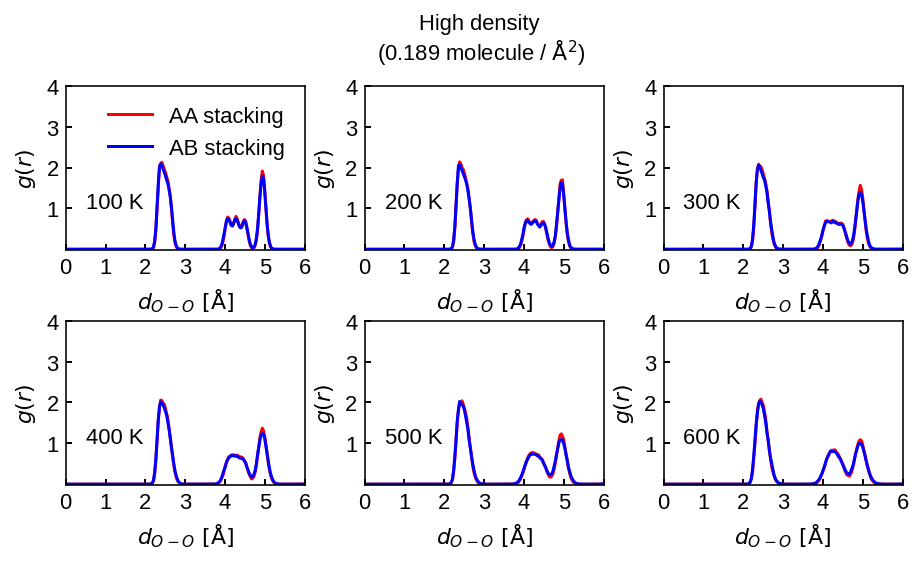}
\caption{\textbf{Temperature dependence of the O--O lateral distribution function in the high-density phase.}
O--O lateral distribution functions, $g(r)$, for water confined in AA- and AB-stacked graphene were computed from PIMD trajectories at temperatures from 100 to 600~K.
Red and blue lines represent AA and AB stacking, respectively.}
\label{fig:LDFhigh}
\end{figure}
\clearpage
\begin{figure}[h]
\centering
\includegraphics[width=1\textwidth]{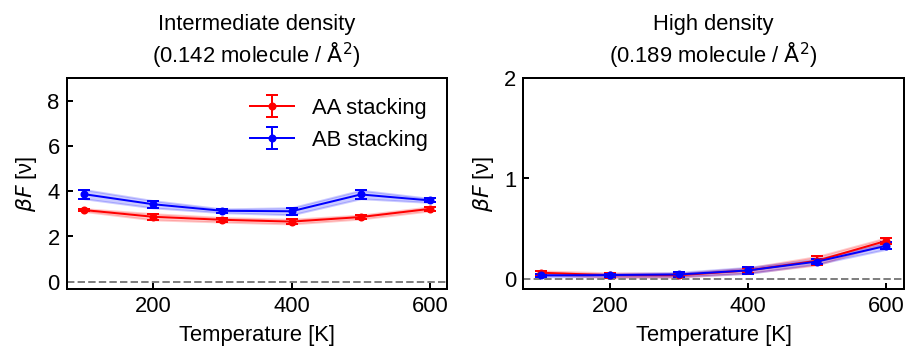}
\caption{\textbf{Temperature dependence of free energy barrier at proton transfer coordinate $\delta = 0$.}
\textbf{a,b}, free energy barrier for water confined in AA- and AB-stacked graphene in the intermediate-density (\textbf{a}) and high-density phases  (\textbf{b}).
Red and blue lines represent AA and AB stacking, respectively.}
\label{fig:PTCbarrier}
\end{figure}
\begin{figure}[h]
\centering
\includegraphics[width=1\textwidth]{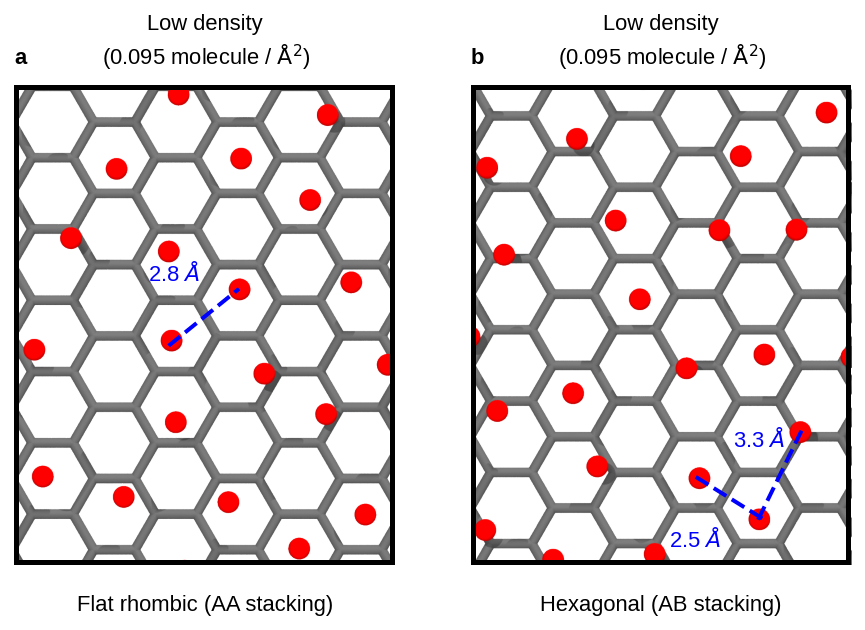}
\caption{\textbf{Neighbouring O--O distances in the low-density liquid phase.
a,b}, Representative liquid-phase snapshots for water confined by AA-stacked  (\textbf{a}) and AB-stacked graphene (\textbf{b}).
Oxygen atoms are shown in red and carbon atoms in dark grey.
Blue dashed lines indicate representative neighbouring O--O distances.
The system contains two graphene layers, but one graphene layer is omitted for clarity.}
\label{fig:OOdistance_liquid}
\end{figure}
\begin{figure}[h]
\centering
\includegraphics[width=1\textwidth]{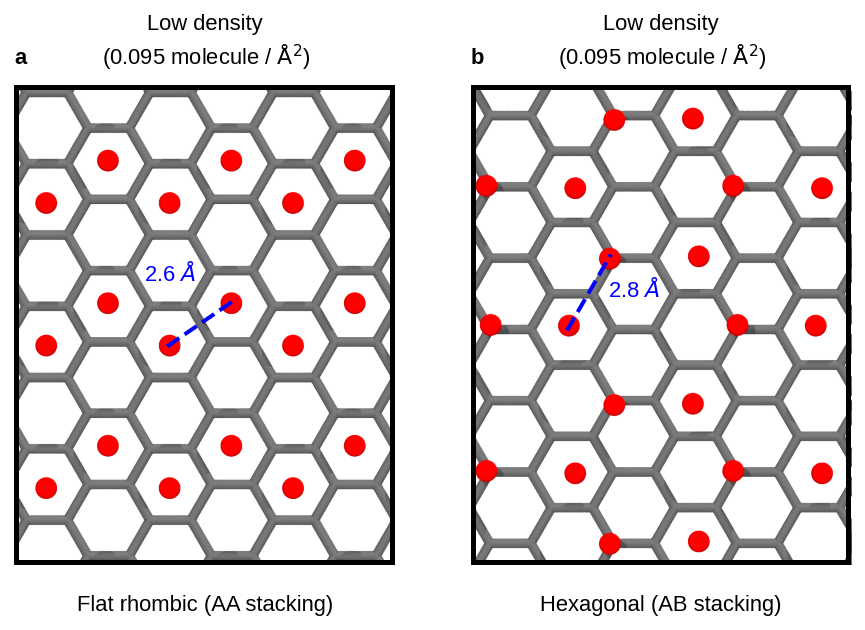}
\caption{\textbf{Neighbouring O--O distances in low-density confined water.
a,b}, Representative neighbouring O--O distances in the flat-rhombic phase under AA stacking (\textbf{a}) and the hexagonal phase under AB stacking (\textbf{b}). Oxygen atoms are shown in red and carbon atoms in dark grey. Blue dashed lines indicate the measured O--O distances. The system contains two graphene layers, but one graphene layer is omitted for clarity.}
\label{fig:OOdistance}
\end{figure}
\begin{figure*}[!t]
\centering
\includegraphics[width=\textwidth]{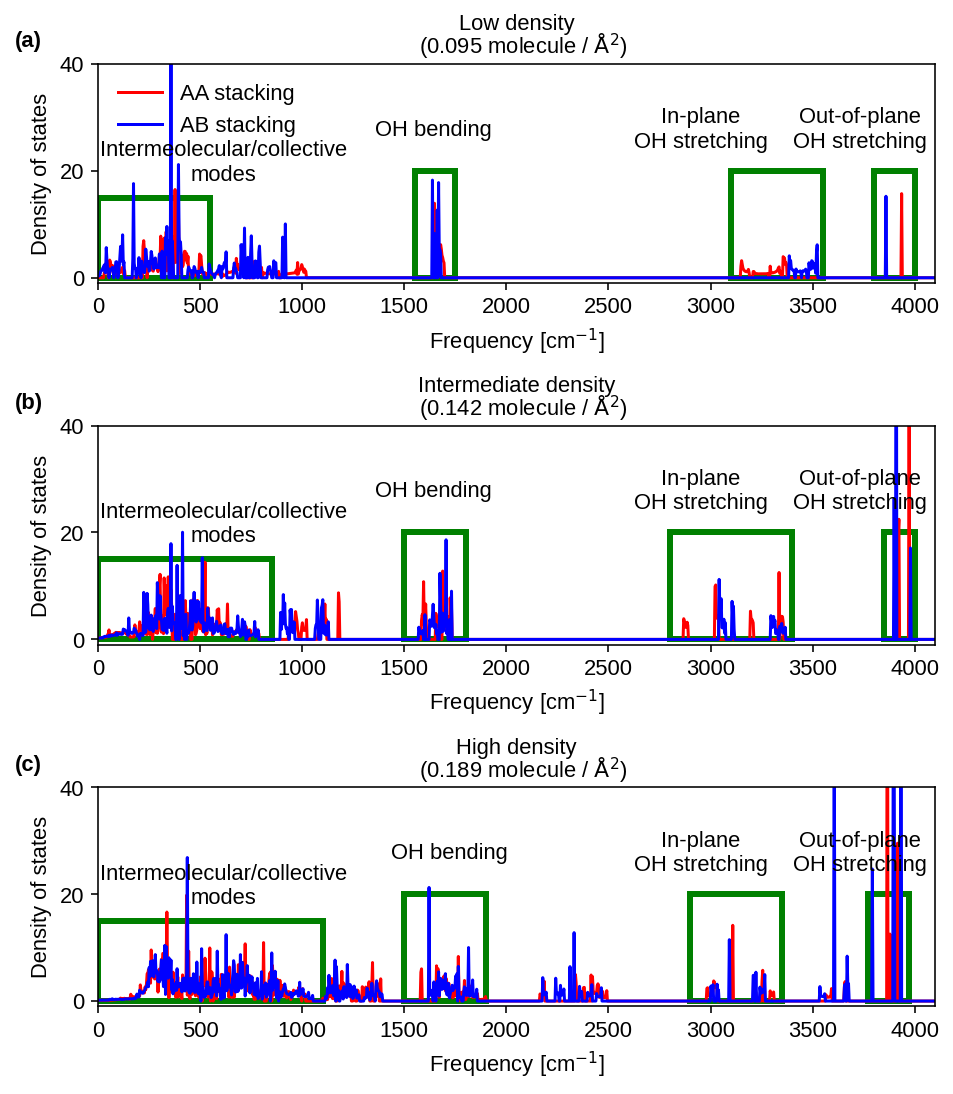}
\caption{\textbf{Phonon density of states of confined water.}
\textbf{a--c}, Phonon density of states (DOS) of water confined by rigid AA- and AB-stacked graphene sheets over the 0--4000~cm$^{-1}$ frequency range for low-density (\textbf{a}), intermediate-density (\textbf{b}) and high-density phases (\textbf{c}).
Red and blue lines correspond to AA and AB stacking, respectively.
Green rectangles indicate the labelled phonon modes, including intermolecular/collective modes, O--H bending, and in-plane and out-of-plane O--H stretching modes.}
\label{fig:phonon_SI}
\end{figure*}
\clearpage
\newpage
\begin{table}[t!]
  \centering
  \begin{tabular}{l|cc|cc|cc}
    \toprule
    Density
    & \multicolumn{2}{c|}{$\Delta E$}
    & \multicolumn{2}{c|}{$\Delta E_{\mathrm{water}}$}
    & \multicolumn{2}{c}{$\Delta E_{\mathrm{conf}}$} \\
    \cline{2-7}
    & DFT & RPA+GWSE & DFT & RPA+GWSE & DFT & RPA+GWSE \\
    \cline{2-7}
    & [eV] & [eV] & [eV] & [eV] & [eV] & [eV]\\
    \midrule
    Low
    & -0.08 & -0.14
    & 0.05  & 0.06
    & -0.12 & -0.20 \\
    Intermediate
    & -0.05 & -0.06
    & 0.03  & 0.03
    & -0.08 & -0.09 \\
    High
    & -0.15 & -0.13
    & -0.01 & -0.01
    & -0.14 & -0.12 \\
    \bottomrule
  \end{tabular}
  \caption{\textbf{DFT and RPA+GWSE energy differences between water confined in AA- and AB-stacked bilayer graphene nanocapillaries.}
  The total energy is decomposed as $E = E_{\text{water}} + E_{\text{conf}} + 2E_{\text{graphene}}$, where $E$ is the single-point energy of the full water--nanocapillary structure, $E_{\text{water}}$ is the single-point energy of the water subsystem obtained by removing graphene from the same structure, and $E_{\text{graphene}}$ is the single-point energy of an isolated graphene sheet used to build the nanocapillary (constant, $E_{\text{graphene}}=-709.92$~eV).
  The confinement energy is defined by subtraction as $E_{\text{conf}} = E - E_{\text{water}} - 2E_{\text{graphene}}$. DFT calculation was performed using revPBE0-D3 functional. Stacking-induced differences are reported per water molecule in eV as $\Delta E_{\_} = E_{\_}^{\mathrm{AA}} - E_{\_}^{\mathrm{AB}}$, where $\_ \in \{\, \text{water}, \text{conf}\}$.}
  \label{tab:rpagwse}
\end{table}
\begin{table}[!t]
  \centering
  \begin{tabular}{l|c|c}
    \toprule
    Density
    & \multicolumn{2}{c}{Cohesive Energy}\\
    \cline{2-3}
    & \multicolumn{1}{c|}{\makebox[4cm][c]{AA stacking [eV]}}
    & \multicolumn{1}{c}{\makebox[4cm][c]{AB stacking [eV]}}  \\
    \midrule
    Low & -0.14 & -0.06 \\
    Intermediate & -0.09 & -0.04 \\
    High & 0.25 & 0.41 \\
    \bottomrule
  \end{tabular}
  \caption{\textbf{Cohesive energy of water confined in AA- and AB-stacked graphene.} Cohesive energies were calculated using DFT at the revPBE0-D3 level and are reported in eV per water molecule. The cohesive energy is defined as $E_{\text{cohesive}} = E - n_{\text{water}}\times E_{\text{single water}} - 2E_{\text{graphene}}$, where $E$ is the single-point energy of the full water--nanocapillary structure, $n_{\text{water}}$ is the number of water molecules confined in nanocapillary, $E_{\text{single water}}$ is single-point energy of single water molecule and $E_{\text{graphene}}$ is the single-point energy of an isolated graphene sheet used to build the nanocapillary.}
  \label{tab:cohesive_energy}
\end{table}
\newpage
\begin{table}[!t]
  \centering
  \begin{tabular}{l|c|c}
    \toprule
    Density
    & \multicolumn{1}{c|}{\makebox[2.6cm][c]{AA stacking}}
    & \multicolumn{1}{c}{\makebox[2.6cm][c]{AB stacking}}  \\
    & [\AA] & [\AA] \\
    \midrule
    Low & 2.61 $\pm$ 0.0005 & 2.77 $\pm$ 0.0009\\
    Intermediate & 2.55 $\pm$ 0.0006 & 2.57 $\pm$ 0.0005 \\
    High & 2.38 $\pm$ 0.0003 & 2.38 $\pm$ 0.0003 \\
    \bottomrule
  \end{tabular}
  \caption{\textbf{Mean hydrogen-bonded O--O distances at 100~K for water confined in AA- and AB-stacked graphene.}
  Distances were calculated from PIMD trajectories by averaging over all hydrogen-bonded O--O pairs in all sampled configurations.
  Values are reported in \AA{}.}
  \label{tab:OO_distance}
\end{table}
\begin{table}[!t]
  \centering
  \begin{tabular}{l|c|c}
    \toprule
    Density
    & \multicolumn{1}{c|}{\makebox[2.6cm][c]{$\Delta \bar{\nu}_{\mathrm{OH}}^{\mathrm{flat}}$}}
    & \multicolumn{1}{c}{\makebox[2.6cm][c]{$\Delta \bar{\nu}_{\mathrm{OH}}^{\mathrm{weighted}}$}} \\
    & [cm$^{-1}$] & [cm$^{-1}$] \\
    \midrule
    Low & -150 & -190 \\
    Intermediate & -2 & -53 \\
    High & -5 & -33 \\
    \bottomrule
  \end{tabular}
  \caption{\textbf{Stacking-induced shifts in the in-plane O--H stretching frequencies of phonon DOS.} Frequency differences are reported as $\Delta \bar{\nu}_{\mathrm{OH}}$ = $\Delta\bar{\nu}_{\mathrm{OH}}^{\mathrm{AA}}$ - $\Delta\bar{\nu}_{\mathrm{OH}}^{\mathrm{AB}}$. Two averaging schemes are used to estimate $\bar{\nu}_{\mathrm{OH}}$ from the phonon DOS. The For the flat average, $\bar{\nu}_{\mathrm{OH}}^{\mathrm{flat}}$ = (${\nu}_{\mathrm{OH}}^{\mathrm{max}}$ + ${\nu}_{\mathrm{OH}}^{\mathrm{min}}$) / 2, where $\nu_{\mathrm{OH}}^{\mathrm{max}}$ and $\nu_{\mathrm{OH}}^{\mathrm{min}}$ are the upper and lower bounds of the in-plane O--H stretching region.
  For the weighted average, $\bar{\nu}_{\mathrm{OH}}^{\mathrm{weighted}}$ is calculated using the phonon DOS intensity as the weight.
  Negative values indicate lower in-plane O--H stretching frequencies under AA stacking than under AB stacking.}
  \label{tab:phonon}
\end{table}
\renewcommand{\arraystretch}{1.2}
\begin{table*}[htbp]
\centering\captionsetup{
  justification=centering,
  singlelinecheck=false
}
\caption{\textbf{Summary of confined water systems, simulation details, and representative illustrations in the low-density phase}. For each system, we report the total number of atoms ($N_{\mathrm{atoms}}$), the number of water molecules ($N_{\mathrm{H_2O}}$), the number of carbon atoms ($N_{\mathrm{C}}$), the simulation times of MD, PIMD, and Te PIGS simulations ($t_{\mathrm{sim\_MD}}$, $t_{\mathrm{sim\_PIMD}}$, and $t_{\mathrm{sim\_Te\ PIGS}}$), the 2D number density of water molecules ($\rho_{\mathrm{H_2O}}^{\mathrm{2D}}$), and the vacuum-slab size.}
\label{tab:info_low_density}
\begin{threeparttable}
\setlength{\tabcolsep}{3pt}
\begin{tabular}{lll}
\toprule
\textbf{System} & \textbf{Simulation details} & \textbf{Illustration} \\
\textbf{(dimensions)} & & \\
\midrule
\parbox[c]{0.28\textwidth}{\centering
Monolayer water confined between\\[0.4em]
AA-stacked graphene sheets\\[0.6em]
$(34.22~\text{\AA} \times 39.52~\text{\AA} \times 30.00~\text{\AA})$
}
&
\parbox[c]{0.28\textwidth}{\centering
$N_{\mathrm{atoms}} = 1408$\\
$N_{\mathrm{C}} = 1024$\\
$N_{\mathrm{H_2O}} = 128$\\
$ t_{\substack{\mathrm{sim\_MD}(100\,\mathrm{K}-240\,\mathrm{K};\\320\,\mathrm{K}-600\,\mathrm{K})}} = 120~\mathrm{ps} $
$t_{\mathrm{sim\_MD\,(260\,\mathrm{K}-300\,\mathrm{K})}} = 500~\mathrm{ps}$\\
$t_{\mathrm{sim\_PIMD}} = 100~\mathrm{ps}$\\
$t_{\mathrm{sim\_Te\ PIGS}} = 500~\mathrm{ps}$\\
$\rho_{\mathrm{H_2O}}^{\mathrm{2D}} = 0.095~\mathrm{molecule}/\text{\AA}^2$\\
$\mathrm{Vacuum~slab} = 25~\text{\AA}$
}
&
\parbox[c]{0.24\textwidth}{\centering
\includegraphics[width=0.8\linewidth]{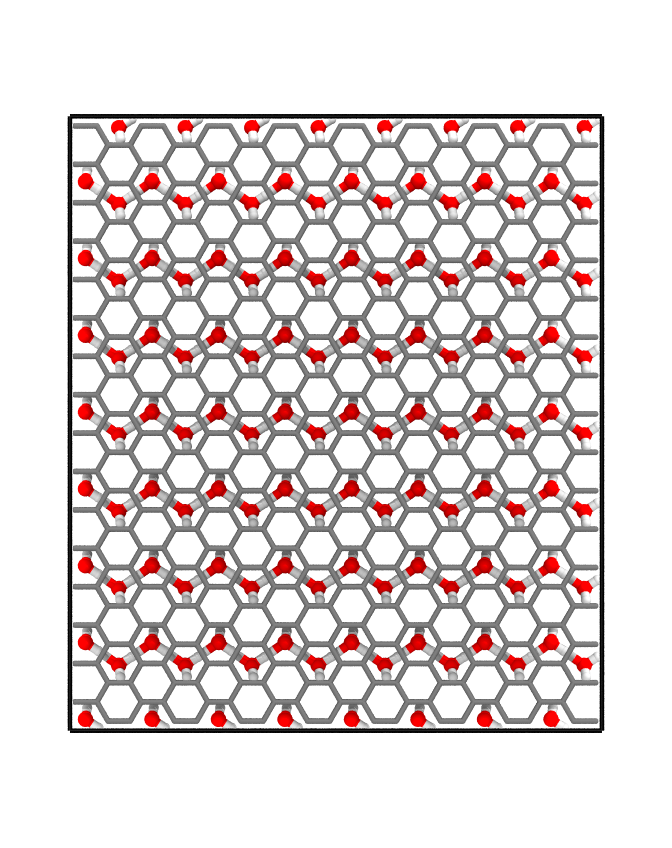}\\[0.6em]
\includegraphics[width=0.8\linewidth]{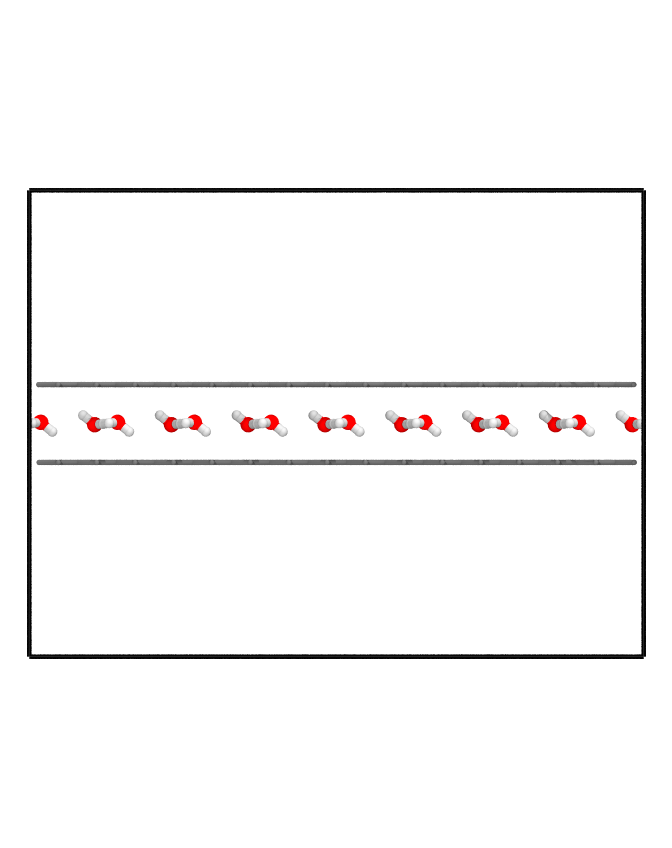}
}
\\
\midrule
\parbox[c]{0.28\textwidth}{\centering
Monolayer water confined between\\[0.4em]
AB-stacked graphene sheets\\[0.6em]
$(34.22~\text{\AA} \times 39.52~\text{\AA} \times 30.00~\text{\AA})$
}
&
\parbox[c]{0.28\textwidth}{\centering
$N_{\mathrm{atoms}} = 1408$\\
$N_{\mathrm{C}} = 1024$\\
$N_{\mathrm{H_2O}} = 128$\\
$t_{\mathrm{sim\_MD\,(100\,\mathrm{K};\,220\,\mathrm{K}-600\,\mathrm{K})}} = 120~\mathrm{ps}$\\
$t_{\mathrm{sim\_MD\,(120\,\mathrm{K}-200\,\mathrm{K})}} = 500~\mathrm{ps}$\\
$t_{\mathrm{sim\_PIMD}} = 100~\mathrm{ps}$\\
$t_{\mathrm{sim\_Te\ PIGS}} = 500~\mathrm{ps}$\\
$\rho_{\mathrm{H_2O}}^{\mathrm{2D}} = 0.095~\mathrm{molecule}/\text{\AA}^2$\\
$\mathrm{Vacuum~slab} = 25~\text{\AA}$
}
&
\parbox[c]{0.24\textwidth}{\centering
\includegraphics[width=0.8\linewidth]{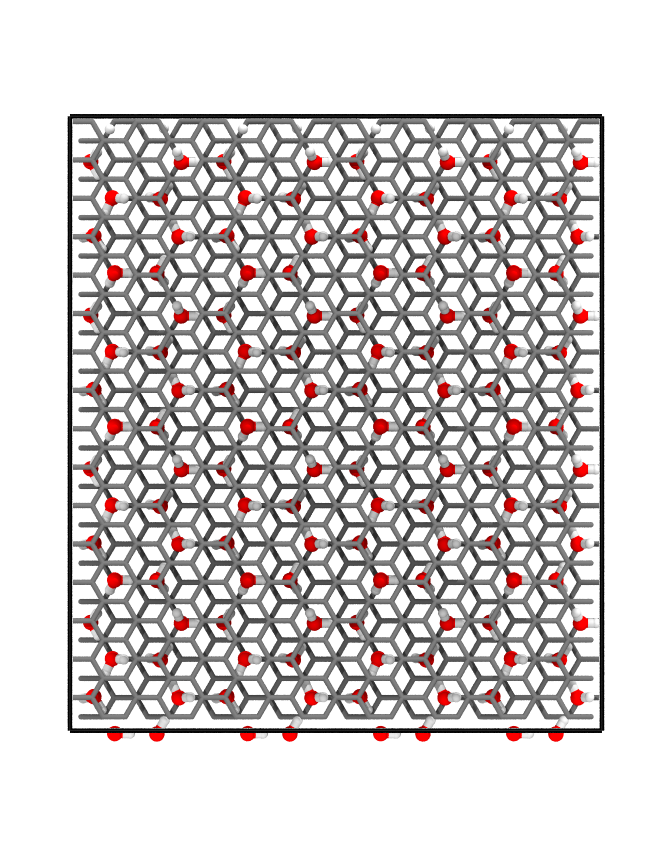}\\[0.6em]
\includegraphics[width=0.8\linewidth]{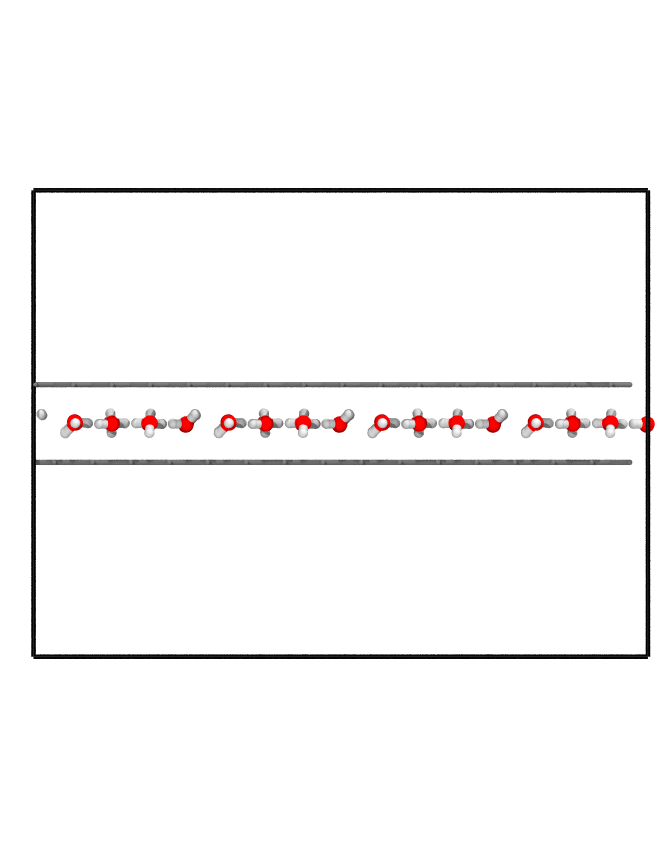}
}
\\
\bottomrule
\end{tabular}
\end{threeparttable}
\end{table*}
\renewcommand{\arraystretch}{1.2}
\begin{table*}[htbp]
\centering
\centering\captionsetup{
  justification=centering,
  singlelinecheck=false
}
\caption{\textbf{Summary of confined water systems, simulation details, and representative illustrations in the intermediate-density phase.} For each system, we report the total number of atoms ($N_{\mathrm{atoms}}$), the number of water molecules ($N_{\mathrm{H_2O}}$), the number of carbon atoms ($N_{\mathrm{C}}$), the simulation times of MD, PIMD, and Te PIGS simulations ($t_{\mathrm{sim\_MD}}$, $t_{\mathrm{sim\_PIMD}}$, and $t_{\mathrm{sim\_Te\ PIGS}}$), the 2D number density of water molecules ($\rho_{\mathrm{H_2O}}^{\mathrm{2D}}$), and the vacuum-slab size.}
\label{tab:info_intermediate_density}
\begin{threeparttable}
\setlength{\tabcolsep}{3pt}
\begin{tabular}{lll}
\toprule
\textbf{System} & \textbf{Simulation details} & \textbf{Illustration} \\
\textbf{(dimensions)} & & \\
\midrule
\parbox[c]{0.28\textwidth}{\centering
Monolayer water confined between\\[0.4em]
AA-stacked graphene sheets\\[0.6em]
$(34.22~\text{\AA} \times 39.52~\text{\AA} \times 30.00~\text{\AA})$
}
&
\parbox[c]{0.28\textwidth}{\centering
$N_{\mathrm{atoms}} = 1600$\\
$N_{\mathrm{C}} = 1024$\\
$N_{\mathrm{H_2O}} = 192$\\
$ t_{\substack{\mathrm{sim\_MD}(100\,\mathrm{K}-380\,\mathrm{K};\\460\,\mathrm{K}-600\,\mathrm{K})}} = 120~\mathrm{ps} $
$= 120~\mathrm{ps}$\\
$t_{\mathrm{sim\_MD\,(400~\mathrm{K} - 440~\mathrm{K})}} = 500~\mathrm{ps}$\\
$t_{\mathrm{sim\_PIMD}} = 100~\mathrm{ps}$\\
$t_{\mathrm{sim\_Te\ PIGS}} = 2~\mathrm{ns}$\\
$\rho_{\mathrm{H_2O}}^{\mathrm{2D}} = 0.142~\mathrm{molecule}/\text{\AA}^2$\\
$\mathrm{Vacuum~slab} = 25~\text{\AA}$
}
&
\parbox[c]{0.24\textwidth}{\centering
\includegraphics[width=0.8\linewidth]{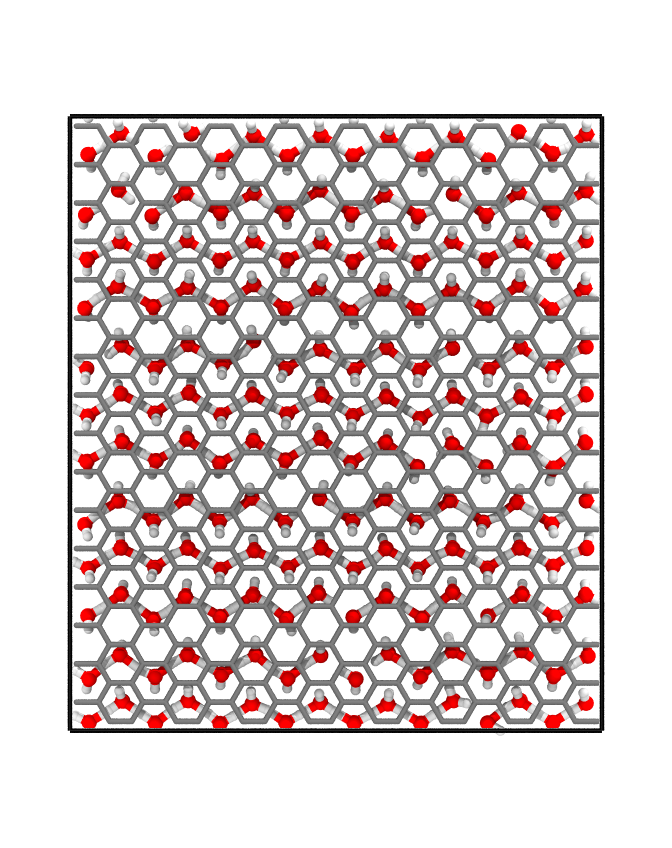}\\[0.6em]
\includegraphics[width=0.8\linewidth]{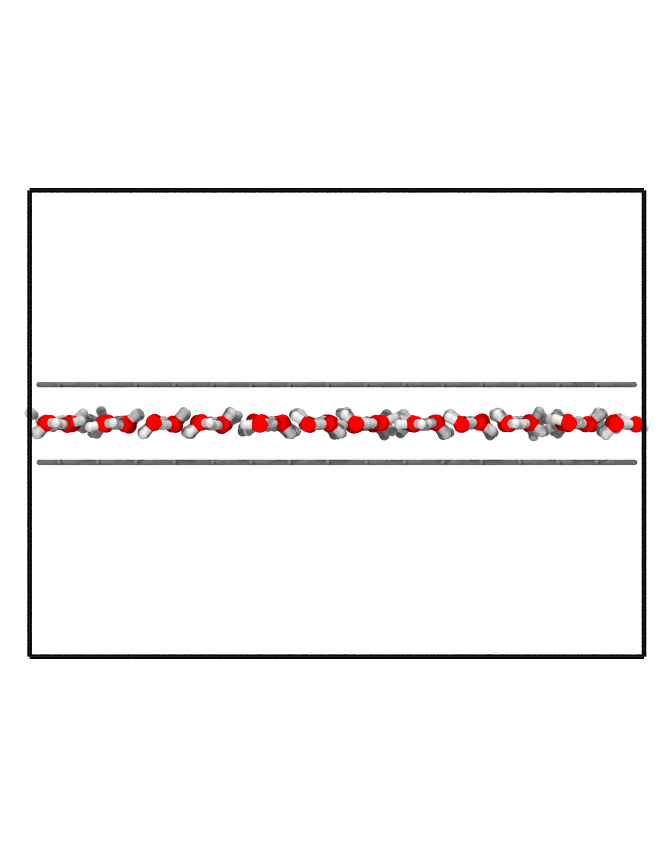}
}
\\
\midrule
\parbox[c]{0.28\textwidth}{\centering
Monolayer water confined between\\[0.4em]
AB-stacked graphene sheets\\[0.6em]
$(34.22~\text{\AA} \times 39.52~\text{\AA} \times 30.00~\text{\AA})$
}
&
\parbox[c]{0.28\textwidth}{\centering
$N_{\mathrm{atoms}} = 1600$\\
$N_{\mathrm{C}} = 1024$\\
$N_{\mathrm{H_2O}} = 192$\\
$ t_{\substack{\mathrm{sim\_MD}(100\,\mathrm{K}-320\,\mathrm{K};\\460\,\mathrm{K}-600\,\mathrm{K})}} = 120~\mathrm{ps} $
$t_{\mathrm{sim\_MD\,(340~\mathrm{K} - 440~\mathrm{K})}} = 500~\mathrm{ps}$\\
$t_{\mathrm{sim\_PIMD}} = 100~\mathrm{ps}$\\
$t_{\mathrm{sim\_Te\ PIGS}} = 2~\mathrm{ns}$\\
$\rho_{\mathrm{H_2O}}^{\mathrm{2D}} = 0.142~\mathrm{molecule}/\text{\AA}^2$\\
$\mathrm{Vacuum~slab} = 25~\text{\AA}$
}
&
\parbox[c]{0.24\textwidth}{\centering
\includegraphics[width=0.8\linewidth]{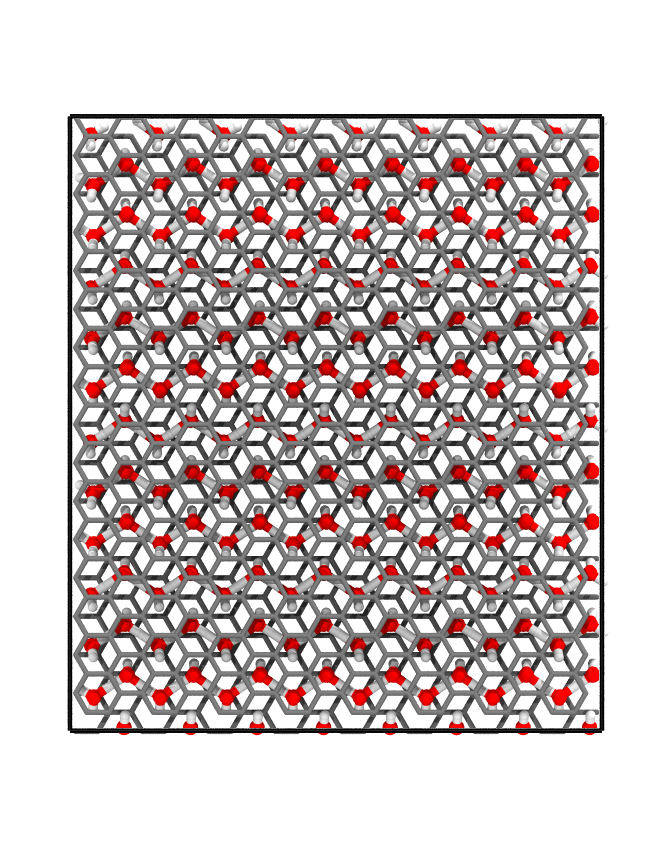}\\[0.6em]
\includegraphics[width=0.8\linewidth]{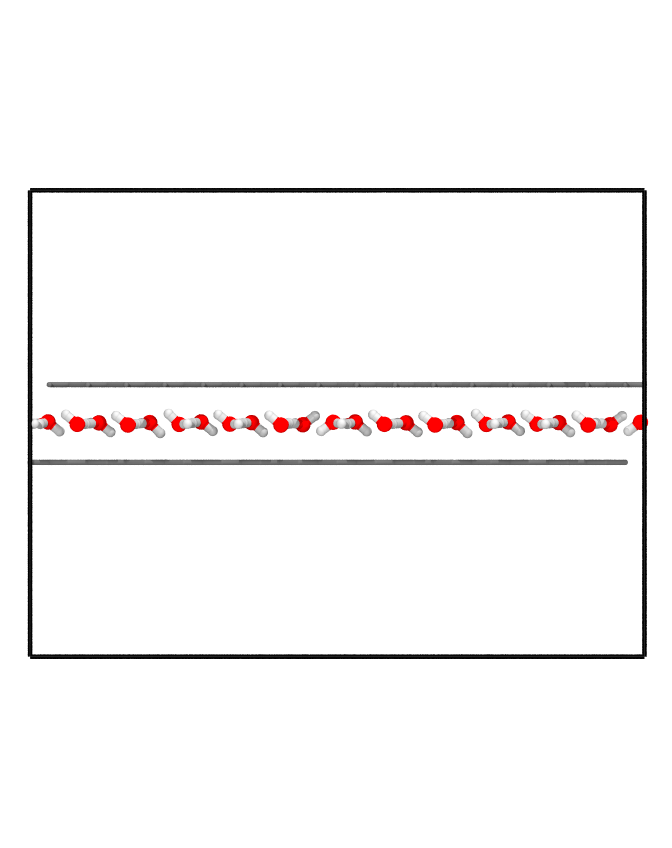}
}
\\
\bottomrule
\end{tabular}
\end{threeparttable}
\end{table*}
\renewcommand{\arraystretch}{1.2}
\begin{table*}[htbp]
\centering
\centering\captionsetup{
  justification=centering,
  singlelinecheck=false
}
\caption{\textbf{Summary of confined water systems, simulation details, and representative illustrations in the high-density phase.} For each system, we report the total number of atoms ($N_{\mathrm{atoms}}$), the number of water molecules ($N_{\mathrm{H_2O}}$), the number of carbon atoms ($N_{\mathrm{C}}$), the simulation times of MD, PIMD, and Te PIGS simulations ($t_{\mathrm{sim\_MD}}$, $t_{\mathrm{sim\_PIMD}}$, and $t_{\mathrm{sim\_Te\ PIGS}}$), the 2D number density of water molecules ($\rho_{\mathrm{H_2O}}^{\mathrm{2D}}$), and the vacuum-slab size.}
\label{tab:info_high_density}
\begin{threeparttable}
\setlength{\tabcolsep}{3pt}
\begin{tabular}{lll}
\toprule
\textbf{System} & \textbf{Simulation details} & \textbf{Illustration} \\
\textbf{(dimensions)} & & \\
\midrule
\parbox[c]{0.28\textwidth}{\centering
Monolayer water confined between\\[0.4em]
AA-stacked graphene sheets\\[0.6em]
$(34.22~\text{\AA} \times 39.52~\text{\AA} \times 30.00~\text{\AA})$
}
&
\parbox[c]{0.28\textwidth}{\centering
$N_{\mathrm{atoms}} = 1792$\\
$N_{\mathrm{C}} = 1024$\\
$N_{\mathrm{H_2O}} = 256$\\
$t_{\mathrm{sim\_MD\,(100~\mathrm{K} - 600~\mathrm{K})}} = 120~\mathrm{ps}$\\
$t_{\mathrm{sim\_PIMD}} = 100~\mathrm{ps}$\\
$t_{\mathrm{sim\_Te\ PIGS}} = 1~\mathrm{ns}$\\
$\rho_{\mathrm{H_2O}}^{\mathrm{2D}} = 0.189~\mathrm{molecule}/\text{\AA}^2$\\
$\mathrm{Vacuum~slab} = 25~\text{\AA}$
}
&
\parbox[c]{0.24\textwidth}{\centering
\includegraphics[width=0.8\linewidth]{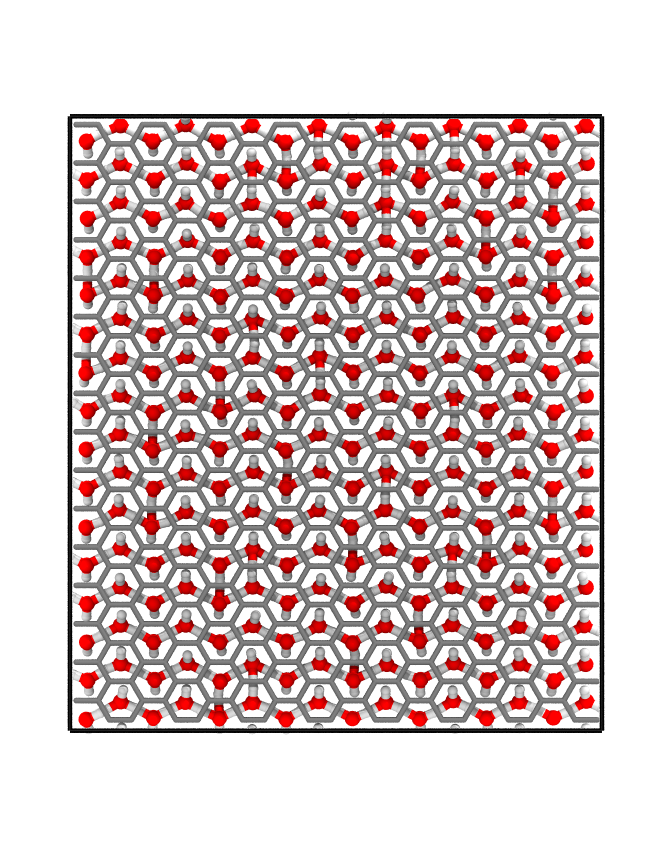}\\[0.6em]
\includegraphics[width=0.8\linewidth]{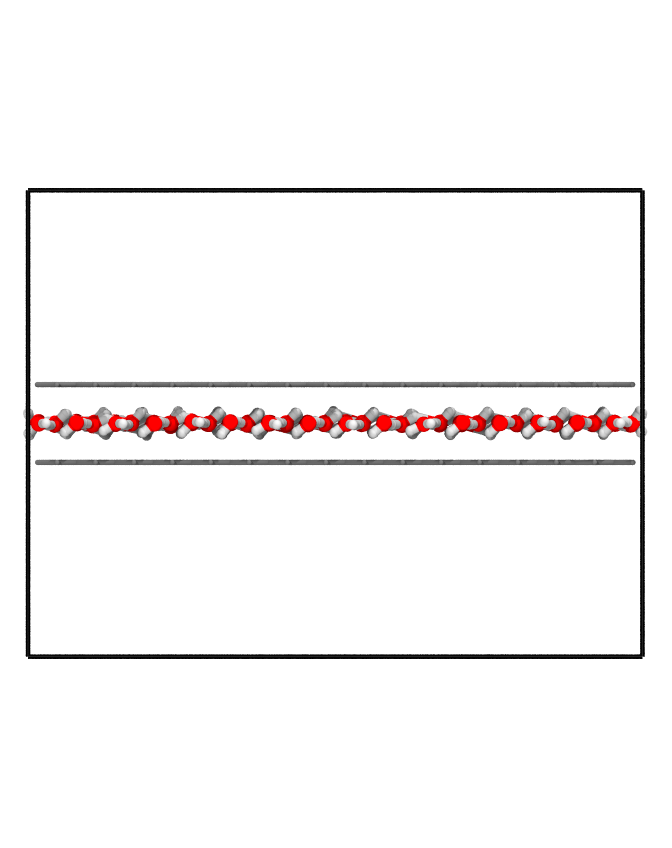}
}
\\
\midrule
\parbox[c]{0.28\textwidth}{\centering
Monolayer water confined between\\[0.4em]
AB-stacked graphene sheets\\[0.6em]
$(34.22~\text{\AA} \times 39.52~\text{\AA} \times 30.00~\text{\AA})$
}
&
\parbox[c]{0.28\textwidth}{\centering
$N_{\mathrm{atoms}} = 1792$\\
$N_{\mathrm{C}} = 1024$\\
$N_{\mathrm{H_2O}} = 256$\\
$t_{\mathrm{sim\_MD\,(100~\mathrm{K} - 600~\mathrm{K})}} = 120~\mathrm{ps}$\\
$t_{\mathrm{sim\_PIMD}} = 100~\mathrm{ps}$\\
$t_{\mathrm{sim\_Te\ PIGS}} = 1~\mathrm{ns}$\\
$\rho_{\mathrm{H_2O}}^{\mathrm{2D}} = 0.189~\mathrm{molecule}/\text{\AA}^2$\\
$\mathrm{Vacuum~slab} = 25~\text{\AA}$
}
&
\parbox[c]{0.24\textwidth}{\centering
\includegraphics[width=0.8\linewidth]{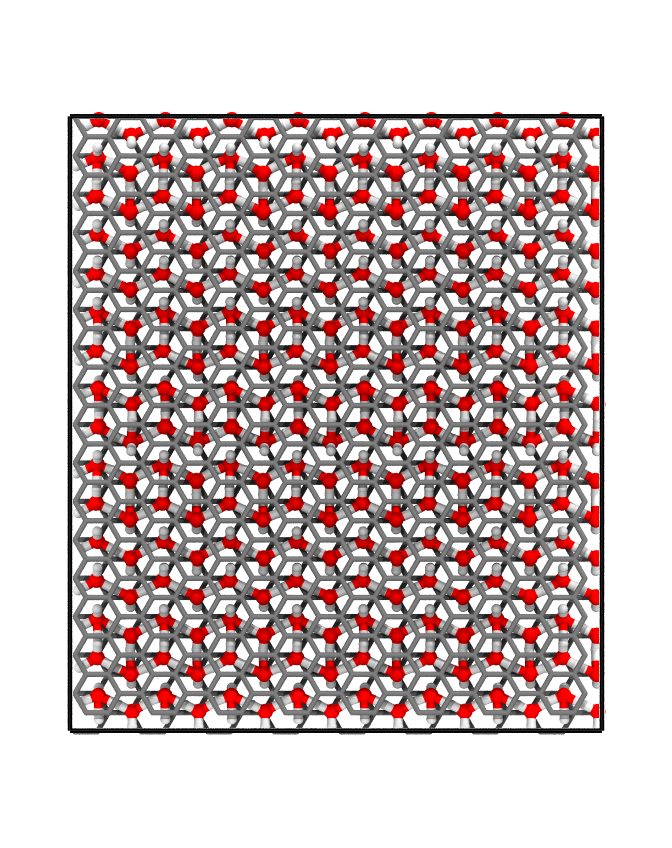}\\[0.6em]
\includegraphics[width=0.8\linewidth]{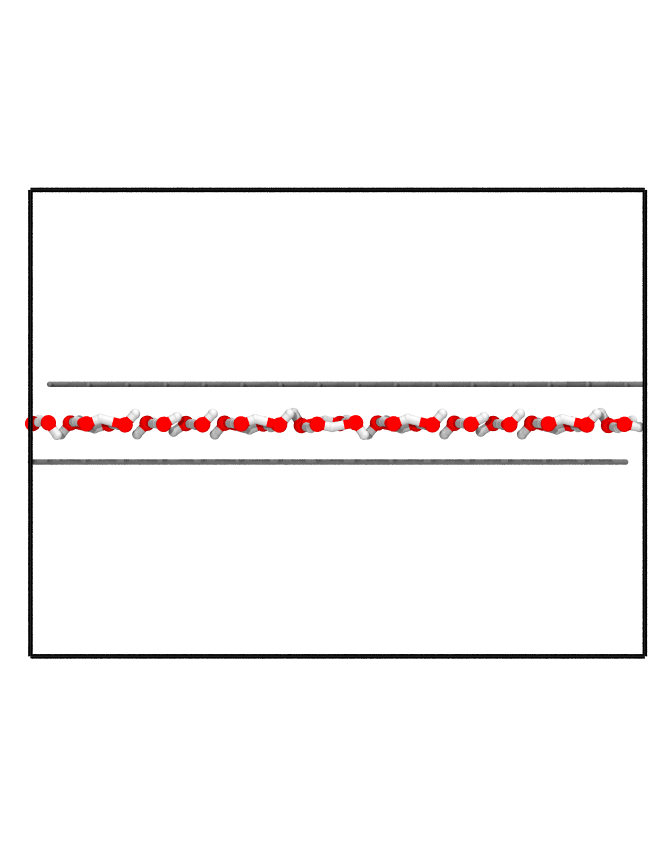}
}
\\
\bottomrule
\end{tabular}
\end{threeparttable}
\end{table*}

\begin{thebibliography}{78}%
\makeatletter
\providecommand \@ifxundefined [1]{%
 \@ifx{#1\undefined}
}%
\providecommand \@ifnum [1]{%
 \ifnum #1\expandafter \@firstoftwo
 \else \expandafter \@secondoftwo
 \fi
}%
\providecommand \@ifx [1]{%
 \ifx #1\expandafter \@firstoftwo
 \else \expandafter \@secondoftwo
 \fi
}%
\providecommand \natexlab [1]{#1}%
\providecommand \enquote  [1]{``#1''}%
\providecommand \bibnamefont  [1]{#1}%
\providecommand \bibfnamefont [1]{#1}%
\providecommand \citenamefont [1]{#1}%
\providecommand \href@noop [0]{\@secondoftwo}%
\providecommand \href [0]{\begingroup \@sanitize@url \@href}%
\providecommand \@href[1]{\@@startlink{#1}\@@href}%
\providecommand \@@href[1]{\endgroup#1\@@endlink}%
\providecommand \@sanitize@url [0]{\catcode `\\12\catcode `\$12\catcode
  `\&12\catcode `\#12\catcode `\^12\catcode `\_12\catcode `\%12\relax}%
\providecommand \@@startlink[1]{}%
\providecommand \@@endlink[0]{}%
\providecommand \url  [0]{\begingroup\@sanitize@url \@url }%
\providecommand \@url [1]{\endgroup\@href {#1}{\urlprefix }}%
\providecommand \urlprefix  [0]{URL }%
\providecommand \Eprint [0]{\href }%
\providecommand \doibase [0]{https://doi.org/}%
\providecommand \selectlanguage [0]{\@gobble}%
\providecommand \bibinfo  [0]{\@secondoftwo}%
\providecommand \bibfield  [0]{\@secondoftwo}%
\providecommand \translation [1]{[#1]}%
\providecommand \BibitemOpen [0]{}%
\providecommand \bibitemStop [0]{}%
\providecommand \bibitemNoStop [0]{.\EOS\space}%
\providecommand \EOS [0]{\spacefactor3000\relax}%
\providecommand \BibitemShut  [1]{\csname bibitem#1\endcsname}%
\let\auto@bib@innerbib\@empty
\bibitem [{\citenamefont {Faucher}\ \emph {et~al.}(2019)\citenamefont
  {Faucher}, \citenamefont {Aluru}, \citenamefont {Bazant}, \citenamefont
  {Blankschtein}, \citenamefont {Brozena}, \citenamefont {Cumings},
  \citenamefont {Pedro~de Souza}, \citenamefont {Elimelech}, \citenamefont
  {Epsztein}, \citenamefont {Fourkas}, \citenamefont {Rajan}, \citenamefont
  {Kulik}, \citenamefont {Levy}, \citenamefont {Majumdar}, \citenamefont
  {Martin}, \citenamefont {McEldrew}, \citenamefont {Misra}, \citenamefont
  {Noy}, \citenamefont {Pham}, \citenamefont {Reed}, \citenamefont {Schwegler},
  \citenamefont {Siwy}, \citenamefont {Wang},\ and\ \citenamefont
  {Strano}}]{faucher_critical_2019}%
  \BibitemOpen
  \bibfield  {author} {\bibinfo {author} {\bibfnamefont {S.}~\bibnamefont
  {Faucher}}, \bibinfo {author} {\bibfnamefont {N.}~\bibnamefont {Aluru}},
  \bibinfo {author} {\bibfnamefont {M.~Z.}\ \bibnamefont {Bazant}}, \bibinfo
  {author} {\bibfnamefont {D.}~\bibnamefont {Blankschtein}}, \bibinfo {author}
  {\bibfnamefont {A.~H.}\ \bibnamefont {Brozena}}, \bibinfo {author}
  {\bibfnamefont {J.}~\bibnamefont {Cumings}}, \bibinfo {author} {\bibfnamefont
  {J.}~\bibnamefont {Pedro~de Souza}}, \bibinfo {author} {\bibfnamefont
  {M.}~\bibnamefont {Elimelech}}, \bibinfo {author} {\bibfnamefont
  {R.}~\bibnamefont {Epsztein}}, \bibinfo {author} {\bibfnamefont {J.~T.}\
  \bibnamefont {Fourkas}}, \bibinfo {author} {\bibfnamefont {A.~G.}\
  \bibnamefont {Rajan}}, \bibinfo {author} {\bibfnamefont {H.~J.}\ \bibnamefont
  {Kulik}}, \bibinfo {author} {\bibfnamefont {A.}~\bibnamefont {Levy}},
  \bibinfo {author} {\bibfnamefont {A.}~\bibnamefont {Majumdar}}, \bibinfo
  {author} {\bibfnamefont {C.}~\bibnamefont {Martin}}, \bibinfo {author}
  {\bibfnamefont {M.}~\bibnamefont {McEldrew}}, \bibinfo {author}
  {\bibfnamefont {R.~P.}\ \bibnamefont {Misra}}, \bibinfo {author}
  {\bibfnamefont {A.}~\bibnamefont {Noy}}, \bibinfo {author} {\bibfnamefont
  {T.~A.}\ \bibnamefont {Pham}}, \bibinfo {author} {\bibfnamefont
  {M.}~\bibnamefont {Reed}}, \bibinfo {author} {\bibfnamefont {E.}~\bibnamefont
  {Schwegler}}, \bibinfo {author} {\bibfnamefont {Z.}~\bibnamefont {Siwy}},
  \bibinfo {author} {\bibfnamefont {Y.}~\bibnamefont {Wang}},\ and\ \bibinfo
  {author} {\bibfnamefont {M.}~\bibnamefont {Strano}},\ }\bibfield  {title}
  {\bibinfo {title} {Critical {Knowledge} {Gaps} in {Mass} {Transport} through
  {Single}-{Digit} {Nanopores}: {A} {Review} and {Perspective}},\ }\href
  {https://doi.org/10.1021/acs.jpcc.9b02178} {\bibfield  {journal} {\bibinfo
  {journal} {The Journal of Physical Chemistry C}\ }\textbf {\bibinfo {volume}
  {123}},\ \bibinfo {pages} {21309} (\bibinfo {year} {2019})}\BibitemShut
  {NoStop}%
\bibitem [{\citenamefont {Aluru}\ \emph {et~al.}(2023)\citenamefont {Aluru},
  \citenamefont {Aydin}, \citenamefont {Bazant}, \citenamefont {Blankschtein},
  \citenamefont {Brozena}, \citenamefont {de~Souza}, \citenamefont {Elimelech},
  \citenamefont {Faucher}, \citenamefont {Fourkas}, \citenamefont {Koman},
  \citenamefont {Kuehne}, \citenamefont {Kulik}, \citenamefont {Li},
  \citenamefont {Li}, \citenamefont {Li}, \citenamefont {Majumdar},
  \citenamefont {Martis}, \citenamefont {Misra}, \citenamefont {Noy},
  \citenamefont {Pham}, \citenamefont {Qu}, \citenamefont {Rayabharam},
  \citenamefont {Reed}, \citenamefont {Ritt}, \citenamefont {Schwegler},
  \citenamefont {Siwy}, \citenamefont {Strano}, \citenamefont {Wang},
  \citenamefont {Yao}, \citenamefont {Zhan},\ and\ \citenamefont
  {Zhang}}]{aluru_fluids_2023}%
  \BibitemOpen
  \bibfield  {author} {\bibinfo {author} {\bibfnamefont {N.~R.}\ \bibnamefont
  {Aluru}}, \bibinfo {author} {\bibfnamefont {F.}~\bibnamefont {Aydin}},
  \bibinfo {author} {\bibfnamefont {M.~Z.}\ \bibnamefont {Bazant}}, \bibinfo
  {author} {\bibfnamefont {D.}~\bibnamefont {Blankschtein}}, \bibinfo {author}
  {\bibfnamefont {A.~H.}\ \bibnamefont {Brozena}}, \bibinfo {author}
  {\bibfnamefont {J.~P.}\ \bibnamefont {de~Souza}}, \bibinfo {author}
  {\bibfnamefont {M.}~\bibnamefont {Elimelech}}, \bibinfo {author}
  {\bibfnamefont {S.}~\bibnamefont {Faucher}}, \bibinfo {author} {\bibfnamefont
  {J.~T.}\ \bibnamefont {Fourkas}}, \bibinfo {author} {\bibfnamefont {V.~B.}\
  \bibnamefont {Koman}}, \bibinfo {author} {\bibfnamefont {M.}~\bibnamefont
  {Kuehne}}, \bibinfo {author} {\bibfnamefont {H.~J.}\ \bibnamefont {Kulik}},
  \bibinfo {author} {\bibfnamefont {H.-K.}\ \bibnamefont {Li}}, \bibinfo
  {author} {\bibfnamefont {Y.}~\bibnamefont {Li}}, \bibinfo {author}
  {\bibfnamefont {Z.}~\bibnamefont {Li}}, \bibinfo {author} {\bibfnamefont
  {A.}~\bibnamefont {Majumdar}}, \bibinfo {author} {\bibfnamefont
  {J.}~\bibnamefont {Martis}}, \bibinfo {author} {\bibfnamefont {R.~P.}\
  \bibnamefont {Misra}}, \bibinfo {author} {\bibfnamefont {A.}~\bibnamefont
  {Noy}}, \bibinfo {author} {\bibfnamefont {T.~A.}\ \bibnamefont {Pham}},
  \bibinfo {author} {\bibfnamefont {H.}~\bibnamefont {Qu}}, \bibinfo {author}
  {\bibfnamefont {A.}~\bibnamefont {Rayabharam}}, \bibinfo {author}
  {\bibfnamefont {M.~A.}\ \bibnamefont {Reed}}, \bibinfo {author}
  {\bibfnamefont {C.~L.}\ \bibnamefont {Ritt}}, \bibinfo {author}
  {\bibfnamefont {E.}~\bibnamefont {Schwegler}}, \bibinfo {author}
  {\bibfnamefont {Z.}~\bibnamefont {Siwy}}, \bibinfo {author} {\bibfnamefont
  {M.~S.}\ \bibnamefont {Strano}}, \bibinfo {author} {\bibfnamefont
  {Y.}~\bibnamefont {Wang}}, \bibinfo {author} {\bibfnamefont {Y.-C.}\
  \bibnamefont {Yao}}, \bibinfo {author} {\bibfnamefont {C.}~\bibnamefont
  {Zhan}},\ and\ \bibinfo {author} {\bibfnamefont {Z.}~\bibnamefont {Zhang}},\
  }\bibfield  {title} {\bibinfo {title} {Fluids and {Electrolytes} under
  {Confinement} in {Single}-{Digit} {Nanopores}},\ }\href
  {https://doi.org/10.1021/acs.chemrev.2c00155} {\bibfield  {journal} {\bibinfo
   {journal} {Chemical Reviews}\ }\textbf {\bibinfo {volume} {123}},\ \bibinfo
  {pages} {2737} (\bibinfo {year} {2023})}\BibitemShut {NoStop}%
\bibitem [{\citenamefont {Trushin}\ \emph {et~al.}(2025)\citenamefont
  {Trushin}, \citenamefont {Andreeva}, \citenamefont {Peeters},\ and\
  \citenamefont {Novoselov}}]{trushin_structure_2025}%
  \BibitemOpen
  \bibfield  {author} {\bibinfo {author} {\bibfnamefont {M.}~\bibnamefont
  {Trushin}}, \bibinfo {author} {\bibfnamefont {D.~V.}\ \bibnamefont
  {Andreeva}}, \bibinfo {author} {\bibfnamefont {F.~M.}\ \bibnamefont
  {Peeters}},\ and\ \bibinfo {author} {\bibfnamefont {K.~S.}\ \bibnamefont
  {Novoselov}},\ }\bibfield  {title} {{\selectlanguage {en}\bibinfo {title}
  {Structure and flow of low-dimensional water}},\ }\href
  {https://doi.org/10.1038/s42254-025-00857-x} {\bibfield  {journal} {\bibinfo
  {journal} {Nature Reviews Physics}\ }\textbf {\bibinfo {volume} {7}},\
  \bibinfo {pages} {502} (\bibinfo {year} {2025})}\BibitemShut {NoStop}%
\bibitem [{\citenamefont {Bocquet}(2020)}]{bocquet_nanofluidics_2020}%
  \BibitemOpen
  \bibfield  {author} {\bibinfo {author} {\bibfnamefont {L.}~\bibnamefont
  {Bocquet}},\ }\bibfield  {title} {{\selectlanguage {en}\bibinfo {title}
  {Nanofluidics coming of age}},\ }\href
  {https://doi.org/10.1038/s41563-020-0625-8} {\bibfield  {journal} {\bibinfo
  {journal} {Nature Materials}\ }\textbf {\bibinfo {volume} {19}},\ \bibinfo
  {pages} {254} (\bibinfo {year} {2020})}\BibitemShut {NoStop}%
\bibitem [{\citenamefont {Majumder}\ \emph {et~al.}(2005)\citenamefont
  {Majumder}, \citenamefont {Chopra}, \citenamefont {Andrews},\ and\
  \citenamefont {Hinds}}]{majumder_enhanced_2005}%
  \BibitemOpen
  \bibfield  {author} {\bibinfo {author} {\bibfnamefont {M.}~\bibnamefont
  {Majumder}}, \bibinfo {author} {\bibfnamefont {N.}~\bibnamefont {Chopra}},
  \bibinfo {author} {\bibfnamefont {R.}~\bibnamefont {Andrews}},\ and\ \bibinfo
  {author} {\bibfnamefont {B.~J.}\ \bibnamefont {Hinds}},\ }\bibfield  {title}
  {{\selectlanguage {en}\bibinfo {title} {Enhanced flow in carbon nanotubes}},\
  }\href {https://doi.org/10.1038/438044a} {\bibfield  {journal} {\bibinfo
  {journal} {Nature}\ }\textbf {\bibinfo {volume} {438}},\ \bibinfo {pages}
  {44} (\bibinfo {year} {2005})}\BibitemShut {NoStop}%
\bibitem [{\citenamefont {Whitby}\ and\ \citenamefont
  {Quirke}(2007)}]{whitby_fluid_2007}%
  \BibitemOpen
  \bibfield  {author} {\bibinfo {author} {\bibfnamefont {M.}~\bibnamefont
  {Whitby}}\ and\ \bibinfo {author} {\bibfnamefont {N.}~\bibnamefont
  {Quirke}},\ }\bibfield  {title} {{\selectlanguage {en}\bibinfo {title} {Fluid
  flow in carbon nanotubes and nanopipes}},\ }\href
  {https://doi.org/10.1038/nnano.2006.175} {\bibfield  {journal} {\bibinfo
  {journal} {Nature Nanotechnology}\ }\textbf {\bibinfo {volume} {2}},\
  \bibinfo {pages} {87} (\bibinfo {year} {2007})}\BibitemShut {NoStop}%
\bibitem [{\citenamefont {Holt}\ \emph {et~al.}(2006)\citenamefont {Holt},
  \citenamefont {Park}, \citenamefont {Wang}, \citenamefont {Stadermann},
  \citenamefont {Artyukhin}, \citenamefont {Grigoropoulos}, \citenamefont
  {Noy},\ and\ \citenamefont {Bakajin}}]{holt_fast_2006}%
  \BibitemOpen
  \bibfield  {author} {\bibinfo {author} {\bibfnamefont {J.~K.}\ \bibnamefont
  {Holt}}, \bibinfo {author} {\bibfnamefont {H.~G.}\ \bibnamefont {Park}},
  \bibinfo {author} {\bibfnamefont {Y.}~\bibnamefont {Wang}}, \bibinfo {author}
  {\bibfnamefont {M.}~\bibnamefont {Stadermann}}, \bibinfo {author}
  {\bibfnamefont {A.~B.}\ \bibnamefont {Artyukhin}}, \bibinfo {author}
  {\bibfnamefont {C.~P.}\ \bibnamefont {Grigoropoulos}}, \bibinfo {author}
  {\bibfnamefont {A.}~\bibnamefont {Noy}},\ and\ \bibinfo {author}
  {\bibfnamefont {O.}~\bibnamefont {Bakajin}},\ }\bibfield  {title}
  {{\selectlanguage {en}\bibinfo {title} {Fast {Mass} {Transport} {Through}
  {Sub}-2-{Nanometer} {Carbon} {Nanotubes}}},\ }\href
  {https://doi.org/10.1126/science.1126298} {\bibfield  {journal} {\bibinfo
  {journal} {Science}\ }\textbf {\bibinfo {volume} {312}},\ \bibinfo {pages}
  {1034} (\bibinfo {year} {2006})}\BibitemShut {NoStop}%
\bibitem [{\citenamefont {Peng}\ \emph {et~al.}(2018)\citenamefont {Peng},
  \citenamefont {Cao}, \citenamefont {He}, \citenamefont {Guo}, \citenamefont
  {Hapala}, \citenamefont {Ma}, \citenamefont {Cheng}, \citenamefont {Chen},
  \citenamefont {Xie}, \citenamefont {Li}, \citenamefont {Jelínek},
  \citenamefont {Xu}, \citenamefont {Gao}, \citenamefont {Wang},\ and\
  \citenamefont {Jiang}}]{peng_publisher_2018}%
  \BibitemOpen
  \bibfield  {author} {\bibinfo {author} {\bibfnamefont {J.}~\bibnamefont
  {Peng}}, \bibinfo {author} {\bibfnamefont {D.}~\bibnamefont {Cao}}, \bibinfo
  {author} {\bibfnamefont {Z.}~\bibnamefont {He}}, \bibinfo {author}
  {\bibfnamefont {J.}~\bibnamefont {Guo}}, \bibinfo {author} {\bibfnamefont
  {P.}~\bibnamefont {Hapala}}, \bibinfo {author} {\bibfnamefont
  {R.}~\bibnamefont {Ma}}, \bibinfo {author} {\bibfnamefont {B.}~\bibnamefont
  {Cheng}}, \bibinfo {author} {\bibfnamefont {J.}~\bibnamefont {Chen}},
  \bibinfo {author} {\bibfnamefont {W.~J.}\ \bibnamefont {Xie}}, \bibinfo
  {author} {\bibfnamefont {X.-Z.}\ \bibnamefont {Li}}, \bibinfo {author}
  {\bibfnamefont {P.}~\bibnamefont {Jelínek}}, \bibinfo {author}
  {\bibfnamefont {L.-M.}\ \bibnamefont {Xu}}, \bibinfo {author} {\bibfnamefont
  {Y.~Q.}\ \bibnamefont {Gao}}, \bibinfo {author} {\bibfnamefont {E.-G.}\
  \bibnamefont {Wang}},\ and\ \bibinfo {author} {\bibfnamefont
  {Y.}~\bibnamefont {Jiang}},\ }\bibfield  {title} {{\selectlanguage
  {en}\bibinfo {title} {Publisher {Correction}: {The} effect of hydration
  number on the interfacial transport of sodium ions}},\ }\href
  {https://doi.org/10.1038/s41586-018-0473-8} {\bibfield  {journal} {\bibinfo
  {journal} {Nature}\ }\textbf {\bibinfo {volume} {563}},\ \bibinfo {pages}
  {E18} (\bibinfo {year} {2018})}\BibitemShut {NoStop}%
\bibitem [{\citenamefont {Falk}\ \emph {et~al.}(2010)\citenamefont {Falk},
  \citenamefont {Sedlmeier}, \citenamefont {Joly}, \citenamefont {Netz},\ and\
  \citenamefont {Bocquet}}]{falk_molecular_2010}%
  \BibitemOpen
  \bibfield  {author} {\bibinfo {author} {\bibfnamefont {K.}~\bibnamefont
  {Falk}}, \bibinfo {author} {\bibfnamefont {F.}~\bibnamefont {Sedlmeier}},
  \bibinfo {author} {\bibfnamefont {L.}~\bibnamefont {Joly}}, \bibinfo {author}
  {\bibfnamefont {R.~R.}\ \bibnamefont {Netz}},\ and\ \bibinfo {author}
  {\bibfnamefont {L.}~\bibnamefont {Bocquet}},\ }\bibfield  {title} {\bibinfo
  {title} {Molecular {Origin} of {Fast} {Water} {Transport} in {Carbon}
  {Nanotube} {Membranes}: {Superlubricity} versus {Curvature} {Dependent}
  {Friction}},\ }\href {https://doi.org/10.1021/nl1021046} {\bibfield
  {journal} {\bibinfo  {journal} {Nano Letters}\ }\textbf {\bibinfo {volume}
  {10}},\ \bibinfo {pages} {4067} (\bibinfo {year} {2010})}\BibitemShut
  {NoStop}%
\bibitem [{\citenamefont {Secchi}\ \emph {et~al.}(2016)\citenamefont {Secchi},
  \citenamefont {Marbach}, \citenamefont {Niguès}, \citenamefont {Stein},
  \citenamefont {Siria},\ and\ \citenamefont {Bocquet}}]{secchi_massive_2016}%
  \BibitemOpen
  \bibfield  {author} {\bibinfo {author} {\bibfnamefont {E.}~\bibnamefont
  {Secchi}}, \bibinfo {author} {\bibfnamefont {S.}~\bibnamefont {Marbach}},
  \bibinfo {author} {\bibfnamefont {A.}~\bibnamefont {Niguès}}, \bibinfo
  {author} {\bibfnamefont {D.}~\bibnamefont {Stein}}, \bibinfo {author}
  {\bibfnamefont {A.}~\bibnamefont {Siria}},\ and\ \bibinfo {author}
  {\bibfnamefont {L.}~\bibnamefont {Bocquet}},\ }\bibfield  {title}
  {{\selectlanguage {en}\bibinfo {title} {Massive radius-dependent flow
  slippage in carbon nanotubes}},\ }\href {https://doi.org/10.1038/nature19315}
  {\bibfield  {journal} {\bibinfo  {journal} {Nature}\ }\textbf {\bibinfo
  {volume} {537}},\ \bibinfo {pages} {210} (\bibinfo {year}
  {2016})}\BibitemShut {NoStop}%
\bibitem [{\citenamefont {Siria}\ \emph {et~al.}(2013)\citenamefont {Siria},
  \citenamefont {Poncharal}, \citenamefont {Biance}, \citenamefont {Fulcrand},
  \citenamefont {Blase}, \citenamefont {Purcell},\ and\ \citenamefont
  {Bocquet}}]{siria_giant_2013}%
  \BibitemOpen
  \bibfield  {author} {\bibinfo {author} {\bibfnamefont {A.}~\bibnamefont
  {Siria}}, \bibinfo {author} {\bibfnamefont {P.}~\bibnamefont {Poncharal}},
  \bibinfo {author} {\bibfnamefont {A.-L.}\ \bibnamefont {Biance}}, \bibinfo
  {author} {\bibfnamefont {R.}~\bibnamefont {Fulcrand}}, \bibinfo {author}
  {\bibfnamefont {X.}~\bibnamefont {Blase}}, \bibinfo {author} {\bibfnamefont
  {S.~T.}\ \bibnamefont {Purcell}},\ and\ \bibinfo {author} {\bibfnamefont
  {L.}~\bibnamefont {Bocquet}},\ }\bibfield  {title} {{\selectlanguage
  {eng}\bibinfo {title} {Giant osmotic energy conversion measured in a single
  transmembrane boron nitride nanotube}},\ }\href
  {https://doi.org/10.1038/nature11876} {\bibfield  {journal} {\bibinfo
  {journal} {Nature}\ }\textbf {\bibinfo {volume} {494}},\ \bibinfo {pages}
  {455} (\bibinfo {year} {2013})}\BibitemShut {NoStop}%
\bibitem [{\citenamefont {Siria}\ \emph {et~al.}(2017)\citenamefont {Siria},
  \citenamefont {Bocquet},\ and\ \citenamefont {Bocquet}}]{siria2017new}%
  \BibitemOpen
  \bibfield  {author} {\bibinfo {author} {\bibfnamefont {A.}~\bibnamefont
  {Siria}}, \bibinfo {author} {\bibfnamefont {M.-L.}\ \bibnamefont {Bocquet}},\
  and\ \bibinfo {author} {\bibfnamefont {L.}~\bibnamefont {Bocquet}},\
  }\bibfield  {title} {\bibinfo {title} {New avenues for the large-scale
  harvesting of blue energy},\ }\href@noop {} {\bibfield  {journal} {\bibinfo
  {journal} {Nature Reviews Chemistry}\ }\textbf {\bibinfo {volume} {1}},\
  \bibinfo {pages} {0091} (\bibinfo {year} {2017})}\BibitemShut {NoStop}%
\bibitem [{\citenamefont {Fumagalli}\ \emph {et~al.}(2018)\citenamefont
  {Fumagalli}, \citenamefont {Esfandiar}, \citenamefont {Fabregas},
  \citenamefont {Hu}, \citenamefont {Ares}, \citenamefont {Janardanan},
  \citenamefont {Yang}, \citenamefont {Radha}, \citenamefont {Taniguchi},
  \citenamefont {Watanabe}, \citenamefont {Gomila}, \citenamefont {Novoselov},\
  and\ \citenamefont {Geim}}]{fumagalli_anomalously_2018}%
  \BibitemOpen
  \bibfield  {author} {\bibinfo {author} {\bibfnamefont {L.}~\bibnamefont
  {Fumagalli}}, \bibinfo {author} {\bibfnamefont {A.}~\bibnamefont
  {Esfandiar}}, \bibinfo {author} {\bibfnamefont {R.}~\bibnamefont {Fabregas}},
  \bibinfo {author} {\bibfnamefont {S.}~\bibnamefont {Hu}}, \bibinfo {author}
  {\bibfnamefont {P.}~\bibnamefont {Ares}}, \bibinfo {author} {\bibfnamefont
  {A.}~\bibnamefont {Janardanan}}, \bibinfo {author} {\bibfnamefont
  {Q.}~\bibnamefont {Yang}}, \bibinfo {author} {\bibfnamefont {B.}~\bibnamefont
  {Radha}}, \bibinfo {author} {\bibfnamefont {T.}~\bibnamefont {Taniguchi}},
  \bibinfo {author} {\bibfnamefont {K.}~\bibnamefont {Watanabe}}, \bibinfo
  {author} {\bibfnamefont {G.}~\bibnamefont {Gomila}}, \bibinfo {author}
  {\bibfnamefont {K.~S.}\ \bibnamefont {Novoselov}},\ and\ \bibinfo {author}
  {\bibfnamefont {A.~K.}\ \bibnamefont {Geim}},\ }\bibfield  {title}
  {{\selectlanguage {en}\bibinfo {title} {Anomalously low dielectric constant
  of confined water}},\ }\href {https://doi.org/10.1126/science.aat4191}
  {\bibfield  {journal} {\bibinfo  {journal} {Science}\ }\textbf {\bibinfo
  {volume} {360}},\ \bibinfo {pages} {1339} (\bibinfo {year}
  {2018})}\BibitemShut {NoStop}%
\bibitem [{\citenamefont {Wang}\ \emph
  {et~al.}(2025{\natexlab{a}})\citenamefont {Wang}, \citenamefont {Souilamas},
  \citenamefont {Esfandiar}, \citenamefont {Fabregas}, \citenamefont
  {Benaglia}, \citenamefont {Nevison-Andrews}, \citenamefont {Yang},
  \citenamefont {Normansell}, \citenamefont {Ares}, \citenamefont {Ferrari},
  \citenamefont {Principi}, \citenamefont {Geim},\ and\ \citenamefont
  {Fumagalli}}]{wang_-plane_2025}%
  \BibitemOpen
  \bibfield  {author} {\bibinfo {author} {\bibfnamefont {R.}~\bibnamefont
  {Wang}}, \bibinfo {author} {\bibfnamefont {M.}~\bibnamefont {Souilamas}},
  \bibinfo {author} {\bibfnamefont {A.}~\bibnamefont {Esfandiar}}, \bibinfo
  {author} {\bibfnamefont {R.}~\bibnamefont {Fabregas}}, \bibinfo {author}
  {\bibfnamefont {S.}~\bibnamefont {Benaglia}}, \bibinfo {author}
  {\bibfnamefont {H.}~\bibnamefont {Nevison-Andrews}}, \bibinfo {author}
  {\bibfnamefont {Q.}~\bibnamefont {Yang}}, \bibinfo {author} {\bibfnamefont
  {J.}~\bibnamefont {Normansell}}, \bibinfo {author} {\bibfnamefont
  {P.}~\bibnamefont {Ares}}, \bibinfo {author} {\bibfnamefont {G.}~\bibnamefont
  {Ferrari}}, \bibinfo {author} {\bibfnamefont {A.}~\bibnamefont {Principi}},
  \bibinfo {author} {\bibfnamefont {A.~K.}\ \bibnamefont {Geim}},\ and\
  \bibinfo {author} {\bibfnamefont {L.}~\bibnamefont {Fumagalli}},\ }\bibfield
  {title} {{\selectlanguage {en}\bibinfo {title} {In-plane dielectric constant
  and conductivity of confined water}},\ }\href
  {https://doi.org/10.1038/s41586-025-09558-y} {\bibfield  {journal} {\bibinfo
  {journal} {Nature}\ }\textbf {\bibinfo {volume} {646}},\ \bibinfo {pages}
  {606} (\bibinfo {year} {2025}{\natexlab{a}})}\BibitemShut {NoStop}%
\bibitem [{\citenamefont {Surwade}\ \emph {et~al.}(2015)\citenamefont
  {Surwade}, \citenamefont {Smirnov}, \citenamefont {Vlassiouk}, \citenamefont
  {Unocic}, \citenamefont {Veith}, \citenamefont {Dai},\ and\ \citenamefont
  {Mahurin}}]{surwade_water_2015}%
  \BibitemOpen
  \bibfield  {author} {\bibinfo {author} {\bibfnamefont {S.~P.}\ \bibnamefont
  {Surwade}}, \bibinfo {author} {\bibfnamefont {S.~N.}\ \bibnamefont
  {Smirnov}}, \bibinfo {author} {\bibfnamefont {I.~V.}\ \bibnamefont
  {Vlassiouk}}, \bibinfo {author} {\bibfnamefont {R.~R.}\ \bibnamefont
  {Unocic}}, \bibinfo {author} {\bibfnamefont {G.~M.}\ \bibnamefont {Veith}},
  \bibinfo {author} {\bibfnamefont {S.}~\bibnamefont {Dai}},\ and\ \bibinfo
  {author} {\bibfnamefont {S.~M.}\ \bibnamefont {Mahurin}},\ }\bibfield
  {title} {{\selectlanguage {en}\bibinfo {title} {Water desalination using
  nanoporous single-layer graphene}},\ }\href
  {https://doi.org/10.1038/nnano.2015.37} {\bibfield  {journal} {\bibinfo
  {journal} {Nature Nanotechnology}\ }\textbf {\bibinfo {volume} {10}},\
  \bibinfo {pages} {459} (\bibinfo {year} {2015})}\BibitemShut {NoStop}%
\bibitem [{\citenamefont {Radha}\ \emph {et~al.}(2016)\citenamefont {Radha},
  \citenamefont {Esfandiar}, \citenamefont {Wang}, \citenamefont {Rooney},
  \citenamefont {Gopinadhan}, \citenamefont {Keerthi}, \citenamefont
  {Mishchenko}, \citenamefont {Janardanan}, \citenamefont {Blake},
  \citenamefont {Fumagalli}, \citenamefont {Lozada-Hidalgo}, \citenamefont
  {Garaj}, \citenamefont {Haigh}, \citenamefont {Grigorieva}, \citenamefont
  {Wu},\ and\ \citenamefont {Geim}}]{radha_molecular_2016}%
  \BibitemOpen
  \bibfield  {author} {\bibinfo {author} {\bibfnamefont {B.}~\bibnamefont
  {Radha}}, \bibinfo {author} {\bibfnamefont {A.}~\bibnamefont {Esfandiar}},
  \bibinfo {author} {\bibfnamefont {F.~C.}\ \bibnamefont {Wang}}, \bibinfo
  {author} {\bibfnamefont {A.~P.}\ \bibnamefont {Rooney}}, \bibinfo {author}
  {\bibfnamefont {K.}~\bibnamefont {Gopinadhan}}, \bibinfo {author}
  {\bibfnamefont {A.}~\bibnamefont {Keerthi}}, \bibinfo {author} {\bibfnamefont
  {A.}~\bibnamefont {Mishchenko}}, \bibinfo {author} {\bibfnamefont
  {A.}~\bibnamefont {Janardanan}}, \bibinfo {author} {\bibfnamefont
  {P.}~\bibnamefont {Blake}}, \bibinfo {author} {\bibfnamefont
  {L.}~\bibnamefont {Fumagalli}}, \bibinfo {author} {\bibfnamefont
  {M.}~\bibnamefont {Lozada-Hidalgo}}, \bibinfo {author} {\bibfnamefont
  {S.}~\bibnamefont {Garaj}}, \bibinfo {author} {\bibfnamefont {S.~J.}\
  \bibnamefont {Haigh}}, \bibinfo {author} {\bibfnamefont {I.~V.}\ \bibnamefont
  {Grigorieva}}, \bibinfo {author} {\bibfnamefont {H.~A.}\ \bibnamefont {Wu}},\
  and\ \bibinfo {author} {\bibfnamefont {A.~K.}\ \bibnamefont {Geim}},\
  }\bibfield  {title} {{\selectlanguage {en}\bibinfo {title} {Molecular
  transport through capillaries made with atomic-scale precision}},\ }\href
  {https://doi.org/10.1038/nature19363} {\bibfield  {journal} {\bibinfo
  {journal} {Nature}\ }\textbf {\bibinfo {volume} {538}},\ \bibinfo {pages}
  {222} (\bibinfo {year} {2016})}\BibitemShut {NoStop}%
\bibitem [{\citenamefont {Esfandiar}\ \emph {et~al.}(2017)\citenamefont
  {Esfandiar}, \citenamefont {Radha}, \citenamefont {Wang}, \citenamefont
  {Yang}, \citenamefont {Hu}, \citenamefont {Garaj}, \citenamefont {Nair},
  \citenamefont {Geim},\ and\ \citenamefont
  {Gopinadhan}}]{esfandiar_size_2017}%
  \BibitemOpen
  \bibfield  {author} {\bibinfo {author} {\bibfnamefont {A.}~\bibnamefont
  {Esfandiar}}, \bibinfo {author} {\bibfnamefont {B.}~\bibnamefont {Radha}},
  \bibinfo {author} {\bibfnamefont {F.~C.}\ \bibnamefont {Wang}}, \bibinfo
  {author} {\bibfnamefont {Q.}~\bibnamefont {Yang}}, \bibinfo {author}
  {\bibfnamefont {S.}~\bibnamefont {Hu}}, \bibinfo {author} {\bibfnamefont
  {S.}~\bibnamefont {Garaj}}, \bibinfo {author} {\bibfnamefont {R.~R.}\
  \bibnamefont {Nair}}, \bibinfo {author} {\bibfnamefont {A.~K.}\ \bibnamefont
  {Geim}},\ and\ \bibinfo {author} {\bibfnamefont {K.}~\bibnamefont
  {Gopinadhan}},\ }\bibfield  {title} {{\selectlanguage {en}\bibinfo {title}
  {Size effect in ion transport through angstrom-scale slits}},\ }\href
  {https://doi.org/10.1126/science.aan5275} {\bibfield  {journal} {\bibinfo
  {journal} {Science}\ }\textbf {\bibinfo {volume} {358}},\ \bibinfo {pages}
  {511} (\bibinfo {year} {2017})}\BibitemShut {NoStop}%
\bibitem [{\citenamefont {Abraham}\ \emph {et~al.}(2017)\citenamefont
  {Abraham}, \citenamefont {Vasu}, \citenamefont {Williams}, \citenamefont
  {Gopinadhan}, \citenamefont {Su}, \citenamefont {Cherian}, \citenamefont
  {Dix}, \citenamefont {Prestat}, \citenamefont {Haigh}, \citenamefont
  {Grigorieva}, \citenamefont {Carbone}, \citenamefont {Geim},\ and\
  \citenamefont {Nair}}]{abraham_tunable_2017}%
  \BibitemOpen
  \bibfield  {author} {\bibinfo {author} {\bibfnamefont {J.}~\bibnamefont
  {Abraham}}, \bibinfo {author} {\bibfnamefont {K.~S.}\ \bibnamefont {Vasu}},
  \bibinfo {author} {\bibfnamefont {C.~D.}\ \bibnamefont {Williams}}, \bibinfo
  {author} {\bibfnamefont {K.}~\bibnamefont {Gopinadhan}}, \bibinfo {author}
  {\bibfnamefont {Y.}~\bibnamefont {Su}}, \bibinfo {author} {\bibfnamefont
  {C.~T.}\ \bibnamefont {Cherian}}, \bibinfo {author} {\bibfnamefont
  {J.}~\bibnamefont {Dix}}, \bibinfo {author} {\bibfnamefont {E.}~\bibnamefont
  {Prestat}}, \bibinfo {author} {\bibfnamefont {S.~J.}\ \bibnamefont {Haigh}},
  \bibinfo {author} {\bibfnamefont {I.~V.}\ \bibnamefont {Grigorieva}},
  \bibinfo {author} {\bibfnamefont {P.}~\bibnamefont {Carbone}}, \bibinfo
  {author} {\bibfnamefont {A.~K.}\ \bibnamefont {Geim}},\ and\ \bibinfo
  {author} {\bibfnamefont {R.~R.}\ \bibnamefont {Nair}},\ }\bibfield  {title}
  {{\selectlanguage {en}\bibinfo {title} {Tunable sieving of ions using
  graphene oxide membranes}},\ }\href {https://doi.org/10.1038/nnano.2017.21}
  {\bibfield  {journal} {\bibinfo  {journal} {Nature Nanotechnology}\ }\textbf
  {\bibinfo {volume} {12}},\ \bibinfo {pages} {546} (\bibinfo {year}
  {2017})}\BibitemShut {NoStop}%
\bibitem [{\citenamefont {Chakravarty}\ \emph {et~al.}(2010)\citenamefont
  {Chakravarty}, \citenamefont {Qian}, \citenamefont {El-Sayed},\ and\
  \citenamefont {Prausnitz}}]{chakravarty_delivery_2010}%
  \BibitemOpen
  \bibfield  {author} {\bibinfo {author} {\bibfnamefont {P.}~\bibnamefont
  {Chakravarty}}, \bibinfo {author} {\bibfnamefont {W.}~\bibnamefont {Qian}},
  \bibinfo {author} {\bibfnamefont {M.~A.}\ \bibnamefont {El-Sayed}},\ and\
  \bibinfo {author} {\bibfnamefont {M.~R.}\ \bibnamefont {Prausnitz}},\
  }\bibfield  {title} {{\selectlanguage {en}\bibinfo {title} {Delivery of
  molecules into cells using carbon nanoparticles activated by femtosecond
  laser pulses}},\ }\href {https://doi.org/10.1038/nnano.2010.126} {\bibfield
  {journal} {\bibinfo  {journal} {Nature Nanotechnology}\ }\textbf {\bibinfo
  {volume} {5}},\ \bibinfo {pages} {607} (\bibinfo {year} {2010})}\BibitemShut
  {NoStop}%
\bibitem [{\citenamefont {Geng}\ \emph {et~al.}(2014)\citenamefont {Geng},
  \citenamefont {Kim}, \citenamefont {Zhang}, \citenamefont {Escalada},
  \citenamefont {Tunuguntla}, \citenamefont {Comolli}, \citenamefont {Allen},
  \citenamefont {Shnyrova}, \citenamefont {Cho}, \citenamefont {Munoz},
  \citenamefont {Wang}, \citenamefont {Grigoropoulos}, \citenamefont
  {Ajo-Franklin}, \citenamefont {Frolov},\ and\ \citenamefont
  {Noy}}]{geng_stochastic_2014}%
  \BibitemOpen
  \bibfield  {author} {\bibinfo {author} {\bibfnamefont {J.}~\bibnamefont
  {Geng}}, \bibinfo {author} {\bibfnamefont {K.}~\bibnamefont {Kim}}, \bibinfo
  {author} {\bibfnamefont {J.}~\bibnamefont {Zhang}}, \bibinfo {author}
  {\bibfnamefont {A.}~\bibnamefont {Escalada}}, \bibinfo {author}
  {\bibfnamefont {R.}~\bibnamefont {Tunuguntla}}, \bibinfo {author}
  {\bibfnamefont {L.~R.}\ \bibnamefont {Comolli}}, \bibinfo {author}
  {\bibfnamefont {F.~I.}\ \bibnamefont {Allen}}, \bibinfo {author}
  {\bibfnamefont {A.~V.}\ \bibnamefont {Shnyrova}}, \bibinfo {author}
  {\bibfnamefont {K.~R.}\ \bibnamefont {Cho}}, \bibinfo {author} {\bibfnamefont
  {D.}~\bibnamefont {Munoz}}, \bibinfo {author} {\bibfnamefont {Y.~M.}\
  \bibnamefont {Wang}}, \bibinfo {author} {\bibfnamefont {C.~P.}\ \bibnamefont
  {Grigoropoulos}}, \bibinfo {author} {\bibfnamefont {C.~M.}\ \bibnamefont
  {Ajo-Franklin}}, \bibinfo {author} {\bibfnamefont {V.~A.}\ \bibnamefont
  {Frolov}},\ and\ \bibinfo {author} {\bibfnamefont {A.}~\bibnamefont {Noy}},\
  }\bibfield  {title} {{\selectlanguage {en}\bibinfo {title} {Stochastic
  transport through carbon nanotubes in lipid bilayers and live cell
  membranes}},\ }\href {https://doi.org/10.1038/nature13817} {\bibfield
  {journal} {\bibinfo  {journal} {Nature}\ }\textbf {\bibinfo {volume} {514}},\
  \bibinfo {pages} {612} (\bibinfo {year} {2014})}\BibitemShut {NoStop}%
\bibitem [{\citenamefont {Dong}\ \emph {et~al.}(2018)\citenamefont {Dong},
  \citenamefont {Pei}, \citenamefont {Zhao}, \citenamefont {Goh}, \citenamefont
  {Qi}, \citenamefont {Xiao}, \citenamefont {Chen}, \citenamefont {Huang},\
  and\ \citenamefont {Fang}}]{dong_situ_2018}%
  \BibitemOpen
  \bibfield  {author} {\bibinfo {author} {\bibfnamefont {B.}~\bibnamefont
  {Dong}}, \bibinfo {author} {\bibfnamefont {Y.}~\bibnamefont {Pei}}, \bibinfo
  {author} {\bibfnamefont {F.}~\bibnamefont {Zhao}}, \bibinfo {author}
  {\bibfnamefont {T.~W.}\ \bibnamefont {Goh}}, \bibinfo {author} {\bibfnamefont
  {Z.}~\bibnamefont {Qi}}, \bibinfo {author} {\bibfnamefont {C.}~\bibnamefont
  {Xiao}}, \bibinfo {author} {\bibfnamefont {K.}~\bibnamefont {Chen}}, \bibinfo
  {author} {\bibfnamefont {W.}~\bibnamefont {Huang}},\ and\ \bibinfo {author}
  {\bibfnamefont {N.}~\bibnamefont {Fang}},\ }\bibfield  {title}
  {{\selectlanguage {en}\bibinfo {title} {In situ quantitative single-molecule
  study of dynamic catalytic processes in nanoconfinement}},\ }\href
  {https://doi.org/10.1038/s41929-017-0021-1} {\bibfield  {journal} {\bibinfo
  {journal} {Nature Catalysis}\ }\textbf {\bibinfo {volume} {1}},\ \bibinfo
  {pages} {135} (\bibinfo {year} {2018})}\BibitemShut {NoStop}%
\bibitem [{\citenamefont {Zuo}\ \emph {et~al.}(2023)\citenamefont {Zuo},
  \citenamefont {Ye}, \citenamefont {Jiao}, \citenamefont {Luo}, \citenamefont
  {Fang}, \citenamefont {Schubert}, \citenamefont {McKeown}, \citenamefont
  {Liu}, \citenamefont {Yang},\ and\ \citenamefont
  {Xu}}]{zuo_near-frictionless_2023}%
  \BibitemOpen
  \bibfield  {author} {\bibinfo {author} {\bibfnamefont {P.}~\bibnamefont
  {Zuo}}, \bibinfo {author} {\bibfnamefont {C.}~\bibnamefont {Ye}}, \bibinfo
  {author} {\bibfnamefont {Z.}~\bibnamefont {Jiao}}, \bibinfo {author}
  {\bibfnamefont {J.}~\bibnamefont {Luo}}, \bibinfo {author} {\bibfnamefont
  {J.}~\bibnamefont {Fang}}, \bibinfo {author} {\bibfnamefont {U.~S.}\
  \bibnamefont {Schubert}}, \bibinfo {author} {\bibfnamefont {N.~B.}\
  \bibnamefont {McKeown}}, \bibinfo {author} {\bibfnamefont {T.~L.}\
  \bibnamefont {Liu}}, \bibinfo {author} {\bibfnamefont {Z.}~\bibnamefont
  {Yang}},\ and\ \bibinfo {author} {\bibfnamefont {T.}~\bibnamefont {Xu}},\
  }\bibfield  {title} {{\selectlanguage {en}\bibinfo {title} {Near-frictionless
  ion transport within triazine framework membranes}},\ }\href
  {https://doi.org/10.1038/s41586-023-05888-x} {\bibfield  {journal} {\bibinfo
  {journal} {Nature}\ }\textbf {\bibinfo {volume} {617}},\ \bibinfo {pages}
  {299} (\bibinfo {year} {2023})}\BibitemShut {NoStop}%
\bibitem [{\citenamefont {Algara-Siller}\ \emph
  {et~al.}(2015{\natexlab{a}})\citenamefont {Algara-Siller}, \citenamefont
  {Lehtinen}, \citenamefont {Wang}, \citenamefont {Nair}, \citenamefont
  {Kaiser}, \citenamefont {Wu}, \citenamefont {Geim},\ and\ \citenamefont
  {Grigorieva}}]{algara-siller_square_2015}%
  \BibitemOpen
  \bibfield  {author} {\bibinfo {author} {\bibfnamefont {G.}~\bibnamefont
  {Algara-Siller}}, \bibinfo {author} {\bibfnamefont {O.}~\bibnamefont
  {Lehtinen}}, \bibinfo {author} {\bibfnamefont {F.~C.}\ \bibnamefont {Wang}},
  \bibinfo {author} {\bibfnamefont {R.~R.}\ \bibnamefont {Nair}}, \bibinfo
  {author} {\bibfnamefont {U.}~\bibnamefont {Kaiser}}, \bibinfo {author}
  {\bibfnamefont {H.~A.}\ \bibnamefont {Wu}}, \bibinfo {author} {\bibfnamefont
  {A.~K.}\ \bibnamefont {Geim}},\ and\ \bibinfo {author} {\bibfnamefont
  {I.~V.}\ \bibnamefont {Grigorieva}},\ }\bibfield  {title} {{\selectlanguage
  {en}\bibinfo {title} {Square ice in graphene nanocapillaries}},\ }\href
  {https://doi.org/10.1038/nature14295} {\bibfield  {journal} {\bibinfo
  {journal} {Nature}\ }\textbf {\bibinfo {volume} {519}},\ \bibinfo {pages}
  {443} (\bibinfo {year} {2015}{\natexlab{a}})}\BibitemShut {NoStop}%
\bibitem [{\citenamefont {Zhou}\ \emph {et~al.}(2015)\citenamefont {Zhou},
  \citenamefont {Yin}, \citenamefont {Wang}, \citenamefont {Zhang},
  \citenamefont {Xu}, \citenamefont {Borisevich}, \citenamefont {Sun},
  \citenamefont {Idrobo}, \citenamefont {Chisholm}, \citenamefont {Pantelides},
  \citenamefont {Klie},\ and\ \citenamefont {Lupini}}]{zhou_observation_2015}%
  \BibitemOpen
  \bibfield  {author} {\bibinfo {author} {\bibfnamefont {W.}~\bibnamefont
  {Zhou}}, \bibinfo {author} {\bibfnamefont {K.}~\bibnamefont {Yin}}, \bibinfo
  {author} {\bibfnamefont {C.}~\bibnamefont {Wang}}, \bibinfo {author}
  {\bibfnamefont {Y.}~\bibnamefont {Zhang}}, \bibinfo {author} {\bibfnamefont
  {T.}~\bibnamefont {Xu}}, \bibinfo {author} {\bibfnamefont {A.}~\bibnamefont
  {Borisevich}}, \bibinfo {author} {\bibfnamefont {L.}~\bibnamefont {Sun}},
  \bibinfo {author} {\bibfnamefont {J.~C.}\ \bibnamefont {Idrobo}}, \bibinfo
  {author} {\bibfnamefont {M.~F.}\ \bibnamefont {Chisholm}}, \bibinfo {author}
  {\bibfnamefont {S.~T.}\ \bibnamefont {Pantelides}}, \bibinfo {author}
  {\bibfnamefont {R.~F.}\ \bibnamefont {Klie}},\ and\ \bibinfo {author}
  {\bibfnamefont {A.~R.}\ \bibnamefont {Lupini}},\ }\bibfield  {title}
  {{\selectlanguage {en}\bibinfo {title} {The observation of square ice in
  graphene questioned}},\ }\href {https://doi.org/10.1038/nature16145}
  {\bibfield  {journal} {\bibinfo  {journal} {Nature}\ }\textbf {\bibinfo
  {volume} {528}},\ \bibinfo {pages} {E1} (\bibinfo {year} {2015})}\BibitemShut
  {NoStop}%
\bibitem [{\citenamefont {Algara-Siller}\ \emph
  {et~al.}(2015{\natexlab{b}})\citenamefont {Algara-Siller}, \citenamefont
  {Lehtinen},\ and\ \citenamefont {Kaiser}}]{algara-siller_algara-siller_2015}%
  \BibitemOpen
  \bibfield  {author} {\bibinfo {author} {\bibfnamefont {G.}~\bibnamefont
  {Algara-Siller}}, \bibinfo {author} {\bibfnamefont {O.}~\bibnamefont
  {Lehtinen}},\ and\ \bibinfo {author} {\bibfnamefont {U.}~\bibnamefont
  {Kaiser}},\ }\bibfield  {title} {{\selectlanguage {en}\bibinfo {title}
  {Algara-{Siller} et al . reply}},\ }\href
  {https://doi.org/10.1038/nature16149} {\bibfield  {journal} {\bibinfo
  {journal} {Nature}\ }\textbf {\bibinfo {volume} {528}},\ \bibinfo {pages}
  {E3} (\bibinfo {year} {2015}{\natexlab{b}})}\BibitemShut {NoStop}%
\bibitem [{\citenamefont {Wang}\ \emph {et~al.}(2015)\citenamefont {Wang},
  \citenamefont {Wu},\ and\ \citenamefont {Geim}}]{wang_wang_2015}%
  \BibitemOpen
  \bibfield  {author} {\bibinfo {author} {\bibfnamefont {F.~C.}\ \bibnamefont
  {Wang}}, \bibinfo {author} {\bibfnamefont {H.~A.}\ \bibnamefont {Wu}},\ and\
  \bibinfo {author} {\bibfnamefont {A.~K.}\ \bibnamefont {Geim}},\ }\bibfield
  {title} {{\selectlanguage {en}\bibinfo {title} {Wang et al. reply}},\ }\href
  {https://doi.org/10.1038/nature16146} {\bibfield  {journal} {\bibinfo
  {journal} {Nature}\ }\textbf {\bibinfo {volume} {528}},\ \bibinfo {pages}
  {E3} (\bibinfo {year} {2015})}\BibitemShut {NoStop}%
\bibitem [{\citenamefont {Zheng}\ \emph {et~al.}(2026)\citenamefont {Zheng},
  \citenamefont {Zhang}, \citenamefont {Jiang}, \citenamefont {He},
  \citenamefont {Stöhr}, \citenamefont {Denisenko}, \citenamefont {Wrachtrup},
  \citenamefont {Zeng}, \citenamefont {Bian}, \citenamefont {Wang},\ and\
  \citenamefont {Jiang}}]{zheng_experimental_2026}%
  \BibitemOpen
  \bibfield  {author} {\bibinfo {author} {\bibfnamefont {W.}~\bibnamefont
  {Zheng}}, \bibinfo {author} {\bibfnamefont {S.}~\bibnamefont {Zhang}},
  \bibinfo {author} {\bibfnamefont {J.}~\bibnamefont {Jiang}}, \bibinfo
  {author} {\bibfnamefont {Y.}~\bibnamefont {He}}, \bibinfo {author}
  {\bibfnamefont {R.}~\bibnamefont {Stöhr}}, \bibinfo {author} {\bibfnamefont
  {A.}~\bibnamefont {Denisenko}}, \bibinfo {author} {\bibfnamefont
  {J.}~\bibnamefont {Wrachtrup}}, \bibinfo {author} {\bibfnamefont {X.~C.}\
  \bibnamefont {Zeng}}, \bibinfo {author} {\bibfnamefont {K.}~\bibnamefont
  {Bian}}, \bibinfo {author} {\bibfnamefont {E.-G.}\ \bibnamefont {Wang}},\
  and\ \bibinfo {author} {\bibfnamefont {Y.}~\bibnamefont {Jiang}},\ }\bibfield
   {title} {{\selectlanguage {en}\bibinfo {title} {Experimental observation of
  liquid–solid transition of nanoconfined water at ambient temperature}},\
  }\href {https://doi.org/10.1038/s41563-025-02456-8} {\bibfield  {journal}
  {\bibinfo  {journal} {Nature Materials}\ ,\ \bibinfo {pages} {1}} (\bibinfo
  {year} {2026})}\BibitemShut {NoStop}%
\bibitem [{\citenamefont {Koga}\ \emph {et~al.}(1997)\citenamefont {Koga},
  \citenamefont {Zeng},\ and\ \citenamefont {Tanaka}}]{koga_freezing_1997}%
  \BibitemOpen
  \bibfield  {author} {\bibinfo {author} {\bibfnamefont {K.}~\bibnamefont
  {Koga}}, \bibinfo {author} {\bibfnamefont {X.~C.}\ \bibnamefont {Zeng}},\
  and\ \bibinfo {author} {\bibfnamefont {H.}~\bibnamefont {Tanaka}},\
  }\bibfield  {title} {{\selectlanguage {en}\bibinfo {title} {Freezing of
  {Confined} {Water}: {A} {Bilayer} {Ice} {Phase} in {Hydrophobic}
  {Nanopores}}},\ }\href {https://doi.org/10.1103/PhysRevLett.79.5262}
  {\bibfield  {journal} {\bibinfo  {journal} {Physical Review Letters}\
  }\textbf {\bibinfo {volume} {79}},\ \bibinfo {pages} {5262} (\bibinfo {year}
  {1997})}\BibitemShut {NoStop}%
\bibitem [{\citenamefont {Han}\ \emph {et~al.}(2010)\citenamefont {Han},
  \citenamefont {Choi}, \citenamefont {Kumar},\ and\ \citenamefont
  {Stanley}}]{han_phase_2010}%
  \BibitemOpen
  \bibfield  {author} {\bibinfo {author} {\bibfnamefont {S.}~\bibnamefont
  {Han}}, \bibinfo {author} {\bibfnamefont {M.~Y.}\ \bibnamefont {Choi}},
  \bibinfo {author} {\bibfnamefont {P.}~\bibnamefont {Kumar}},\ and\ \bibinfo
  {author} {\bibfnamefont {H.~E.}\ \bibnamefont {Stanley}},\ }\bibfield
  {title} {{\selectlanguage {en}\bibinfo {title} {Phase transitions in confined
  water nanofilms}},\ }\href {https://doi.org/10.1038/nphys1708} {\bibfield
  {journal} {\bibinfo  {journal} {Nature Physics}\ }\textbf {\bibinfo {volume}
  {6}},\ \bibinfo {pages} {685} (\bibinfo {year} {2010})}\BibitemShut {NoStop}%
\bibitem [{\citenamefont {Zhao}\ \emph
  {et~al.}(2014{\natexlab{a}})\citenamefont {Zhao}, \citenamefont {Wang},
  \citenamefont {Bai}, \citenamefont {Yuan}, \citenamefont {Yang},\ and\
  \citenamefont {Zeng}}]{zhao_highly_2014}%
  \BibitemOpen
  \bibfield  {author} {\bibinfo {author} {\bibfnamefont {W.-H.}\ \bibnamefont
  {Zhao}}, \bibinfo {author} {\bibfnamefont {L.}~\bibnamefont {Wang}}, \bibinfo
  {author} {\bibfnamefont {J.}~\bibnamefont {Bai}}, \bibinfo {author}
  {\bibfnamefont {L.-F.}\ \bibnamefont {Yuan}}, \bibinfo {author}
  {\bibfnamefont {J.}~\bibnamefont {Yang}},\ and\ \bibinfo {author}
  {\bibfnamefont {X.~C.}\ \bibnamefont {Zeng}},\ }\bibfield  {title} {\bibinfo
  {title} {Highly {Confined} {Water}: {Two}-{Dimensional} {Ice}, {Amorphous}
  {Ice}, and {Clathrate} {Hydrates}},\ }\href
  {https://doi.org/10.1021/ar5001549} {\bibfield  {journal} {\bibinfo
  {journal} {Accounts of Chemical Research}\ }\textbf {\bibinfo {volume}
  {47}},\ \bibinfo {pages} {2505} (\bibinfo {year}
  {2014}{\natexlab{a}})}\BibitemShut {NoStop}%
\bibitem [{\citenamefont {Zhao}\ \emph
  {et~al.}(2014{\natexlab{b}})\citenamefont {Zhao}, \citenamefont {Bai},
  \citenamefont {Yuan}, \citenamefont {Yang},\ and\ \citenamefont
  {Zeng}}]{zhao_ferroelectric_2014}%
  \BibitemOpen
  \bibfield  {author} {\bibinfo {author} {\bibfnamefont {W.-H.}\ \bibnamefont
  {Zhao}}, \bibinfo {author} {\bibfnamefont {J.}~\bibnamefont {Bai}}, \bibinfo
  {author} {\bibfnamefont {L.-F.}\ \bibnamefont {Yuan}}, \bibinfo {author}
  {\bibfnamefont {J.}~\bibnamefont {Yang}},\ and\ \bibinfo {author}
  {\bibfnamefont {X.~C.}\ \bibnamefont {Zeng}},\ }\bibfield  {title}
  {{\selectlanguage {en}\bibinfo {title} {Ferroelectric hexagonal and rhombic
  monolayer ice phases}},\ }\href {https://doi.org/10.1039/C3SC53368A}
  {\bibfield  {journal} {\bibinfo  {journal} {Chemical Science}\ }\textbf
  {\bibinfo {volume} {5}},\ \bibinfo {pages} {1757} (\bibinfo {year}
  {2014}{\natexlab{b}})}\BibitemShut {NoStop}%
\bibitem [{\citenamefont {Sobrino Fernandez~Mario}\ \emph
  {et~al.}(2015)\citenamefont {Sobrino Fernandez~Mario}, \citenamefont
  {Neek-Amal},\ and\ \citenamefont
  {Peeters}}]{sobrino_fernandez_mario_aa-stacked_2015}%
  \BibitemOpen
  \bibfield  {author} {\bibinfo {author} {\bibfnamefont {M.}~\bibnamefont
  {Sobrino Fernandez~Mario}}, \bibinfo {author} {\bibfnamefont
  {M.}~\bibnamefont {Neek-Amal}},\ and\ \bibinfo {author} {\bibfnamefont
  {F.~M.}\ \bibnamefont {Peeters}},\ }\bibfield  {title} {{\selectlanguage
  {en}\bibinfo {title} {{AA}-stacked bilayer square ice between graphene
  layers}},\ }\href {https://doi.org/10.1103/PhysRevB.92.245428} {\bibfield
  {journal} {\bibinfo  {journal} {Physical Review B}\ }\textbf {\bibinfo
  {volume} {92}},\ \bibinfo {pages} {245428} (\bibinfo {year}
  {2015})}\BibitemShut {NoStop}%
\bibitem [{\citenamefont {Zhu}\ \emph {et~al.}(2015)\citenamefont {Zhu},
  \citenamefont {Wang}, \citenamefont {Bai}, \citenamefont {Zeng},\ and\
  \citenamefont {Wu}}]{zhu_compression_2015}%
  \BibitemOpen
  \bibfield  {author} {\bibinfo {author} {\bibfnamefont {Y.}~\bibnamefont
  {Zhu}}, \bibinfo {author} {\bibfnamefont {F.}~\bibnamefont {Wang}}, \bibinfo
  {author} {\bibfnamefont {J.}~\bibnamefont {Bai}}, \bibinfo {author}
  {\bibfnamefont {X.~C.}\ \bibnamefont {Zeng}},\ and\ \bibinfo {author}
  {\bibfnamefont {H.}~\bibnamefont {Wu}},\ }\bibfield  {title}
  {{\selectlanguage {en}\bibinfo {title} {Compression {Limit} of
  {Two}-{Dimensional} {Water} {Constrained} in {Graphene} {Nanocapillaries}}},\
  }\href {https://doi.org/10.1021/acsnano.5b06572} {\bibfield  {journal}
  {\bibinfo  {journal} {ACS Nano}\ }\textbf {\bibinfo {volume} {9}},\ \bibinfo
  {pages} {12197} (\bibinfo {year} {2015})}\BibitemShut {NoStop}%
\bibitem [{\citenamefont {Zhu}\ \emph {et~al.}(2016)\citenamefont {Zhu},
  \citenamefont {Zhao}, \citenamefont {Wang}, \citenamefont {Yin},
  \citenamefont {Jia}, \citenamefont {Yang}, \citenamefont {Zeng},\ and\
  \citenamefont {Yuan}}]{zhu_two-dimensional_2016}%
  \BibitemOpen
  \bibfield  {author} {\bibinfo {author} {\bibfnamefont {W.}~\bibnamefont
  {Zhu}}, \bibinfo {author} {\bibfnamefont {W.-H.}\ \bibnamefont {Zhao}},
  \bibinfo {author} {\bibfnamefont {L.}~\bibnamefont {Wang}}, \bibinfo {author}
  {\bibfnamefont {D.}~\bibnamefont {Yin}}, \bibinfo {author} {\bibfnamefont
  {M.}~\bibnamefont {Jia}}, \bibinfo {author} {\bibfnamefont {J.}~\bibnamefont
  {Yang}}, \bibinfo {author} {\bibfnamefont {X.~C.}\ \bibnamefont {Zeng}},\
  and\ \bibinfo {author} {\bibfnamefont {L.-F.}\ \bibnamefont {Yuan}},\
  }\bibfield  {title} {{\selectlanguage {en}\bibinfo {title} {Two-dimensional
  interlocked pentagonal bilayer ice: how do water molecules form a hydrogen
  bonding network?}},\ }\href {https://doi.org/10.1039/C5CP07524F} {\bibfield
  {journal} {\bibinfo  {journal} {Physical Chemistry Chemical Physics}\
  }\textbf {\bibinfo {volume} {18}},\ \bibinfo {pages} {14216} (\bibinfo {year}
  {2016})}\BibitemShut {NoStop}%
\bibitem [{\citenamefont {Corsetti}\ \emph
  {et~al.}(2016{\natexlab{a}})\citenamefont {Corsetti}, \citenamefont
  {Zubeltzu},\ and\ \citenamefont {Artacho}}]{corsetti_enhanced_2016}%
  \BibitemOpen
  \bibfield  {author} {\bibinfo {author} {\bibfnamefont {F.}~\bibnamefont
  {Corsetti}}, \bibinfo {author} {\bibfnamefont {J.}~\bibnamefont {Zubeltzu}},\
  and\ \bibinfo {author} {\bibfnamefont {E.}~\bibnamefont {Artacho}},\
  }\bibfield  {title} {{\selectlanguage {en}\bibinfo {title} {Enhanced
  {Configurational} {Entropy} in {High}-{Density} {Nanoconfined} {Bilayer}
  {Ice}}},\ }\href {https://doi.org/10.1103/PhysRevLett.116.085901} {\bibfield
  {journal} {\bibinfo  {journal} {Physical Review Letters}\ }\textbf {\bibinfo
  {volume} {116}},\ \bibinfo {pages} {085901} (\bibinfo {year}
  {2016}{\natexlab{a}})}\BibitemShut {NoStop}%
\bibitem [{\citenamefont {Zubeltzu}\ \emph {et~al.}(2016)\citenamefont
  {Zubeltzu}, \citenamefont {Corsetti}, \citenamefont {Fernández-Serra},\ and\
  \citenamefont {Artacho}}]{zubeltzu_continuous_2016}%
  \BibitemOpen
  \bibfield  {author} {\bibinfo {author} {\bibfnamefont {J.}~\bibnamefont
  {Zubeltzu}}, \bibinfo {author} {\bibfnamefont {F.}~\bibnamefont {Corsetti}},
  \bibinfo {author} {\bibfnamefont {M.~V.}\ \bibnamefont {Fernández-Serra}},\
  and\ \bibinfo {author} {\bibfnamefont {E.}~\bibnamefont {Artacho}},\
  }\bibfield  {title} {\bibinfo {title} {Continuous melting through a hexatic
  phase in confined bilayer water},\ }\href
  {https://doi.org/10.1103/PhysRevE.93.062137} {\bibfield  {journal} {\bibinfo
  {journal} {Physical Review E}\ }\textbf {\bibinfo {volume} {93}},\ \bibinfo
  {pages} {062137} (\bibinfo {year} {2016})}\BibitemShut {NoStop}%
\bibitem [{\citenamefont {Raju}\ \emph {et~al.}(2018)\citenamefont {Raju},
  \citenamefont {van Duin},\ and\ \citenamefont {Ihme}}]{raju_phase_2018}%
  \BibitemOpen
  \bibfield  {author} {\bibinfo {author} {\bibfnamefont {M.}~\bibnamefont
  {Raju}}, \bibinfo {author} {\bibfnamefont {A.}~\bibnamefont {van Duin}},\
  and\ \bibinfo {author} {\bibfnamefont {M.}~\bibnamefont {Ihme}},\ }\bibfield
  {title} {{\selectlanguage {en}\bibinfo {title} {Phase transitions of ordered
  ice in graphene nanocapillaries and carbon nanotubes}},\ }\href
  {https://doi.org/10.1038/s41598-018-22201-3} {\bibfield  {journal} {\bibinfo
  {journal} {Scientific Reports}\ }\textbf {\bibinfo {volume} {8}},\ \bibinfo
  {pages} {3851} (\bibinfo {year} {2018})}\BibitemShut {NoStop}%
\bibitem [{\citenamefont {Li}\ and\ \citenamefont
  {Schmidt}(2019)}]{li_replica_2019}%
  \BibitemOpen
  \bibfield  {author} {\bibinfo {author} {\bibfnamefont {S.}~\bibnamefont
  {Li}}\ and\ \bibinfo {author} {\bibfnamefont {B.}~\bibnamefont {Schmidt}},\
  }\bibfield  {title} {{\selectlanguage {en}\bibinfo {title} {Replica exchange
  {MD} simulations of two-dimensional water in graphene nanocapillaries:
  rhombic versus square structures, proton ordering, and phase transitions}},\
  }\href {https://doi.org/10.1039/C9CP00849G} {\bibfield  {journal} {\bibinfo
  {journal} {Physical Chemistry Chemical Physics}\ }\textbf {\bibinfo {volume}
  {21}},\ \bibinfo {pages} {17640} (\bibinfo {year} {2019})}\BibitemShut
  {NoStop}%
\bibitem [{\citenamefont {Chen}\ \emph
  {et~al.}(2016{\natexlab{a}})\citenamefont {Chen}, \citenamefont
  {Schusteritsch}, \citenamefont {Pickard}, \citenamefont {Salzmann},\ and\
  \citenamefont {Michaelides}}]{chen_two_2016}%
  \BibitemOpen
  \bibfield  {author} {\bibinfo {author} {\bibfnamefont {J.}~\bibnamefont
  {Chen}}, \bibinfo {author} {\bibfnamefont {G.}~\bibnamefont {Schusteritsch}},
  \bibinfo {author} {\bibfnamefont {C.~J.}\ \bibnamefont {Pickard}}, \bibinfo
  {author} {\bibfnamefont {C.~G.}\ \bibnamefont {Salzmann}},\ and\ \bibinfo
  {author} {\bibfnamefont {A.}~\bibnamefont {Michaelides}},\ }\bibfield
  {title} {\bibinfo {title} {Two {Dimensional} {Ice} from {First} {Principles}:
  {Structures} and {Phase} {Transitions}},\ }\href
  {https://doi.org/10.1103/PhysRevLett.116.025501} {\bibfield  {journal}
  {\bibinfo  {journal} {Physical Review Letters}\ }\textbf {\bibinfo {volume}
  {116}},\ \bibinfo {pages} {025501} (\bibinfo {year}
  {2016}{\natexlab{a}})}\BibitemShut {NoStop}%
\bibitem [{\citenamefont {Corsetti}\ \emph
  {et~al.}(2016{\natexlab{b}})\citenamefont {Corsetti}, \citenamefont
  {Matthews},\ and\ \citenamefont {Artacho}}]{corsetti_structural_2016}%
  \BibitemOpen
  \bibfield  {author} {\bibinfo {author} {\bibfnamefont {F.}~\bibnamefont
  {Corsetti}}, \bibinfo {author} {\bibfnamefont {P.}~\bibnamefont {Matthews}},\
  and\ \bibinfo {author} {\bibfnamefont {E.}~\bibnamefont {Artacho}},\
  }\bibfield  {title} {{\selectlanguage {en}\bibinfo {title} {Structural and
  configurational properties of nanoconfined monolayer ice from first
  principles}},\ }\href {https://doi.org/10.1038/srep18651} {\bibfield
  {journal} {\bibinfo  {journal} {Scientific Reports}\ }\textbf {\bibinfo
  {volume} {6}},\ \bibinfo {pages} {18651} (\bibinfo {year}
  {2016}{\natexlab{b}})}\BibitemShut {NoStop}%
\bibitem [{\citenamefont {Chen}\ \emph
  {et~al.}(2016{\natexlab{b}})\citenamefont {Chen}, \citenamefont {Zen},
  \citenamefont {Brandenburg}, \citenamefont {Alfè},\ and\ \citenamefont
  {Michaelides}}]{chen_evidence_2016}%
  \BibitemOpen
  \bibfield  {author} {\bibinfo {author} {\bibfnamefont {J.}~\bibnamefont
  {Chen}}, \bibinfo {author} {\bibfnamefont {A.}~\bibnamefont {Zen}}, \bibinfo
  {author} {\bibfnamefont {J.~G.}\ \bibnamefont {Brandenburg}}, \bibinfo
  {author} {\bibfnamefont {D.}~\bibnamefont {Alfè}},\ and\ \bibinfo {author}
  {\bibfnamefont {A.}~\bibnamefont {Michaelides}},\ }\bibfield  {title}
  {\bibinfo {title} {Evidence for stable square ice from quantum {Monte}
  {Carlo}},\ }\href {https://doi.org/10.1103/PhysRevB.94.220102} {\bibfield
  {journal} {\bibinfo  {journal} {Physical Review B}\ }\textbf {\bibinfo
  {volume} {94}},\ \bibinfo {pages} {220102} (\bibinfo {year}
  {2016}{\natexlab{b}})}\BibitemShut {NoStop}%
\bibitem [{\citenamefont {Kapil}\ \emph {et~al.}(2022)\citenamefont {Kapil},
  \citenamefont {Schran}, \citenamefont {Zen}, \citenamefont {Chen},
  \citenamefont {Pickard},\ and\ \citenamefont
  {Michaelides}}]{kapil_first-principles_2022}%
  \BibitemOpen
  \bibfield  {author} {\bibinfo {author} {\bibfnamefont {V.}~\bibnamefont
  {Kapil}}, \bibinfo {author} {\bibfnamefont {C.}~\bibnamefont {Schran}},
  \bibinfo {author} {\bibfnamefont {A.}~\bibnamefont {Zen}}, \bibinfo {author}
  {\bibfnamefont {J.}~\bibnamefont {Chen}}, \bibinfo {author} {\bibfnamefont
  {C.~J.}\ \bibnamefont {Pickard}},\ and\ \bibinfo {author} {\bibfnamefont
  {A.}~\bibnamefont {Michaelides}},\ }\bibfield  {title} {{\selectlanguage
  {en}\bibinfo {title} {The first-principles phase diagram of monolayer
  nanoconfined water}},\ }\href {https://doi.org/10.1038/s41586-022-05036-x}
  {\bibfield  {journal} {\bibinfo  {journal} {Nature}\ }\textbf {\bibinfo
  {volume} {609}},\ \bibinfo {pages} {512} (\bibinfo {year}
  {2022})}\BibitemShut {NoStop}%
\bibitem [{\citenamefont {Lin}\ \emph {et~al.}(2023)\citenamefont {Lin},
  \citenamefont {Jiang}, \citenamefont {Zeng},\ and\ \citenamefont
  {Li}}]{lin_temperature-pressure_2023}%
  \BibitemOpen
  \bibfield  {author} {\bibinfo {author} {\bibfnamefont {B.}~\bibnamefont
  {Lin}}, \bibinfo {author} {\bibfnamefont {J.}~\bibnamefont {Jiang}}, \bibinfo
  {author} {\bibfnamefont {X.~C.}\ \bibnamefont {Zeng}},\ and\ \bibinfo
  {author} {\bibfnamefont {L.}~\bibnamefont {Li}},\ }\bibfield  {title}
  {{\selectlanguage {en}\bibinfo {title} {Temperature-pressure phase diagram of
  confined monolayer water/ice at first-principles accuracy with a
  machine-learning force field}},\ }\href
  {https://doi.org/10.1038/s41467-023-39829-z} {\bibfield  {journal} {\bibinfo
  {journal} {Nature Communications}\ }\textbf {\bibinfo {volume} {14}},\
  \bibinfo {pages} {4110} (\bibinfo {year} {2023})}\BibitemShut {NoStop}%
\bibitem [{\citenamefont {Jiang}\ \emph {et~al.}(2024)\citenamefont {Jiang},
  \citenamefont {Gao}, \citenamefont {Li}, \citenamefont {Liu}, \citenamefont
  {Zhu}, \citenamefont {Zhu}, \citenamefont {Francisco},\ and\ \citenamefont
  {Zeng}}]{jiang_rich_2024}%
  \BibitemOpen
  \bibfield  {author} {\bibinfo {author} {\bibfnamefont {J.}~\bibnamefont
  {Jiang}}, \bibinfo {author} {\bibfnamefont {Y.}~\bibnamefont {Gao}}, \bibinfo
  {author} {\bibfnamefont {L.}~\bibnamefont {Li}}, \bibinfo {author}
  {\bibfnamefont {Y.}~\bibnamefont {Liu}}, \bibinfo {author} {\bibfnamefont
  {W.}~\bibnamefont {Zhu}}, \bibinfo {author} {\bibfnamefont {C.}~\bibnamefont
  {Zhu}}, \bibinfo {author} {\bibfnamefont {J.~S.}\ \bibnamefont {Francisco}},\
  and\ \bibinfo {author} {\bibfnamefont {X.~C.}\ \bibnamefont {Zeng}},\
  }\bibfield  {title} {{\selectlanguage {en}\bibinfo {title} {Rich proton
  dynamics and phase behaviours of nanoconfined ices}},\ }\href
  {https://doi.org/10.1038/s41567-023-02341-8} {\bibfield  {journal} {\bibinfo
  {journal} {Nature Physics}\ ,\ \bibinfo {pages} {1}} (\bibinfo {year}
  {2024})}\BibitemShut {NoStop}%
\bibitem [{\citenamefont {Kapil}\ \emph {et~al.}(2025)\citenamefont {Kapil},
  \citenamefont {Schran}, \citenamefont {Zen}, \citenamefont {Chen},
  \citenamefont {Pickard},\ and\ \citenamefont
  {Michaelides}}]{kapil_author_2025}%
  \BibitemOpen
  \bibfield  {author} {\bibinfo {author} {\bibfnamefont {V.}~\bibnamefont
  {Kapil}}, \bibinfo {author} {\bibfnamefont {C.}~\bibnamefont {Schran}},
  \bibinfo {author} {\bibfnamefont {A.}~\bibnamefont {Zen}}, \bibinfo {author}
  {\bibfnamefont {J.}~\bibnamefont {Chen}}, \bibinfo {author} {\bibfnamefont
  {C.~J.}\ \bibnamefont {Pickard}},\ and\ \bibinfo {author} {\bibfnamefont
  {A.}~\bibnamefont {Michaelides}},\ }\bibfield  {title} {{\selectlanguage
  {en}\bibinfo {title} {Author {Correction}: {The} first-principles phase
  diagram of monolayer nanoconfined water}},\ }\href
  {https://doi.org/10.1038/s41586-025-09872-5} {\bibfield  {journal} {\bibinfo
  {journal} {Nature}\ }\textbf {\bibinfo {volume} {648}},\ \bibinfo {pages}
  {E15} (\bibinfo {year} {2025})}\BibitemShut {NoStop}%
\bibitem [{\citenamefont {Ravindra}\ \emph {et~al.}(2024)\citenamefont
  {Ravindra}, \citenamefont {Advincula}, \citenamefont {Schran}, \citenamefont
  {Michaelides},\ and\ \citenamefont
  {Kapil}}]{ravindra_quasi-one-dimensional_2024}%
  \BibitemOpen
  \bibfield  {author} {\bibinfo {author} {\bibfnamefont {P.}~\bibnamefont
  {Ravindra}}, \bibinfo {author} {\bibfnamefont {X.~R.}\ \bibnamefont
  {Advincula}}, \bibinfo {author} {\bibfnamefont {C.}~\bibnamefont {Schran}},
  \bibinfo {author} {\bibfnamefont {A.}~\bibnamefont {Michaelides}},\ and\
  \bibinfo {author} {\bibfnamefont {V.}~\bibnamefont {Kapil}},\ }\bibfield
  {title} {{\selectlanguage {en}\bibinfo {title} {Quasi-one-dimensional
  hydrogen bonding in nanoconfined ice}},\ }\href
  {https://doi.org/10.1038/s41467-024-51124-z} {\bibfield  {journal} {\bibinfo
  {journal} {Nature Communications}\ }\textbf {\bibinfo {volume} {15}},\
  \bibinfo {pages} {7301} (\bibinfo {year} {2024})}\BibitemShut {NoStop}%
\bibitem [{\citenamefont {Dufils}\ \emph {et~al.}(2024)\citenamefont {Dufils},
  \citenamefont {Schran}, \citenamefont {Chen}, \citenamefont {K. Geim},
  \citenamefont {Fumagalli},\ and\ \citenamefont
  {Michaelides}}]{dufils_origin_2024}%
  \BibitemOpen
  \bibfield  {author} {\bibinfo {author} {\bibfnamefont {T.}~\bibnamefont
  {Dufils}}, \bibinfo {author} {\bibfnamefont {C.}~\bibnamefont {Schran}},
  \bibinfo {author} {\bibfnamefont {J.}~\bibnamefont {Chen}}, \bibinfo {author}
  {\bibfnamefont {A.}~\bibnamefont {K. Geim}}, \bibinfo {author}
  {\bibfnamefont {L.}~\bibnamefont {Fumagalli}},\ and\ \bibinfo {author}
  {\bibfnamefont {A.}~\bibnamefont {Michaelides}},\ }\bibfield  {title}
  {{\selectlanguage {en}\bibinfo {title} {Origin of dielectric polarization
  suppression in confined water from first principles}},\ }\href
  {https://doi.org/10.1039/D3SC04740G} {\bibfield  {journal} {\bibinfo
  {journal} {Chemical Science}\ }\textbf {\bibinfo {volume} {15}},\ \bibinfo
  {pages} {516} (\bibinfo {year} {2024})}\BibitemShut {NoStop}%
\bibitem [{\citenamefont {Advincula}\ \emph
  {et~al.}(2025{\natexlab{a}})\citenamefont {Advincula}, \citenamefont
  {Litman}, \citenamefont {Fong}, \citenamefont {Witt}, \citenamefont
  {Schran},\ and\ \citenamefont {Michaelides}}]{advincula_how_2025}%
  \BibitemOpen
  \bibfield  {author} {\bibinfo {author} {\bibfnamefont {X.~R.}\ \bibnamefont
  {Advincula}}, \bibinfo {author} {\bibfnamefont {Y.}~\bibnamefont {Litman}},
  \bibinfo {author} {\bibfnamefont {K.~D.}\ \bibnamefont {Fong}}, \bibinfo
  {author} {\bibfnamefont {W.~C.}\ \bibnamefont {Witt}}, \bibinfo {author}
  {\bibfnamefont {C.}~\bibnamefont {Schran}},\ and\ \bibinfo {author}
  {\bibfnamefont {A.}~\bibnamefont {Michaelides}},\ }\href
  {https://doi.org/10.48550/arXiv.2508.13034} {\bibinfo {title} {How reactive
  is water at the nanoscale and how to control it?}} (\bibinfo {year}
  {2025}{\natexlab{a}}),\ \bibinfo {note} {arXiv:2508.13034}\BibitemShut
  {NoStop}%
\bibitem [{\citenamefont {Ravindra}\ \emph {et~al.}(2026)\citenamefont
  {Ravindra}, \citenamefont {Advincula}, \citenamefont {Shi}, \citenamefont
  {Coles}, \citenamefont {Michaelides},\ and\ \citenamefont
  {Kapil}}]{ravindra_nuclear_2026}%
  \BibitemOpen
  \bibfield  {author} {\bibinfo {author} {\bibfnamefont {P.}~\bibnamefont
  {Ravindra}}, \bibinfo {author} {\bibfnamefont {X.~R.}\ \bibnamefont
  {Advincula}}, \bibinfo {author} {\bibfnamefont {B.~X.}\ \bibnamefont {Shi}},
  \bibinfo {author} {\bibfnamefont {S.~W.}\ \bibnamefont {Coles}}, \bibinfo
  {author} {\bibfnamefont {A.}~\bibnamefont {Michaelides}},\ and\ \bibinfo
  {author} {\bibfnamefont {V.}~\bibnamefont {Kapil}},\ }\bibfield  {title}
  {{\selectlanguage {en}\bibinfo {title} {Nuclear quantum effects amplify
  autoionization-driven superionic behaviour in nanoconfined monolayer
  water}},\ }\bibfield  {journal} {\bibinfo  {journal} {Chemical Science}\
  }\href {https://doi.org/10.1039/D6SC00138F} {10.1039/D6SC00138F} (\bibinfo
  {year} {2026})\BibitemShut {NoStop}%
\bibitem [{\citenamefont {Björneholm}\ \emph {et~al.}(2016)\citenamefont
  {Björneholm}, \citenamefont {Hansen}, \citenamefont {Hodgson}, \citenamefont
  {Liu}, \citenamefont {Limmer}, \citenamefont {Michaelides}, \citenamefont
  {Pedevilla}, \citenamefont {Rossmeisl}, \citenamefont {Shen}, \citenamefont
  {Tocci}, \citenamefont {Tyrode}, \citenamefont {Walz}, \citenamefont
  {Werner},\ and\ \citenamefont {Bluhm}}]{bjorneholm_water_2016}%
  \BibitemOpen
  \bibfield  {author} {\bibinfo {author} {\bibfnamefont {O.}~\bibnamefont
  {Björneholm}}, \bibinfo {author} {\bibfnamefont {M.~H.}\ \bibnamefont
  {Hansen}}, \bibinfo {author} {\bibfnamefont {A.}~\bibnamefont {Hodgson}},
  \bibinfo {author} {\bibfnamefont {L.-M.}\ \bibnamefont {Liu}}, \bibinfo
  {author} {\bibfnamefont {D.~T.}\ \bibnamefont {Limmer}}, \bibinfo {author}
  {\bibfnamefont {A.}~\bibnamefont {Michaelides}}, \bibinfo {author}
  {\bibfnamefont {P.}~\bibnamefont {Pedevilla}}, \bibinfo {author}
  {\bibfnamefont {J.}~\bibnamefont {Rossmeisl}}, \bibinfo {author}
  {\bibfnamefont {H.}~\bibnamefont {Shen}}, \bibinfo {author} {\bibfnamefont
  {G.}~\bibnamefont {Tocci}}, \bibinfo {author} {\bibfnamefont
  {E.}~\bibnamefont {Tyrode}}, \bibinfo {author} {\bibfnamefont {M.-M.}\
  \bibnamefont {Walz}}, \bibinfo {author} {\bibfnamefont {J.}~\bibnamefont
  {Werner}},\ and\ \bibinfo {author} {\bibfnamefont {H.}~\bibnamefont
  {Bluhm}},\ }\bibfield  {title} {{\selectlanguage {en}\bibinfo {title} {Water
  at {Interfaces}}},\ }\href {https://doi.org/10.1021/acs.chemrev.6b00045}
  {\bibfield  {journal} {\bibinfo  {journal} {Chemical Reviews}\ }\textbf
  {\bibinfo {volume} {116}},\ \bibinfo {pages} {7698} (\bibinfo {year}
  {2016})}\BibitemShut {NoStop}%
\bibitem [{\citenamefont {Wang}\ \emph
  {et~al.}(2025{\natexlab{b}})\citenamefont {Wang}, \citenamefont {Tang},
  \citenamefont {Yu}, \citenamefont {Chiang}, \citenamefont {Yu}, \citenamefont
  {Ohto}, \citenamefont {Chen}, \citenamefont {Nagata},\ and\ \citenamefont
  {Bonn}}]{wang_interfaces_2025}%
  \BibitemOpen
  \bibfield  {author} {\bibinfo {author} {\bibfnamefont {Y.}~\bibnamefont
  {Wang}}, \bibinfo {author} {\bibfnamefont {F.}~\bibnamefont {Tang}}, \bibinfo
  {author} {\bibfnamefont {X.}~\bibnamefont {Yu}}, \bibinfo {author}
  {\bibfnamefont {K.-Y.}\ \bibnamefont {Chiang}}, \bibinfo {author}
  {\bibfnamefont {C.-C.}\ \bibnamefont {Yu}}, \bibinfo {author} {\bibfnamefont
  {T.}~\bibnamefont {Ohto}}, \bibinfo {author} {\bibfnamefont {Y.}~\bibnamefont
  {Chen}}, \bibinfo {author} {\bibfnamefont {Y.}~\bibnamefont {Nagata}},\ and\
  \bibinfo {author} {\bibfnamefont {M.}~\bibnamefont {Bonn}},\ }\bibfield
  {title} {{\selectlanguage {en}\bibinfo {title} {Interfaces govern the
  structure of angstrom-scale confined water solutions}},\ }\href
  {https://doi.org/10.1038/s41467-025-62625-w} {\bibfield  {journal} {\bibinfo
  {journal} {Nature Communications}\ }\textbf {\bibinfo {volume} {16}},\
  \bibinfo {pages} {7288} (\bibinfo {year} {2025}{\natexlab{b}})}\BibitemShut
  {NoStop}%
\bibitem [{\citenamefont {Frisenda}\ \emph {et~al.}(2018)\citenamefont
  {Frisenda}, \citenamefont {Navarro-Moratalla}, \citenamefont {Gant},
  \citenamefont {Lara}, \citenamefont {Jarillo-Herrero}, \citenamefont
  {Gorbachev},\ and\ \citenamefont {Castellanos-Gomez}}]{frisenda_recent_2018}%
  \BibitemOpen
  \bibfield  {author} {\bibinfo {author} {\bibfnamefont {R.}~\bibnamefont
  {Frisenda}}, \bibinfo {author} {\bibfnamefont {E.}~\bibnamefont
  {Navarro-Moratalla}}, \bibinfo {author} {\bibfnamefont {P.}~\bibnamefont
  {Gant}}, \bibinfo {author} {\bibfnamefont {D.~P.~D.}\ \bibnamefont {Lara}},
  \bibinfo {author} {\bibfnamefont {P.}~\bibnamefont {Jarillo-Herrero}},
  \bibinfo {author} {\bibfnamefont {R.~V.}\ \bibnamefont {Gorbachev}},\ and\
  \bibinfo {author} {\bibfnamefont {A.}~\bibnamefont {Castellanos-Gomez}},\
  }\bibfield  {title} {{\selectlanguage {en}\bibinfo {title} {Recent progress
  in the assembly of nanodevices and van der {Waals} heterostructures by
  deterministic placement of {2D} materials}},\ }\href
  {https://doi.org/10.1039/C7CS00556C} {\bibfield  {journal} {\bibinfo
  {journal} {Chemical Society Reviews}\ }\textbf {\bibinfo {volume} {47}},\
  \bibinfo {pages} {53} (\bibinfo {year} {2018})}\BibitemShut {NoStop}%
\bibitem [{\citenamefont {Zhang}\ \emph {et~al.}(2021)\citenamefont {Zhang},
  \citenamefont {Lin}, \citenamefont {Lei}, \citenamefont {Qi}, \citenamefont
  {Ban}, \citenamefont {Vinu}, \citenamefont {Yi},\ and\ \citenamefont
  {Liu}}]{zhang_twist_2021}%
  \BibitemOpen
  \bibfield  {author} {\bibinfo {author} {\bibfnamefont {E.}~\bibnamefont
  {Zhang}}, \bibinfo {author} {\bibfnamefont {F.}~\bibnamefont {Lin}}, \bibinfo
  {author} {\bibfnamefont {Z.}~\bibnamefont {Lei}}, \bibinfo {author}
  {\bibfnamefont {S.}~\bibnamefont {Qi}}, \bibinfo {author} {\bibfnamefont
  {S.}~\bibnamefont {Ban}}, \bibinfo {author} {\bibfnamefont {A.}~\bibnamefont
  {Vinu}}, \bibinfo {author} {\bibfnamefont {J.}~\bibnamefont {Yi}},\ and\
  \bibinfo {author} {\bibfnamefont {Y.}~\bibnamefont {Liu}},\ }\bibfield
  {title} {{\selectlanguage {en}\bibinfo {title} {Twist the doorknob to open
  the electronic properties of graphene-based van der {Waals} structure}},\
  }\href {https://doi.org/10.1016/j.matt.2021.08.020} {\bibfield  {journal}
  {\bibinfo  {journal} {Matter}\ }\textbf {\bibinfo {volume} {4}},\ \bibinfo
  {pages} {3444} (\bibinfo {year} {2021})}\BibitemShut {NoStop}%
\bibitem [{\citenamefont {Fox}\ \emph {et~al.}(2024)\citenamefont {Fox},
  \citenamefont {Mao}, \citenamefont {Zhang}, \citenamefont {Wang},\ and\
  \citenamefont {Xiao}}]{fox_stacking_2024}%
  \BibitemOpen
  \bibfield  {author} {\bibinfo {author} {\bibfnamefont {C.}~\bibnamefont
  {Fox}}, \bibinfo {author} {\bibfnamefont {Y.}~\bibnamefont {Mao}}, \bibinfo
  {author} {\bibfnamefont {X.}~\bibnamefont {Zhang}}, \bibinfo {author}
  {\bibfnamefont {Y.}~\bibnamefont {Wang}},\ and\ \bibinfo {author}
  {\bibfnamefont {J.}~\bibnamefont {Xiao}},\ }\bibfield  {title}
  {{\selectlanguage {en}\bibinfo {title} {Stacking {Order} {Engineering} of
  {Two}-{Dimensional} {Materials} and {Device} {Applications}}},\ }\href
  {https://doi.org/10.1021/acs.chemrev.3c00618} {\bibfield  {journal} {\bibinfo
   {journal} {Chemical Reviews}\ }\textbf {\bibinfo {volume} {124}},\ \bibinfo
  {pages} {1862} (\bibinfo {year} {2024})}\BibitemShut {NoStop}%
\bibitem [{\citenamefont {Pickard}\ and\ \citenamefont
  {Needs}(2011{\natexlab{a}})}]{pickard_ab_2011}%
  \BibitemOpen
  \bibfield  {author} {\bibinfo {author} {\bibfnamefont {C.~J.}\ \bibnamefont
  {Pickard}}\ and\ \bibinfo {author} {\bibfnamefont {R.~J.}\ \bibnamefont
  {Needs}},\ }\bibfield  {title} {{\selectlanguage {en}\bibinfo {title} {Ab
  initiorandom structure searching}},\ }\href
  {https://doi.org/10.1088/0953-8984/23/5/053201} {\bibfield  {journal}
  {\bibinfo  {journal} {Journal of Physics: Condensed Matter}\ }\textbf
  {\bibinfo {volume} {23}},\ \bibinfo {pages} {053201} (\bibinfo {year}
  {2011}{\natexlab{a}})}\BibitemShut {NoStop}%
\bibitem [{\citenamefont {Rahman}(1964)}]{rahman_correlations_1964}%
  \BibitemOpen
  \bibfield  {author} {\bibinfo {author} {\bibfnamefont {A.}~\bibnamefont
  {Rahman}},\ }\bibfield  {title} {\bibinfo {title} {Correlations in the
  {Motion} of {Atoms} in {Liquid} {Argon}},\ }\href
  {https://doi.org/10.1103/PhysRev.136.A405} {\bibfield  {journal} {\bibinfo
  {journal} {Physical Review}\ }\textbf {\bibinfo {volume} {136}},\ \bibinfo
  {pages} {A405} (\bibinfo {year} {1964})}\BibitemShut {NoStop}%
\bibitem [{\citenamefont {Parrinello}\ and\ \citenamefont
  {Rahman}(1984)}]{parrinello_study_1984}%
  \BibitemOpen
  \bibfield  {author} {\bibinfo {author} {\bibfnamefont {M.}~\bibnamefont
  {Parrinello}}\ and\ \bibinfo {author} {\bibfnamefont {A.}~\bibnamefont
  {Rahman}},\ }\bibfield  {title} {\bibinfo {title} {Study of an {F} center in
  molten {KCl}},\ }\href {https://doi.org/10.1063/1.446740} {\bibfield
  {journal} {\bibinfo  {journal} {The Journal of Chemical Physics}\ }\textbf
  {\bibinfo {volume} {80}},\ \bibinfo {pages} {860} (\bibinfo {year}
  {1984})}\BibitemShut {NoStop}%
\bibitem [{\citenamefont {Musil}\ \emph {et~al.}(2022)\citenamefont {Musil},
  \citenamefont {Zaporozhets}, \citenamefont {Noé}, \citenamefont {Clementi},\
  and\ \citenamefont {Kapil}}]{musil_quantum_2022}%
  \BibitemOpen
  \bibfield  {author} {\bibinfo {author} {\bibfnamefont {F.}~\bibnamefont
  {Musil}}, \bibinfo {author} {\bibfnamefont {I.}~\bibnamefont {Zaporozhets}},
  \bibinfo {author} {\bibfnamefont {F.}~\bibnamefont {Noé}}, \bibinfo {author}
  {\bibfnamefont {C.}~\bibnamefont {Clementi}},\ and\ \bibinfo {author}
  {\bibfnamefont {V.}~\bibnamefont {Kapil}},\ }\bibfield  {title} {\bibinfo
  {title} {Quantum dynamics using path integral coarse-graining},\ }\bibfield
  {journal} {\bibinfo  {journal} {The Journal of Chemical Physics}\ }\href
  {https://doi.org/10.1063/5.0120386} {10.1063/5.0120386} (\bibinfo {year}
  {2022}),\ \bibinfo {note} {publisher: American Institute of
  Physics}\BibitemShut {NoStop}%
\bibitem [{\citenamefont {Coles}\ \emph {et~al.}(2026)\citenamefont {Coles},
  \citenamefont {Hajibabaei}, \citenamefont {Kapil}, \citenamefont {Advincula},
  \citenamefont {Schran}, \citenamefont {Cox},\ and\ \citenamefont
  {Michaelides}}]{coles_nanoconfined_2026}%
  \BibitemOpen
  \bibfield  {author} {\bibinfo {author} {\bibfnamefont {S.~W.}\ \bibnamefont
  {Coles}}, \bibinfo {author} {\bibfnamefont {A.}~\bibnamefont {Hajibabaei}},
  \bibinfo {author} {\bibfnamefont {V.}~\bibnamefont {Kapil}}, \bibinfo
  {author} {\bibfnamefont {X.~R.}\ \bibnamefont {Advincula}}, \bibinfo {author}
  {\bibfnamefont {C.}~\bibnamefont {Schran}}, \bibinfo {author} {\bibfnamefont
  {S.~J.}\ \bibnamefont {Cox}},\ and\ \bibinfo {author} {\bibfnamefont
  {A.}~\bibnamefont {Michaelides}},\ }\bibfield  {title} {{\selectlanguage
  {en}\bibinfo {title} {Nanoconfined superionic water is a molecular
  superionic}},\ }\href {https://doi.org/10.1126/sciadv.adz6392} {\bibfield
  {journal} {\bibinfo  {journal} {Science Advances}\ }\textbf {\bibinfo
  {volume} {12}},\ \bibinfo {pages} {eadz6392} (\bibinfo {year}
  {2026})}\BibitemShut {NoStop}%
\bibitem [{\citenamefont {Cao}\ \emph {et~al.}(2018)\citenamefont {Cao},
  \citenamefont {Fatemi}, \citenamefont {Fang}, \citenamefont {Watanabe},
  \citenamefont {Taniguchi}, \citenamefont {Kaxiras},\ and\ \citenamefont
  {Jarillo-Herrero}}]{cao_unconventional_2018}%
  \BibitemOpen
  \bibfield  {author} {\bibinfo {author} {\bibfnamefont {Y.}~\bibnamefont
  {Cao}}, \bibinfo {author} {\bibfnamefont {V.}~\bibnamefont {Fatemi}},
  \bibinfo {author} {\bibfnamefont {S.}~\bibnamefont {Fang}}, \bibinfo {author}
  {\bibfnamefont {K.}~\bibnamefont {Watanabe}}, \bibinfo {author}
  {\bibfnamefont {T.}~\bibnamefont {Taniguchi}}, \bibinfo {author}
  {\bibfnamefont {E.}~\bibnamefont {Kaxiras}},\ and\ \bibinfo {author}
  {\bibfnamefont {P.}~\bibnamefont {Jarillo-Herrero}},\ }\bibfield  {title}
  {{\selectlanguage {en}\bibinfo {title} {Unconventional superconductivity in
  magic-angle graphene superlattices}},\ }\href
  {https://doi.org/10.1038/nature26160} {\bibfield  {journal} {\bibinfo
  {journal} {Nature}\ }\textbf {\bibinfo {volume} {556}},\ \bibinfo {pages}
  {43} (\bibinfo {year} {2018})}\BibitemShut {NoStop}%
\bibitem [{\citenamefont {Kim}\ \emph {et~al.}(2017)\citenamefont {Kim},
  \citenamefont {DaSilva}, \citenamefont {Huang}, \citenamefont {Fallahazad},
  \citenamefont {Larentis}, \citenamefont {Taniguchi}, \citenamefont
  {Watanabe}, \citenamefont {LeRoy}, \citenamefont {MacDonald},\ and\
  \citenamefont {Tutuc}}]{kim_tunable_2017}%
  \BibitemOpen
  \bibfield  {author} {\bibinfo {author} {\bibfnamefont {K.}~\bibnamefont
  {Kim}}, \bibinfo {author} {\bibfnamefont {A.}~\bibnamefont {DaSilva}},
  \bibinfo {author} {\bibfnamefont {S.}~\bibnamefont {Huang}}, \bibinfo
  {author} {\bibfnamefont {B.}~\bibnamefont {Fallahazad}}, \bibinfo {author}
  {\bibfnamefont {S.}~\bibnamefont {Larentis}}, \bibinfo {author}
  {\bibfnamefont {T.}~\bibnamefont {Taniguchi}}, \bibinfo {author}
  {\bibfnamefont {K.}~\bibnamefont {Watanabe}}, \bibinfo {author}
  {\bibfnamefont {B.~J.}\ \bibnamefont {LeRoy}}, \bibinfo {author}
  {\bibfnamefont {A.~H.}\ \bibnamefont {MacDonald}},\ and\ \bibinfo {author}
  {\bibfnamefont {E.}~\bibnamefont {Tutuc}},\ }\bibfield  {title}
  {{\selectlanguage {en}\bibinfo {title} {Tunable moiré bands and strong
  correlations in small-twist-angle bilayer graphene}},\ }\href
  {https://doi.org/10.1073/pnas.1620140114} {\bibfield  {journal} {\bibinfo
  {journal} {Proceedings of the National Academy of Sciences}\ }\textbf
  {\bibinfo {volume} {114}},\ \bibinfo {pages} {3364} (\bibinfo {year}
  {2017})}\BibitemShut {NoStop}%
\bibitem [{\citenamefont {Zhang}\ and\ \citenamefont
  {Yang}(1998)}]{zhang_comment_1998}%
  \BibitemOpen
  \bibfield  {author} {\bibinfo {author} {\bibfnamefont {Y.}~\bibnamefont
  {Zhang}}\ and\ \bibinfo {author} {\bibfnamefont {W.}~\bibnamefont {Yang}},\
  }\bibfield  {title} {\bibinfo {title} {Comment on ``{Generalized} {Gradient}
  {Approximation} {Made} {Simple}''},\ }\href
  {https://doi.org/10.1103/PhysRevLett.80.890} {\bibfield  {journal} {\bibinfo
  {journal} {Physical Review Letters}\ }\textbf {\bibinfo {volume} {80}},\
  \bibinfo {pages} {890} (\bibinfo {year} {1998})}\BibitemShut {NoStop}%
\bibitem [{\citenamefont {Adamo}\ and\ \citenamefont
  {Barone}(1999)}]{adamo_toward_1999}%
  \BibitemOpen
  \bibfield  {author} {\bibinfo {author} {\bibfnamefont {C.}~\bibnamefont
  {Adamo}}\ and\ \bibinfo {author} {\bibfnamefont {V.}~\bibnamefont {Barone}},\
  }\bibfield  {title} {\bibinfo {title} {Toward reliable density functional
  methods without adjustable parameters: {The} {PBE0} model},\ }\href
  {https://doi.org/10.1063/1.478522} {\bibfield  {journal} {\bibinfo  {journal}
  {The Journal of Chemical Physics}\ }\textbf {\bibinfo {volume} {110}},\
  \bibinfo {pages} {6158} (\bibinfo {year} {1999})}\BibitemShut {NoStop}%
\bibitem [{\citenamefont {Grimme}\ \emph {et~al.}(2010)\citenamefont {Grimme},
  \citenamefont {Antony}, \citenamefont {Ehrlich},\ and\ \citenamefont
  {Krieg}}]{grimme_consistent_2010}%
  \BibitemOpen
  \bibfield  {author} {\bibinfo {author} {\bibfnamefont {S.}~\bibnamefont
  {Grimme}}, \bibinfo {author} {\bibfnamefont {J.}~\bibnamefont {Antony}},
  \bibinfo {author} {\bibfnamefont {S.}~\bibnamefont {Ehrlich}},\ and\ \bibinfo
  {author} {\bibfnamefont {H.}~\bibnamefont {Krieg}},\ }\bibfield  {title}
  {\bibinfo {title} {A consistent and accurate ab initio parametrization of
  density functional dispersion correction ({DFT}-{D}) for the 94 elements
  {H}-{Pu}},\ }\href {https://doi.org/10.1063/1.3382344} {\bibfield  {journal}
  {\bibinfo  {journal} {The Journal of Chemical Physics}\ }\textbf {\bibinfo
  {volume} {132}},\ \bibinfo {pages} {154104} (\bibinfo {year}
  {2010})}\BibitemShut {NoStop}%
\bibitem [{\citenamefont {Advincula}\ \emph
  {et~al.}(2025{\natexlab{b}})\citenamefont {Advincula}, \citenamefont {Fong},
  \citenamefont {Michaelides},\ and\ \citenamefont
  {Schran}}]{advincula_protons_2025}%
  \BibitemOpen
  \bibfield  {author} {\bibinfo {author} {\bibfnamefont {X.~R.}\ \bibnamefont
  {Advincula}}, \bibinfo {author} {\bibfnamefont {K.~D.}\ \bibnamefont {Fong}},
  \bibinfo {author} {\bibfnamefont {A.}~\bibnamefont {Michaelides}},\ and\
  \bibinfo {author} {\bibfnamefont {C.}~\bibnamefont {Schran}},\ }\bibfield
  {title} {{\selectlanguage {en}\bibinfo {title} {Protons {Accumulate} at the
  {Graphene}–{Water} {Interface}}},\ }\href
  {https://doi.org/10.1021/acsnano.5c02053} {\bibfield  {journal} {\bibinfo
  {journal} {ACS Nano}\ }\textbf {\bibinfo {volume} {19}},\ \bibinfo {pages}
  {17728} (\bibinfo {year} {2025}{\natexlab{b}})}\BibitemShut {NoStop}%
\bibitem [{\citenamefont {Brandenburg}\ \emph {et~al.}(2019)\citenamefont
  {Brandenburg}, \citenamefont {Zen}, \citenamefont {Alfè},\ and\
  \citenamefont {Michaelides}}]{brandenburg_interaction_2019}%
  \BibitemOpen
  \bibfield  {author} {\bibinfo {author} {\bibfnamefont {J.~G.}\ \bibnamefont
  {Brandenburg}}, \bibinfo {author} {\bibfnamefont {A.}~\bibnamefont {Zen}},
  \bibinfo {author} {\bibfnamefont {D.}~\bibnamefont {Alfè}},\ and\ \bibinfo
  {author} {\bibfnamefont {A.}~\bibnamefont {Michaelides}},\ }\bibfield
  {title} {\bibinfo {title} {Interaction between water and carbon
  nanostructures: {How} good are current density functional approximations?},\
  }\href {https://doi.org/10.1063/1.5121370} {\bibfield  {journal} {\bibinfo
  {journal} {The Journal of Chemical Physics}\ }\textbf {\bibinfo {volume}
  {151}},\ \bibinfo {pages} {164702} (\bibinfo {year} {2019})}\BibitemShut
  {NoStop}%
\bibitem [{\citenamefont {Marsalek}\ and\ \citenamefont
  {Markland}(2017)}]{marsalek_quantum_2017}%
  \BibitemOpen
  \bibfield  {author} {\bibinfo {author} {\bibfnamefont {O.}~\bibnamefont
  {Marsalek}}\ and\ \bibinfo {author} {\bibfnamefont {T.~E.}\ \bibnamefont
  {Markland}},\ }\bibfield  {title} {\bibinfo {title} {Quantum {Dynamics} and
  {Spectroscopy} of {Ab} {Initio} {Liquid} {Water}: {The} {Interplay} of
  {Nuclear} and {Electronic} {Quantum} {Effects}},\ }\href
  {https://doi.org/10.1021/acs.jpclett.7b00391} {\bibfield  {journal} {\bibinfo
   {journal} {The Journal of Physical Chemistry Letters}\ }\textbf {\bibinfo
  {volume} {8}},\ \bibinfo {pages} {1545} (\bibinfo {year} {2017})}\BibitemShut
  {NoStop}%
\bibitem [{\citenamefont {Cheng}\ \emph {et~al.}(2019)\citenamefont {Cheng},
  \citenamefont {Engel}, \citenamefont {Behler}, \citenamefont {Dellago},\ and\
  \citenamefont {Ceriotti}}]{cheng_ab_2019}%
  \BibitemOpen
  \bibfield  {author} {\bibinfo {author} {\bibfnamefont {B.}~\bibnamefont
  {Cheng}}, \bibinfo {author} {\bibfnamefont {E.~A.}\ \bibnamefont {Engel}},
  \bibinfo {author} {\bibfnamefont {J.}~\bibnamefont {Behler}}, \bibinfo
  {author} {\bibfnamefont {C.}~\bibnamefont {Dellago}},\ and\ \bibinfo {author}
  {\bibfnamefont {M.}~\bibnamefont {Ceriotti}},\ }\bibfield  {title}
  {{\selectlanguage {en}\bibinfo {title} {Ab initio thermodynamics of liquid
  and solid water}},\ }\href {https://doi.org/10.1073/pnas.1815117116}
  {\bibfield  {journal} {\bibinfo  {journal} {Proceedings of the National
  Academy of Sciences}\ }\textbf {\bibinfo {volume} {116}},\ \bibinfo {pages}
  {1110} (\bibinfo {year} {2019})}\BibitemShut {NoStop}%
\bibitem [{\citenamefont {Kapil}\ \emph {et~al.}(2024)\citenamefont {Kapil},
  \citenamefont {Kovács}, \citenamefont {Csányi},\ and\ \citenamefont
  {Michaelides}}]{kapil_first-principles_2024}%
  \BibitemOpen
  \bibfield  {author} {\bibinfo {author} {\bibfnamefont {V.}~\bibnamefont
  {Kapil}}, \bibinfo {author} {\bibfnamefont {D.~P.}\ \bibnamefont {Kovács}},
  \bibinfo {author} {\bibfnamefont {G.}~\bibnamefont {Csányi}},\ and\ \bibinfo
  {author} {\bibfnamefont {A.}~\bibnamefont {Michaelides}},\ }\bibfield
  {title} {{\selectlanguage {en}\bibinfo {title} {First-principles spectroscopy
  of aqueous interfaces using machine-learned electronic and quantum nuclear
  effects}},\ }\href {https://doi.org/10.1039/D3FD00113J} {\bibfield  {journal}
  {\bibinfo  {journal} {Faraday Discussions}\ }\textbf {\bibinfo {volume}
  {249}},\ \bibinfo {pages} {50} (\bibinfo {year} {2024})}\BibitemShut
  {NoStop}%
\bibitem [{\citenamefont {Shi}\ \emph {et~al.}(2022)\citenamefont {Shi},
  \citenamefont {Kapil}, \citenamefont {Zen}, \citenamefont {Chen},
  \citenamefont {Alavi},\ and\ \citenamefont {Michaelides}}]{shi_general_2022}%
  \BibitemOpen
  \bibfield  {author} {\bibinfo {author} {\bibfnamefont {B.~X.}\ \bibnamefont
  {Shi}}, \bibinfo {author} {\bibfnamefont {V.}~\bibnamefont {Kapil}}, \bibinfo
  {author} {\bibfnamefont {A.}~\bibnamefont {Zen}}, \bibinfo {author}
  {\bibfnamefont {J.}~\bibnamefont {Chen}}, \bibinfo {author} {\bibfnamefont
  {A.}~\bibnamefont {Alavi}},\ and\ \bibinfo {author} {\bibfnamefont
  {A.}~\bibnamefont {Michaelides}},\ }\bibfield  {title} {\bibinfo {title}
  {General embedded cluster protocol for accurate modeling of oxygen vacancies
  in metal-oxides},\ }\href {https://doi.org/10.1063/5.0087031} {\bibfield
  {journal} {\bibinfo  {journal} {The Journal of Chemical Physics}\ }\textbf
  {\bibinfo {volume} {156}},\ \bibinfo {pages} {124704} (\bibinfo {year}
  {2022})}\BibitemShut {NoStop}%
\bibitem [{\citenamefont {Klimeš}(2016)}]{klimes_lattice_2016}%
  \BibitemOpen
  \bibfield  {author} {\bibinfo {author} {\bibfnamefont {J.}~\bibnamefont
  {Klimeš}},\ }\bibfield  {title} {\bibinfo {title} {Lattice energies of
  molecular solids from the random phase approximation with singles
  corrections},\ }\href {https://doi.org/10.1063/1.4962188} {\bibfield
  {journal} {\bibinfo  {journal} {The Journal of Chemical Physics}\ }\textbf
  {\bibinfo {volume} {145}},\ \bibinfo {pages} {094506} (\bibinfo {year}
  {2016})}\BibitemShut {NoStop}%
\bibitem [{\citenamefont {Batatia}\ \emph {et~al.}(2022)\citenamefont
  {Batatia}, \citenamefont {Kovacs}, \citenamefont {Simm}, \citenamefont
  {Ortner},\ and\ \citenamefont {Csanyi}}]{batatia_mace_2022}%
  \BibitemOpen
  \bibfield  {author} {\bibinfo {author} {\bibfnamefont {I.}~\bibnamefont
  {Batatia}}, \bibinfo {author} {\bibfnamefont {D.~P.}\ \bibnamefont {Kovacs}},
  \bibinfo {author} {\bibfnamefont {G.}~\bibnamefont {Simm}}, \bibinfo {author}
  {\bibfnamefont {C.}~\bibnamefont {Ortner}},\ and\ \bibinfo {author}
  {\bibfnamefont {G.}~\bibnamefont {Csanyi}},\ }\bibfield  {title}
  {{\selectlanguage {en}\bibinfo {title} {{MACE}: {Higher} {Order}
  {Equivariant} {Message} {Passing} {Neural} {Networks} for {Fast} and
  {Accurate} {Force} {Fields}}},\ }\href
  {https://proceedings.neurips.cc/paper\_files/paper/2022/hash/4a36c3c51af11ed9f34615b81edb5bbc-Abstract-Conference.html}
  {\bibfield  {journal} {\bibinfo  {journal} {Advances in Neural Information
  Processing Systems}\ }\textbf {\bibinfo {volume} {35}},\ \bibinfo {pages}
  {11423} (\bibinfo {year} {2022})}\BibitemShut {NoStop}%
\bibitem [{\citenamefont {Pickard}\ and\ \citenamefont
  {Needs}(2011{\natexlab{b}})}]{Pickard_2011}%
  \BibitemOpen
  \bibfield  {author} {\bibinfo {author} {\bibfnamefont {C.~J.}\ \bibnamefont
  {Pickard}}\ and\ \bibinfo {author} {\bibfnamefont {R.~J.}\ \bibnamefont
  {Needs}},\ }\bibfield  {title} {\bibinfo {title} {Ab initio random structure
  searching},\ }\href {https://doi.org/10.1088/0953-8984/23/5/053201}
  {\bibfield  {journal} {\bibinfo  {journal} {Journal of Physics: Condensed
  Matter}\ }\textbf {\bibinfo {volume} {23}},\ \bibinfo {pages} {053201}
  (\bibinfo {year} {2011}{\natexlab{b}})}\BibitemShut {NoStop}%
\bibitem [{\citenamefont {Hjorth~Larsen}\ \emph {et~al.}(2017)\citenamefont
  {Hjorth~Larsen}, \citenamefont {Jørgen~Mortensen}, \citenamefont
  {Blomqvist}, \citenamefont {Castelli}, \citenamefont {Christensen},
  \citenamefont {Dułak}, \citenamefont {Friis}, \citenamefont {Groves},
  \citenamefont {Hammer}, \citenamefont {Hargus}, \citenamefont {Hermes},
  \citenamefont {Jennings}, \citenamefont {Bjerre~Jensen}, \citenamefont
  {Kermode}, \citenamefont {Kitchin}, \citenamefont {Leonhard~Kolsbjerg},
  \citenamefont {Kubal}, \citenamefont {Kaasbjerg}, \citenamefont {Lysgaard},
  \citenamefont {Bergmann~Maronsson}, \citenamefont {Maxson}, \citenamefont
  {Olsen}, \citenamefont {Pastewka}, \citenamefont {Peterson}, \citenamefont
  {Rostgaard}, \citenamefont {Schiøtz}, \citenamefont {Schütt}, \citenamefont
  {Strange}, \citenamefont {Thygesen}, \citenamefont {Vegge}, \citenamefont
  {Vilhelmsen}, \citenamefont {Walter}, \citenamefont {Zeng},\ and\
  \citenamefont {Jacobsen}}]{hjorth_larsen_atomic_2017}%
  \BibitemOpen
  \bibfield  {author} {\bibinfo {author} {\bibfnamefont {A.}~\bibnamefont
  {Hjorth~Larsen}}, \bibinfo {author} {\bibfnamefont {J.}~\bibnamefont
  {Jørgen~Mortensen}}, \bibinfo {author} {\bibfnamefont {J.}~\bibnamefont
  {Blomqvist}}, \bibinfo {author} {\bibfnamefont {I.~E.}\ \bibnamefont
  {Castelli}}, \bibinfo {author} {\bibfnamefont {R.}~\bibnamefont
  {Christensen}}, \bibinfo {author} {\bibfnamefont {M.}~\bibnamefont {Dułak}},
  \bibinfo {author} {\bibfnamefont {J.}~\bibnamefont {Friis}}, \bibinfo
  {author} {\bibfnamefont {M.~N.}\ \bibnamefont {Groves}}, \bibinfo {author}
  {\bibfnamefont {B.}~\bibnamefont {Hammer}}, \bibinfo {author} {\bibfnamefont
  {C.}~\bibnamefont {Hargus}}, \bibinfo {author} {\bibfnamefont {E.~D.}\
  \bibnamefont {Hermes}}, \bibinfo {author} {\bibfnamefont {P.~C.}\
  \bibnamefont {Jennings}}, \bibinfo {author} {\bibfnamefont {P.}~\bibnamefont
  {Bjerre~Jensen}}, \bibinfo {author} {\bibfnamefont {J.}~\bibnamefont
  {Kermode}}, \bibinfo {author} {\bibfnamefont {J.~R.}\ \bibnamefont
  {Kitchin}}, \bibinfo {author} {\bibfnamefont {E.}~\bibnamefont
  {Leonhard~Kolsbjerg}}, \bibinfo {author} {\bibfnamefont {J.}~\bibnamefont
  {Kubal}}, \bibinfo {author} {\bibfnamefont {K.}~\bibnamefont {Kaasbjerg}},
  \bibinfo {author} {\bibfnamefont {S.}~\bibnamefont {Lysgaard}}, \bibinfo
  {author} {\bibfnamefont {J.}~\bibnamefont {Bergmann~Maronsson}}, \bibinfo
  {author} {\bibfnamefont {T.}~\bibnamefont {Maxson}}, \bibinfo {author}
  {\bibfnamefont {T.}~\bibnamefont {Olsen}}, \bibinfo {author} {\bibfnamefont
  {L.}~\bibnamefont {Pastewka}}, \bibinfo {author} {\bibfnamefont
  {A.}~\bibnamefont {Peterson}}, \bibinfo {author} {\bibfnamefont
  {C.}~\bibnamefont {Rostgaard}}, \bibinfo {author} {\bibfnamefont
  {J.}~\bibnamefont {Schiøtz}}, \bibinfo {author} {\bibfnamefont
  {O.}~\bibnamefont {Schütt}}, \bibinfo {author} {\bibfnamefont
  {M.}~\bibnamefont {Strange}}, \bibinfo {author} {\bibfnamefont {K.~S.}\
  \bibnamefont {Thygesen}}, \bibinfo {author} {\bibfnamefont {T.}~\bibnamefont
  {Vegge}}, \bibinfo {author} {\bibfnamefont {L.}~\bibnamefont {Vilhelmsen}},
  \bibinfo {author} {\bibfnamefont {M.}~\bibnamefont {Walter}}, \bibinfo
  {author} {\bibfnamefont {Z.}~\bibnamefont {Zeng}},\ and\ \bibinfo {author}
  {\bibfnamefont {K.~W.}\ \bibnamefont {Jacobsen}},\ }\bibfield  {title}
  {\bibinfo {title} {The atomic simulation environment—a {Python} library for
  working with atoms},\ }\href {https://doi.org/10.1088/1361-648X/aa680e}
  {\bibfield  {journal} {\bibinfo  {journal} {Journal of Physics: Condensed
  Matter}\ }\textbf {\bibinfo {volume} {29}},\ \bibinfo {pages} {273002}
  (\bibinfo {year} {2017})}\BibitemShut {NoStop}%
\bibitem [{\citenamefont {Kapil}\ \emph {et~al.}(2019)\citenamefont {Kapil},
  \citenamefont {Rossi}, \citenamefont {Marsalek}, \citenamefont {Petraglia},
  \citenamefont {Litman}, \citenamefont {Spura}, \citenamefont {Cheng},
  \citenamefont {Cuzzocrea}, \citenamefont {Meißner}, \citenamefont {Wilkins},
  \citenamefont {Helfrecht}, \citenamefont {Juda}, \citenamefont {Bienvenue},
  \citenamefont {Fang}, \citenamefont {Kessler}, \citenamefont {Poltavsky},
  \citenamefont {Vandenbrande}, \citenamefont {Wieme}, \citenamefont
  {Corminboeuf}, \citenamefont {Kühne}, \citenamefont {Manolopoulos},
  \citenamefont {Markland}, \citenamefont {Richardson}, \citenamefont
  {Tkatchenko}, \citenamefont {Tribello}, \citenamefont {{Van Speybroeck}},\
  and\ \citenamefont {Ceriotti}}]{KAPIL2019214}%
  \BibitemOpen
  \bibfield  {author} {\bibinfo {author} {\bibfnamefont {V.}~\bibnamefont
  {Kapil}}, \bibinfo {author} {\bibfnamefont {M.}~\bibnamefont {Rossi}},
  \bibinfo {author} {\bibfnamefont {O.}~\bibnamefont {Marsalek}}, \bibinfo
  {author} {\bibfnamefont {R.}~\bibnamefont {Petraglia}}, \bibinfo {author}
  {\bibfnamefont {Y.}~\bibnamefont {Litman}}, \bibinfo {author} {\bibfnamefont
  {T.}~\bibnamefont {Spura}}, \bibinfo {author} {\bibfnamefont
  {B.}~\bibnamefont {Cheng}}, \bibinfo {author} {\bibfnamefont
  {A.}~\bibnamefont {Cuzzocrea}}, \bibinfo {author} {\bibfnamefont {R.~H.}\
  \bibnamefont {Meißner}}, \bibinfo {author} {\bibfnamefont {D.~M.}\
  \bibnamefont {Wilkins}}, \bibinfo {author} {\bibfnamefont {B.~A.}\
  \bibnamefont {Helfrecht}}, \bibinfo {author} {\bibfnamefont {P.}~\bibnamefont
  {Juda}}, \bibinfo {author} {\bibfnamefont {S.~P.}\ \bibnamefont {Bienvenue}},
  \bibinfo {author} {\bibfnamefont {W.}~\bibnamefont {Fang}}, \bibinfo {author}
  {\bibfnamefont {J.}~\bibnamefont {Kessler}}, \bibinfo {author} {\bibfnamefont
  {I.}~\bibnamefont {Poltavsky}}, \bibinfo {author} {\bibfnamefont
  {S.}~\bibnamefont {Vandenbrande}}, \bibinfo {author} {\bibfnamefont
  {J.}~\bibnamefont {Wieme}}, \bibinfo {author} {\bibfnamefont
  {C.}~\bibnamefont {Corminboeuf}}, \bibinfo {author} {\bibfnamefont {T.~D.}\
  \bibnamefont {Kühne}}, \bibinfo {author} {\bibfnamefont {D.~E.}\
  \bibnamefont {Manolopoulos}}, \bibinfo {author} {\bibfnamefont {T.~E.}\
  \bibnamefont {Markland}}, \bibinfo {author} {\bibfnamefont {J.~O.}\
  \bibnamefont {Richardson}}, \bibinfo {author} {\bibfnamefont
  {A.}~\bibnamefont {Tkatchenko}}, \bibinfo {author} {\bibfnamefont {G.~A.}\
  \bibnamefont {Tribello}}, \bibinfo {author} {\bibfnamefont {V.}~\bibnamefont
  {{Van Speybroeck}}},\ and\ \bibinfo {author} {\bibfnamefont {M.}~\bibnamefont
  {Ceriotti}},\ }\bibfield  {title} {\bibinfo {title} {i-pi 2.0: A universal
  force engine for advanced molecular simulations},\ }\href
  {https://doi.org/https://doi.org/10.1016/j.cpc.2018.09.020} {\bibfield
  {journal} {\bibinfo  {journal} {Computer Physics Communications}\ }\textbf
  {\bibinfo {volume} {236}},\ \bibinfo {pages} {214} (\bibinfo {year}
  {2019})}\BibitemShut {NoStop}%
\bibitem [{\citenamefont {Toraman}\ \emph
  {et~al.}(2025{\natexlab{a}})\citenamefont {Toraman}, \citenamefont
  {Fauconnier},\ and\ \citenamefont {Verstraelen}}]{toraman_stable_2025}%
  \BibitemOpen
  \bibfield  {author} {\bibinfo {author} {\bibfnamefont {G.}~\bibnamefont
  {Toraman}}, \bibinfo {author} {\bibfnamefont {D.}~\bibnamefont
  {Fauconnier}},\ and\ \bibinfo {author} {\bibfnamefont {T.}~\bibnamefont
  {Verstraelen}},\ }\bibfield  {title} {{\selectlanguage {en}\bibinfo {title}
  {{STable} {AutoCorrelation} {Integral} {Estimator}: {Robust} and {Accurate}
  {Transport} {Properties} from {Molecular} {Dynamics} {Simulations}}},\ }\href
  {https://doi.org/10.1021/acs.jcim.5c01475} {\bibfield  {journal} {\bibinfo
  {journal} {Journal of Chemical Information and Modeling}\ }\textbf {\bibinfo
  {volume} {65}},\ \bibinfo {pages} {10445} (\bibinfo {year}
  {2025}{\natexlab{a}})}\BibitemShut {NoStop}%
\bibitem [{\citenamefont {Toraman}\ \emph
  {et~al.}(2025{\natexlab{b}})\citenamefont {Toraman}, \citenamefont
  {Fauconnier},\ and\ \citenamefont {Verstraelen}}]{toraman2025stable}%
  \BibitemOpen
  \bibfield  {author} {\bibinfo {author} {\bibfnamefont {G.}~\bibnamefont
  {Toraman}}, \bibinfo {author} {\bibfnamefont {D.}~\bibnamefont
  {Fauconnier}},\ and\ \bibinfo {author} {\bibfnamefont {T.}~\bibnamefont
  {Verstraelen}},\ }\bibfield  {title} {\bibinfo {title} {Stable
  autocorrelation integral estimator (stacie): Robust and accurate transport
  properties from molecular dynamics simulations},\ }\href@noop {} {\bibfield
  {journal} {\bibinfo  {journal} {arXiv preprint arXiv:2506.20438}\ } (\bibinfo
  {year} {2025}{\natexlab{b}})}\BibitemShut {NoStop}%
\bibitem [{\citenamefont {Togo}\ \emph {et~al.}(2023)\citenamefont {Togo},
  \citenamefont {Chaput}, \citenamefont {Tadano},\ and\ \citenamefont
  {Tanaka}}]{togo_implementation_2023}%
  \BibitemOpen
  \bibfield  {author} {\bibinfo {author} {\bibfnamefont {A.}~\bibnamefont
  {Togo}}, \bibinfo {author} {\bibfnamefont {L.}~\bibnamefont {Chaput}},
  \bibinfo {author} {\bibfnamefont {T.}~\bibnamefont {Tadano}},\ and\ \bibinfo
  {author} {\bibfnamefont {I.}~\bibnamefont {Tanaka}},\ }\bibfield  {title}
  {\bibinfo {title} {Implementation strategies in phonopy and phono3py},\
  }\href {https://doi.org/10.1088/1361-648X/acd831} {\bibfield  {journal}
  {\bibinfo  {journal} {Journal of Physics: Condensed Matter}\ }\textbf
  {\bibinfo {volume} {35}},\ \bibinfo {pages} {353001} (\bibinfo {year}
  {2023})}\BibitemShut {NoStop}%
\end{thebibliography}
\end{document}